\documentclass[twocolumn,superscriptaddress,aps,prb,floatfix]{revtex4-2}
\usepackage[T1]{fontenc}
\usepackage{amsfonts}
\usepackage{comment}
\usepackage{amsmath,amssymb,epsfig,color}
\usepackage{float}
\usepackage{amsmath}
\usepackage{amssymb}
\usepackage{tabularx}
\usepackage{url}
\usepackage{graphicx}
\usepackage{dcolumn}
\usepackage{bbold}
\usepackage{physics}
\usepackage{multirow}
\usepackage{pifont}
\usepackage{siunitx}
\usepackage[normalem]{ulem}

\usepackage{standalone}

\newcommand{\xmark}{\text{\ding{55}}}
\usepackage{hyperref}
\begin{document}

\title{Tuning of altermagnetism by strain}

\author{M. Khodas}
\affiliation{Racah Institute of Physics, Hebrew University of Jerusalem, Jerusalem 91904, Israel}

\author{Sai Mu}
\affiliation{Department of Physics and Astronomy and SmartState Center for Experimental Nanoscale Physics, University of South Carolina, Columbia, South Carolina 29208, USA}

\author{I. I. Mazin}
\affiliation{Department of Physics and Astronomy, George Mason University, Fairfax, Virginia 22030, USA}
\affiliation{Quantum Science and Engineering Center, George Mason University, Fairfax, Virginia 22030, USA}
\author{K. D. Belashchenko}
\affiliation{Department of Physics and Astronomy and Nebraska Center for Materials and Nanoscience, University of Nebraska-Lincoln, Lincoln, Nebraska 68588, USA}
\date{\today }

\begin{abstract}
For all collinear altermagnets, we sort out piezomagnetic free-energy invariants allowed in the nonrelativistic limit and relativistic piezomagnetic invariants bilinear in the N\'eel vector $\mathbf{L}$ and magnetization $\mathbf{M}$, which include strain-induced Dzyaloshinskii-Moriya interaction. The symmetry-allowed responses are fully determined by the nonrelativistic spin Laue group. In the nonrelativistic limit, two distinct mechanisms are discussed: the band-filling mechanism, which exists in metals and is illustrated using the simple two-dimensional Lieb lattice model, and the temperature-dependent exchange-driven mechanism, which is illustrated using first-principles calculations for transition-metal fluorides. The leading second-order nonrelativistic term in the strain-induced magnetization is also obtained for CrSb. Piezomagnetism due to the strain-induced Dzyaloshinskii-Moriya interaction is calculated from first principles for transition-metal fluorides, MnTe, and CrSb.
Finally, we discuss triplet superconducting correlations supported by altermagnets and protected by inversion rather than time-reversal symmetry. We apply the nonrelativistic classification of Cooper pairs to describe the interplay between strain and superconductivity in the two-dimensional Lieb lattice and in bulk rutile structures. We show that triplet superconductivity is, on average, unitary in an unstrained altermagnet, but becomes non-unitary under piezomagnetically active strain.
\end{abstract}
\date{\today}

\maketitle

\section{Introduction}
\label{sec:Intro}
Altermagnetism is a novel type of magnetic order that shares some properties with ferromagnets and some with antiferromagnets 
(see reviews \cite{Smejkal2022a,Mazin2022,Bai2024,Tamang2024} and references therein).
Similar to antiferromagnets, an altermagnet has zero net magnetization ($\mathbf{M}$) in the nonrelativistic limit.
At the same time, altermagnets exhibit a finite anomalous Hall effect normally observed in ferromagnets.
This implies that the magnetic symmetry group does not include elements combining time reversal $\mathcal{T}$ with inversion $\mathcal{P}$ or with spatial translation $T_\mathbf{t}$ by some vector $\mathbf{t}$.

The breaking of $\mathcal{T}$ and $\mathcal{T} \mathcal{P}$ symmetries gives rise to a finite spin splitting even in the absence of spin-orbit coupling (SOC), similar to ferromagnets.
The spin texture alternates in both real and reciprocal space and is controlled by the symmetries combining $\mathcal{T}$ with real-space rotations and reflections. 

Tuning of altermagnets by strain \cite{Junwei2021,Aoyama2024,Chakraborty2024a,Belashchenko2025,Takahashi2025,Karetta2025} can enable useful functionalities for spintronic applications, such as the ability to manipulate and switch altermagnetic domains via magnetostrictive or piezomagnetic Zeeman coupling. The structure of the third-rank linear piezomagnetic response tensor is determined by the magnetic point group of the material and fully tabulated \cite{Birss1964}. 
Upon contraction with the Levi-Civita symbol, the same tensor structure describes strain-induced anomalous Hall conductivity \cite{Takahashi2025}. 
However, magnetic point groups do not specifically distinguish altermagnets, which are defined \cite{Smejkal2022} by their nonrelativistic spin groups \cite{Litvin1974,Litvin:a14103}. It is, therefore, of interest to determine what can be learned about strain-induced magnetization in altermagnets and its mechanisms from their spin group structure and to connect such symmetry considerations with specific materials via first-principles calculations. 

The distinctive spin-polarization pattern of altermagnetic electronic bands in reciprocal space has nontrivial implications for superconductivity, particularly for the symmetry of Cooper pair wave functions~\cite{Mazin2022a}. 
The electron energy, $E_\sigma(\mathbf{k})$, is determined by the Bloch momentum $\mathbf{k}$ and the spin projection $\sigma$ along the N\'eel vector $\mathbf{L}$. 
In the nonrelativistic limit, the combination of time-reversal symmetry $\mathcal{T}$ and a twofold spin rotation around any axis orthogonal to $\mathbf{L}$ constitutes a symmetry operation that ensures the band dispersion is even in momentum: $E_\sigma(\mathbf{k}) = E_\sigma(-\mathbf{k})$~\cite{Smejkal2022}. 
Equal-spin triplet Cooper pairs are protected by this symmetry. 
On the other hand, the broken $\mathcal{T}$ symmetry leads to $E_\sigma(\mathbf{k}) \ne E_{-\sigma}(-\mathbf{k})$, which is detrimental to conventional singlet pairing. 
Unconventional singlet superconductivity has been proposed theoretically in multi-orbital altermagnets~\cite{Bose2024}, and may also coexist with altermagnetism in the form of finite-momentum pairing~\cite{Sumita2023,Chakraborty2024}. 
Moreover, spin--orbit coupling (SOC) may induce a mixed singlet--triplet order parameter that transforms nontrivially under $\mathcal{T}$~\cite{Carvalho2024}. 
Here, we present a nonrelativistic symmetry classification of Cooper pairs that emphasizes the role of lattice degrees of freedom, and apply it to study the effect of strain.

Piezomagnetically active strain strongly affects equal-spin Cooper pairs supported by an altermagnet. 
To describe the interplay between strain and triplet superconductivity, one must classify equal-spin triplets in terms of the nonrelativistic spin symmetry, assuming negligible SOC. 
The spin-only symmetry of spin rotations around the N\'eel vector $\mathbf{L}$ often plays a limited role in constructing effective tight-binding models \cite{Brekke2023,Roig2024,Parshukov2024,Attias2024,Rao2024,Antonenko2025}. 
However, in the triplet symmetry sector it distinguishes two sets of oppositely polarized Cooper pairs and necessitates the introduction of a multi-component superconducting order parameter. Physically, these two sets of triplets reside on distinct sublattices in real space, and the sublattice degree of freedom is essential for their classification. This degree of freedom is also crucial for describing the optical response of altermagnets \cite{Vila2024} and the formation of in-gap bound states at local impurity sites \cite{Gondolf2025}. Its relevance in the context of superconductivity has been recently emphasized \cite{Chakraborty2024}. Spin symmetry considerations allow us to describe the coupling of strain, as well as longitudinal and transverse magnetization, to the superconducting order parameter.

The rest of this paper is organized as follows.
In Section \ref{sec:nonrel} we consider piezomagnetism in altermagnets in the nonrelativistic limit. First, in Section \ref{sec:symmetry} we derive the lowest-order free-energy invariants allowed by all altermagnetic spin groups. Then, in Sections \ref{sec:Lieb} and \ref{sec:TD-fluorides} we describe two different mechanisms contributing to this response: the band filling effect, which is active in altermagnetic metals, and the temperature-dependent exchange-driven contribution, which is also present in insulators. The band-filling effect is illustrated using the minimal model for the Lieb-lattice altermagnet, and the exchange-driven mechanism is studied using first-principles calculations for transition-metal fluorides. We also calculate the leading (quadratic) order strain-induced magnetization in CrSb, which is a $g$-wave altermagnetic metal, in Section \ref{sec:CrSb-nonlinear}.
Further, in Section \ref{sec:rel-piezo} we consider relativistic effects. In Section \ref{rel-symmetry} we tabulate the strain-induced bilinear $L$-$M$ invariants allowed in all altermagnetic spin groups, and in Section \ref{sec:strain_DMI} we calculate the allowed transverse piezomagnetic coefficients, due to strain-induced Dzyaloshinskii-Moriya interaction (DMI), in transition-metal fluorides, MnTe, and CrSb.
Lastly, in Section \ref{sec:superconductivity} we discuss the ramifications of strain-induced magnetization for the hypothetical superconducting altermagnets. We show that while an altermagnet can only support non-unitary Cooper pairs (except at isolated lines or points), the superconducting order parameter integrated over the entire Fermi surface is unitary. This curious feature disappears under strain, and the superconductor becomes fully non-unitary. Section \ref{sec:conclusions} concludes the paper.

Some of the results of Sections \ref{sec:TD-fluorides} and \ref{sec:strain_DMI} have been previously reported in a dissertation \cite{Mu-thesis}.

\section{Piezomagnetism in the nonrelativistic limit}
\label{sec:nonrel}

\subsection{Symmetry considerations}
\label{sec:symmetry}

Here we analyze the piezomagnetic response in collinear altermagnets in the nonrelativistic limit from the symmetry point of view.
In this limit, the direction of the  N\'eel vector $\mathbf{L}$ is not fixed relative to the lattice, and symmetry operations act differently on the lattice and spin degrees of freedom, forming the so-called spin group \cite{Litvin1974,Chen2024,Jiang2024,Xiao2024}.
A generic element of the spin point group has a form
$g = [g_s||g_l]$, where $g_s$ and $g_l$ act on the spin and lattice degrees of freedom, respectively.

Because the strain tensor is invariant under spatial inversion, the presence of an inversion center has no effect on the allowed nonrelativistic piezomagnetic terms. Therefore, it is sufficient to consider nontrivial spin Laue groups (SLG), which are obtained by adding inversion to the nontrivial spin group $\mathbf{R}_s$.

The nontrivial spin Laue groups supporting collinear altermagnetism have the structure
\begin{align}\label{eq:sg13}
    \mathbf{R}_s &= [E||\mathbf{H}]+[C_{2\hat{\mathbf{L}}_\perp}||\mathcal{A}][E||\mathbf{H}] \nonumber\\&= [E||\mathbf{H}]+[C_{2\hat{\mathbf{L}}_\perp}||\mathcal{A}\mathbf{H}]\, ,
\end{align}
where the site group $\mathbf{H}$ does not exchange the two sublattices.
It contains half of the full crystallographic Laue group of a given crystal.
The other half can be written as $\mathcal{A} \mathbf{H}$, where 
$\mathcal{A}$ maps one sublattice to another.
The spin group of the structure \eqref{eq:sg13} is known as a non-trivial, or third type magnetic group \cite{Lifshitz2005}.

In the present context, the spin-flip operation $C_{2\hat{\mathbf{L}}_\perp}$ is an antisymmetry~\cite{Heesch1930,Shubnikov1951}. Therefore, following Ref. \cite{Litvin:a14103,Smejkal2022}, we use a compact notation in which the elements $[E||h]$ are denoted ${}^1h$, and $[C_{2\hat{\mathbf{L}}_\perp}||g_l]$ (where $g_l=\mathcal{A}h$) are denoted ${}^2g_l$.
For example, a fourfold rotation is labeled ${}^14$ if it leaves the magnetic sublattices invariant or ${}^24$ if it interchanges them. All nontrivial spin Laue groups (SLG) are listed in Tab.~\ref{tab:nrpiezo} using this notation.

\begin{table*}[htb]
\caption{Leading order functions $f_{\mathrm{P}}(\boldsymbol{\varepsilon})$ allowed in the nonrelativistic free energy invariants $\Delta F = \mathbf{L}\cdot \mathbf{M} f_{\mathrm{P}}(\boldsymbol{\varepsilon})$.}
\begin{tabular}{|l|l|l|}
    \hline
         Type & SLG & Prefactors in the free energy invariants \\
         \hline
         \multirow{4}{*}{$d$-wave} &
         ${}^2m_z{}^2m_y{}^1m_x$ \rule{0pt}{2.3ex} & $\varepsilon_{yz}$\\
         & ${}^24/{}^1m$ & $\varepsilon_{xx}-\varepsilon_{yy}$, $\varepsilon_{xy}$\\
         & ${}^24/{}^1m{}^2m_y{}^1m_d$ & $\varepsilon_{xy}$ \\
         & ${}^22/{}^2m$ & $\varepsilon_{xz}$, $\varepsilon_{yz}$\\
\hline
         \multirow{4}{*}{$g$-wave} & ${}^14/{}^1m{}^2m{}^2m$ \rule{0pt}{2.3ex} & $\varepsilon_{xy}(\varepsilon_{xx}-\varepsilon_{yy})$ \\
         & ${}^1\bar3{}^2m_y$ & $(\varepsilon_{xx}-\varepsilon_{yy})\varepsilon_{yz}+2\varepsilon_{xy}\varepsilon_{xz}$ \\
         & ${}^26/{}^2m$ & $(\varepsilon_{xx}-\varepsilon_{yy})\varepsilon_{yz}+2\varepsilon_{xy}\varepsilon_{xz}$, $(\varepsilon_{xx}-\varepsilon_{yy})\varepsilon_{xz}-2\varepsilon_{xy}\varepsilon_{yz}$ \\
         & ${}^26/{}^2m{}^2m_y{}^1m_x$ & $(\varepsilon_{xx}-\varepsilon_{yy})\varepsilon_{yz}+2\varepsilon_{xy}\varepsilon_{xz}$ \\
\hline
        \multirow{2}{*}{$i$-wave} & ${}^16/{}^1m{}^2m{}^2m$ \rule{0pt}{2.3ex} & $\varepsilon_{xy}[3(\varepsilon_{xx}-\varepsilon_{yy})^2-4\varepsilon_{xy}^2]$, $\varepsilon_{xy}(\varepsilon_{yz}^2-\varepsilon_{xz}^2)+(\varepsilon_{xx}-\varepsilon_{yy})\varepsilon_{xz}\varepsilon_{yz}$\\
         & ${}^1m{}^1\bar3{}^2m$ & $(\varepsilon_{xx}-\varepsilon_{yy})(\varepsilon_{yy}-\varepsilon_{zz})(\varepsilon_{zz}-\varepsilon_{xx})$, $(\varepsilon_{xx}-\varepsilon_{yy})\varepsilon_{xy}^2+(\varepsilon_{yy}-\varepsilon_{zz})\varepsilon_{yz}^2+(\varepsilon_{zz}-\varepsilon_{xx})\varepsilon_{xz}^2 $\\
\hline
\end{tabular}
\label{tab:nrpiezo}
\end{table*}

Piezomagnetism arises due to the contribution $\Delta F$ to the free energy that is linear in $\mathbf{M}$ at finite strain tensor $\boldsymbol{\varepsilon}$.
Due to the spin-only symmetry, the magnetization enters the free energy as a scalar product: $\Delta F = \mathbf{L}\cdot \mathbf{M} f_{\mathrm{P}}(\boldsymbol{\varepsilon})$.
At zero strain, $\Delta F$ changes sign under the $[C_{2\hat{\mathbf{L}}_\perp}||\mathcal{A}]$ operation.
The scalar product $\mathbf{L}\cdot \mathbf{M}$ is invariant under any global spin rotation such as $[C_{2\hat{\mathbf{L}}_\perp}||E]$.
Since the space operations are decoupled from spins, $\mathbf{L}\cdot \mathbf{M}$ is invariant under spatial transformations forming $\mathbf{H}$ and changes sign under $\mathcal{A} \mathbf{H}$.
This defines the one-dimensional representation $\Gamma_{\mathbf{L}}$ of the crystallographic point group, $\mathbf{H}+\mathcal{A}\mathbf{H}$.
As $\Gamma_{\mathbf{L}}$ is not trivial, $\Delta F$ is not allowed at zero strain \cite{McClarty2024}.
The $\Gamma_{\mathbf{L}}$ also fixes the momentum space symmetry of non-relativistic spin splitting \cite{Roig2024}.
Therefore, to the leading order in strain, $f_{\mathrm{P}}(\boldsymbol{\varepsilon})$ parametrizing the piezomagnetic response is determined by the altermagnetic spin-momentum correlation pattern: linear in $d$-wave, quadratic in $g$-wave, and cubic in $i$-wave altermagnets.

Table \ref{tab:nrpiezo} lists the nonrelativistic piezomagnetic terms for all altermagnetic SLG, to the leading  order in strain. 
The strain components inducing finite magnetization form combinations that are odd under the $[C_{2\hat{\mathbf{L}}_{\perp}}||\mathcal{A}]$ symmetry, effectively disconnecting the two sublattices.
Therefore, such strain components lift the band degeneracies at high-symmetry momenta protected by the spin symmetries at zero strain. 
In particular, the splitting at the $\Gamma$-point, $\Delta E_\Gamma$ is generally expected.
One can view $\Delta E_\Gamma$ as an alternative measure of piezomagnetism in collinear altermagnets.

In metallic systems, a possible strain-induced inequivalence of the two magnetic sublattices automatically results in a net magnetization in the ground state. This band-filling mechanism of piezomagnetism is illustrated in the next section for the simple two-dimensional Lieb lattice model. 

In insulators this band-filling effect is not possible. However, the sublattice inequivalence also results in the difference in intra-sublattice exchange interaction for the two sublattices. This difference causes a temperature-dependent exchange-driven contribution to the longitudinal piezomagnetic response \cite{Mu-thesis,Consoli2021,Yershov2024}. This mechanism is similar to the exchange-driven longitudinal magnetoelectric effect \cite{Mostovoy2010,Mu2014}, where the inequivalence of the magnetic sublattices is induced by the electric field instead of strain. This mechanism is illustrated in Section \ref{sec:TD-fluorides} for transition-metal fluorides.

\subsection{Band filling effect: Metallic Lieb lattice}
\label{sec:Lieb}

In this section we consider the band filling mechanism of piezomagnetism in metallic or semiconducting altermagnets.
The magnetization in this case is due to the spins of itinerant electrons in a metallic altermagnet. 
The $\mathbf{C}_{\infty}$ makes the spin projection of an electron, $\sigma$ on $\mathbf{L}$, a good quantum number \cite{Smejkal2022}. 
The electronic bands are either spin up ($\uparrow$) or spin down ($\downarrow$) for  $\sigma =\pm 1/2$, respectively.
As the strain breaks the $[C_{2\hat{\mathbf{L}}_{\perp}}||\mathcal{A}]$ symmetry the two spin populations, $n_\uparrow$ and $n_\downarrow$ become imbalanced with the resulting magnetization, 
$\mathbf{M} = \mu_B \hat{\mathbf{L}} (n_\uparrow - n_\downarrow)$, where $\mu_B$ is the Bohr magneton.

For multi-valley altermagnets, the strain-induced magnetization arising from band filling is expressed as a contraction of the strain and deformation potential tensors~\cite{Junwei2021}. This band-filling contribution to the magnetization, which originates from all occupied states, depends on microscopic details and is non-universal. 
Here we illustrate this mechanism using a two-band, single-valley, two-dimensional Lieb lattice model.

\begin{figure}[htb]
\begin{center}
\centering
     \includegraphics[width=1.0\columnwidth]{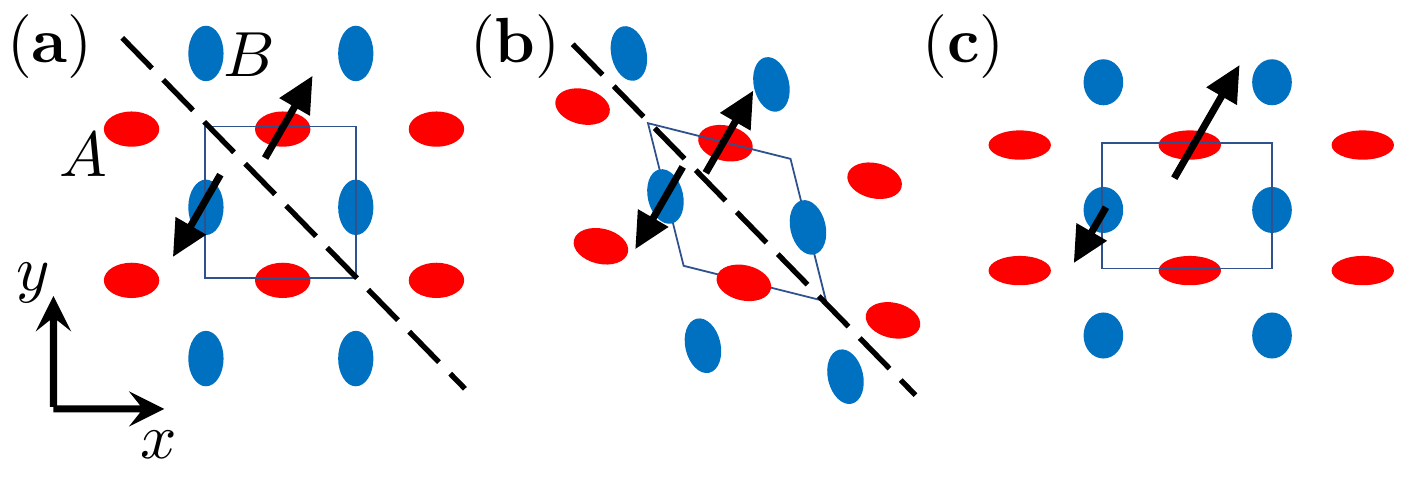}
     \caption{(a) The 2D Lieb lattice model of a $d$-wave altermagnet. 
     The horizontal (red) [vertical (blue)] ovals form an A (B) sublatice connected by $^2m_d$ symmetry operation. 
     The localized magnetic moments satisfy $\mathbf{m}_A = - \mathbf{m}_B$.
     The SLG is ${}^24/{}^1m{}^2m_d{}^1m_y$. (b) The application of the $[110]$ preserves the equivalence of the two sublattices.
     The  $^2m_d$ mirror (dashed) remains the symmetry.
     (c) The $[100]$ strain makes the two sublattices inequivalent, causing the transition to the ferrimagnetic state.
     }
         \label{fig:Lieb}
         \end{center}
\end{figure}

The Lieb lattice shown in Fig.~\ref{fig:Lieb} is an example of a two-dimensional altermagnet with reciprocal space group 6 corresponding to the layer groups 53-64, according to \cite{Zeng2024}.
This model is directly relevant to oxyselenide altermagnets \cite{Junwei2021,Cui2023,KV2Se2O,RbV2Te2O}.
To study piezomagnetism for this lattice, we use
the minimal tight binding model of Ref.~\cite{Antonenko2025} suitably modified by strain.

The Lieb lattice has SLG ${}^24/{}^1m{}^1m{}^2m$ which we adopt in the setting ${}^24/{}^1m{}^2m_d{}^1m_y$ (see Fig.~\ref{fig:Lieb}), obtained by a $\pi/4$ rotation from ${}^24/{}^1m{}^2m_y{}^1m_d$ listed in Table~\ref{tab:nrpiezo}. This is a $d$-wave altermagnet; with the chosen reference frame, the spin-momentum locking pattern is $d_{x^2-y^2}$.
With the $\pi/4$ rotation taken into account, we read from Table \ref{tab:nrpiezo} 
that the nonrelativistic strain-induced magnetization has the form 
$\mathbf{M} = \Lambda \hat{\mathbf{L}} ( \varepsilon_{xx} - \varepsilon_{yy})$, where $\Lambda$ is a constant. This piezomagnetic effect is illustrated in Fig.~\ref{fig:Lieb}c.
Physically, it arises because the $\varepsilon_{xx} - \varepsilon_{yy}$ strain component breaks the equivalence of the two magnetic sublattices. In contrast, as seen in Fig.~\ref{fig:Lieb}b, the $\varepsilon_{xy}$ strain preserves the $[C_{2\hat{\mathbf{L}}_{\perp}}||\mathcal{A}]={}^2m_d$ mirror planes that exchange the two sublattices.

\begin{figure}[htb]
\begin{center}
\centering
     \includegraphics[width=1.0\columnwidth]{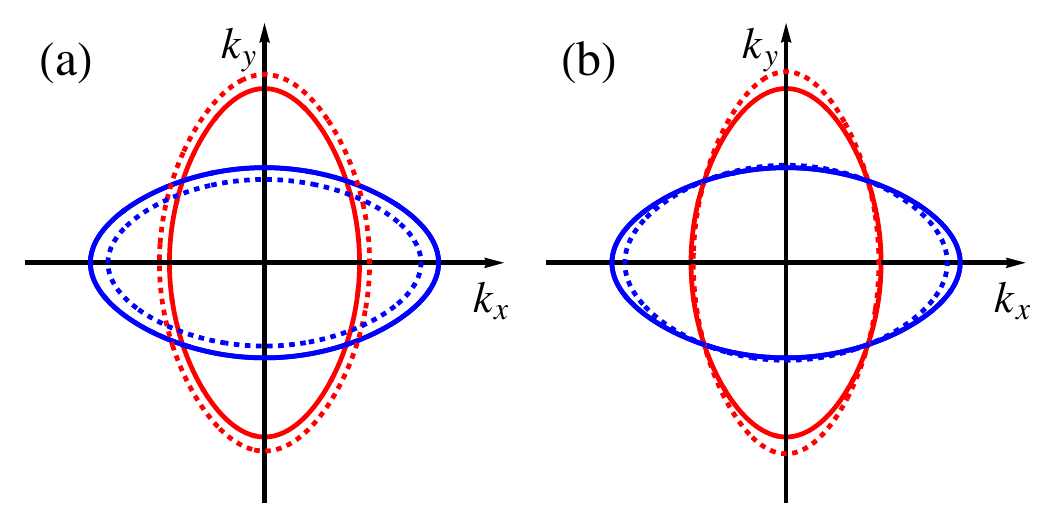}
     \caption{Fermi contours for 
     $E^1_{1/2}(\mathbf{k})$ (red) and $E^{-1}_{-1/2}(\mathbf{k})$ (blue) of the 2D Lieb model defined by Eq. \eqref{eq:Lieb2} with $t_2=0.5$ and $t_{2d}=0.25$ a.u.
     Unequal areas of the $\sigma=\pm1/2$ pockets leads to net magnetization.
     Solid lines correspond to zero strain ($\bar{\delta} t = \delta t_2 =0$).
     The two pockets are related to each other by the non-trivial spin symmetry ${}^2m_d$, which protects the spin degeneracy along the $k_x = \pm k_y$ lines.
     Dashed curves correspond to a small $\varepsilon_{xx}-\varepsilon_{yy}$ strain (see Fig.~\ref{fig:Lieb}c).
     (a) Band splitting effect with $\bar{\delta} t = 0.025$, $\delta t_2 =0$.
     (b) Strain-induced effective mass anisotropy with $\bar{\delta} t = 0$, $\delta t_2 =0.035$.}
         \label{fig:LiebPockets}
         \end{center}
\end{figure}

We will illustrate the effect using a generalization of the minimal model of Ref. ~\onlinecite{Antonenko2025}.
At a finite strain $\varepsilon/2 = \varepsilon_{xx} = -\varepsilon_{yy}$, the original square lattice with the basis $\mathbf{a}_{x}= (1,0)$, $\mathbf{a}_{y}= (0,1)$ is deformed into a rectangular lattice, with the basis,  $\mathbf{a}'_{x}= (1+\varepsilon/2 ,0)/\sqrt{1-(\varepsilon/2)^2}$, $\mathbf{a}'_{y}= (0,1-\varepsilon/2)/\sqrt{1-(\varepsilon/2)^2}$.
We formulate the effective band Hamiltonian, $H_b$ as a function of the components, $k'_{x,y}$ of a Bloch vector $\mathbf{k}$ in the basis modified by strain, $\mathbf{b}'_{x}= (1-\varepsilon/2 ,0)/\sqrt{1-(\varepsilon/2)^2}$, $\mathbf{b}'_{y}= (0,1+\varepsilon/2)/\sqrt{1-(\varepsilon/2)^2}$ formally restoring the Brillouin zone to a square.
As a result, we obtained an extension of the model of Ref.~\cite{Antonenko2025} in the presence of strain:
\begin{align}\label{eq:Hk_Lieb}
    H_b &(\mathbf{k}')  = - 4 t_1 \cos \frac{k'_x}{2} \cos \frac{k'_y}{2} \tau_x +  \frac{J}{2} \mathbf{L}\cdot\boldsymbol{\sigma} \tau_z 
    \notag \\
 + & \frac{1}{2}(1 + \tau_z)[(t_{2b} + \delta t_b)\cos k'_y +  (t_{2a} + \delta t_a)\cos k'_x]
\notag \\
 + &
\frac{1}{2}(1 - \tau_z)[(t_{2a} - \delta t_a)\cos k'_y +  (t_{2b} - \delta t_b)\cos k'_x]\, ,
\end{align}
where the Pauli matrices $\boldsymbol{\tau}$ and $\boldsymbol{\sigma}$ operate in the sublattice and spin spaces, respectively.
The $t_1$ is the nearest neighbour hopping amplitude, the $t_{2a}$ and $t_{2b}$ are the two types of the next nearest neighbour hopping amplitudes in the unstrained lattice.
Application of strain changes these amplitudes by $\delta t_{a,b} \propto \varepsilon$, respectively.
In what follows, we change the notation of the rescaled momenta $\mathbf{k}'$ to $\mathbf{k}$ omitting the prime for clarity.

To simplify the presentation we will assume that the exchange splitting is the dominant energy scale, 
$J  \gg t_{1}$. 
In this regime, the nearest neighbor hopping can be neglected.
Let us rearrange the remaining Hamiltonian \eqref{eq:Hk_Lieb} 
\begin{align}\label{eq:Hk_Lieb1}
    H(\mathbf{k}) & =  \frac{J}{2} \mathbf{L}\cdot\boldsymbol{\sigma} \tau_z 
 +[ t_2 + \tau_z \bar{\delta} t] (\cos k_x + \cos k_y)
  \notag \\
 & +[\tau_z t_{2d} + \delta t_2](\cos k_x - \cos k_y) 
\end{align}
using the notations, $t_2 = (t_{2a} + t_{2b})/2$, $t_{2d} = (t_{2a} - t_{2b})/2$, 
$\bar{\delta} t = (\delta t_{2a} + \delta t_{2b})/2$, and $\delta t_2 = (\delta t_{2a} - \delta t_{2b})/2$.
The four bands given by the Hamiltonian, \eqref{eq:Hk_Lieb1} are
\begin{align}\label{eq:Lieb2}
    E^\alpha_{\sigma}(\mathbf{k}) = & \frac{J}{2}L \sigma \alpha + [ t_2 + \alpha \bar{\delta} t] (\cos k_x + \cos k_y)
    \notag \\
    &+[\alpha t_{2d} + \delta t_2](\cos k_x - \cos k_y) \, ,
\end{align}
where $\alpha = \pm 1$ for the bands residing at $A$ and $B$ sublattices.
At zero strain the bands $ E^\alpha_{\sigma}$ and $ E^{-\alpha}_{-\sigma}$ are degenerate at the symmetry protected lines, $k_x = \pm k_y$, see Fig~\ref{fig:LiebPockets}.
Strain lifts the degeneracy and causes spin splitting, $\Delta E_\Gamma  = 4 \bar{\delta} t$.
In addition, strain introduces the inequivalence of the effective mass tensor for the two spin split bands.
This effect scales with $\delta t_2$.
As the Fermi pockets shrink we expect the magnetization to follow mostly from the spin splitting, while $\delta t_2$ has nearly no effect.

The strain-induced magnetization can be easily estimated for a small hole pockets at $\Gamma$ point formed by two pairs of bands, Eq.~\eqref{eq:Lieb2}.
For definiteness we consider $E^1_{1/2}(\mathbf{k})$ and $E^{-1}_{-1/2}(\mathbf{k})$ pair of bands, and assume that other two bands do not cross the Fermi surface.
At zero strain, the two bands are degenerate at $\Gamma$, $E^{\pm 1}_{\pm 1/2}(\mathbf{k}=0) = J L/2 + 2 t_2 = E_{\mathrm{top}}$.
The hole pockets is small if the chemical potential, $\mu$ is slightly below the top of the band to make the Fermi energy, $E_F = E_{\mathrm{top}}-\mu$ sufficiently small, $E_F \ll t_2$. 
In this limit, we expand the dispersion relation \eqref{eq:Lieb2} to second order in $\mathbf{k}$. 
The Fermi pockets are elliptical, as shown in Fig.~\ref{fig:LiebPockets}, and the net magnetization is determined by the difference in the areas enclosed by the two Fermi contours:
\begin{align}\label{eq:M_Lieb}
   \mathbf{M}=  \mu_B \hat{\mathbf{L}} \left[\frac{2 \bar{\delta} t }{\pi (t_2^2 -t_{2d}^2)^{1/2}}
+   \frac{E_F (t_{2d} \delta t_2 - t_2 \bar{\delta}t)}{\pi (t_2^2 -t_{2d}^2)^{3/2}} \right]\, .
\end{align}
This expression holds at $E_F \ll t_2$ and far from the Lifshitz transition, $t_{2d} \ll t_{2}$.
Under these conditions, the Fermi contours look as shown in Fig.~\ref{fig:LiebPockets}a, and the induced magnetization is dominated by the strain-induced band splitting, 
$M \approx \Delta E_\Gamma \, \hat{\mathbf{L}} / (2\pi t_2)$, similarly to the two-valley model considered in Ref.~\onlinecite{Junwei2021}. 
In general, however, the second term in Eq.~\eqref{eq:M_Lieb} includes an additional contribution arising from the strain-induced effective mass anisotropy proportional to $\delta t_{2d}$, which manifests itself as shown in Fig.~\ref{fig:LiebPockets}b.

\begin{figure*}[htb]
    \includegraphics[width=0.85\textwidth]{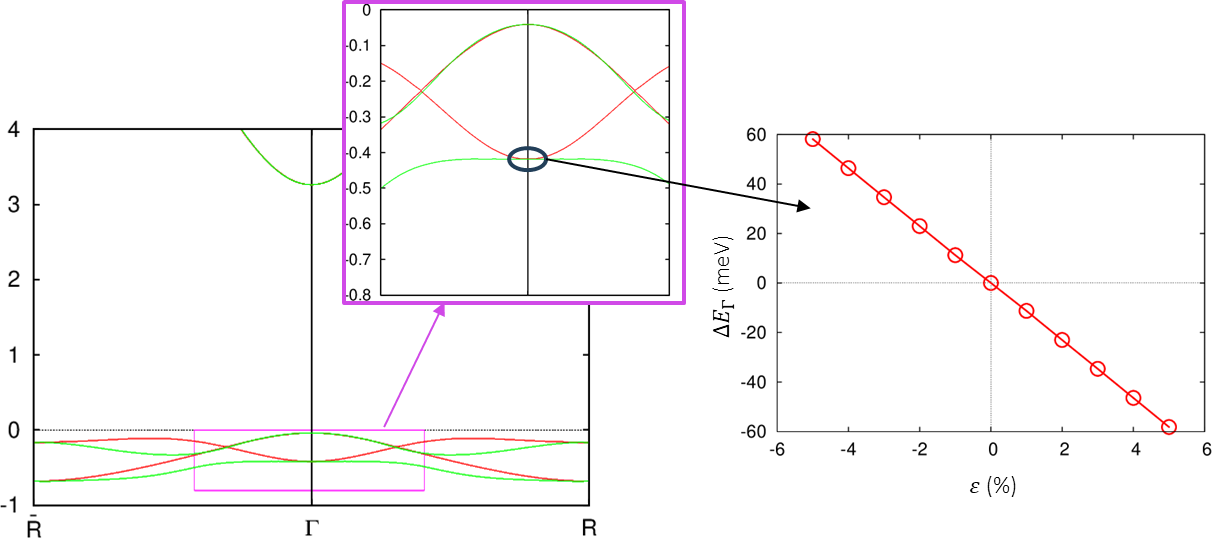}
    \caption{Band structure and spin splitting at the $\Gamma$ point ($\Delta E_\Gamma$) induced by the shear strain $\varepsilon_{xy}=\varepsilon_{yx}=\varepsilon/2$ in MnF$_2$.}
    \label{fig:MnFe-split}
\end{figure*}

The nonrelativistic band-filling mechanism of piezomagnetism described in this section is inactive in insulating altermagnets. However, strain-induced splittings of the Kramers doublets at the $\Gamma$ point are described by the same invariants, listed in Table \ref{tab:nrpiezo}, as the magnetization in the metallic case. For example, in rutile MnF$_2$, which has the same ${}^24/{}^1m{}^2m_d{}^1m_y$ SLG as the Lieb lattice, the $\varepsilon_{xy}$ strain linearly splits the Kramers doublets at $\Gamma$, which can be observed spectroscopically. Figure \ref{fig:MnFe-split} illustrates this splitting by a first-principles calculation for MnF$_2$ for the valence band lying about 0.4 eV below the valence band maximum. Essentially, finite $\varepsilon_{xy}$ strain turns MnF$_2$ into a Luttinger-compensated \cite{Mazin2022} ferrimagnetic insulator where spin splitting at the $\Gamma$ point is always allowed \cite{Yuan2024}.

\subsection{Temperature-dependent contribution: Insulating transition-metal fluorides}
\label{sec:TD-fluorides}

The temperature-dependent exchange-driven contribution to the longitudinal (i.e., $\mathbf{M}\parallel\mathbf{L}$) piezomagnetic response \cite{Mu-thesis,Consoli2021,Yershov2024} is due to the strain-induced breaking of the symmetry elements ($\mathcal{A}\mathbf{H}$ in the notation of Section \ref{sec:symmetry}) that map the two magnetic sublattices onto each other. As a result, effective magnetic fields for the spins on the two sublattices become unequal, inducing finite magnetization. This mechanism is absent at zero temperature at the mean-field level considered here, but quantum fluctuations make it finite even at $T=0$.

Altermagnetic transition-metal fluorides MnF$_2$, FeF$_2$, and CoF$_2$ crystallize in the rutile structure and have the same ${}^{2}4/{}^1m{}^2m{}^1m$ SLG as the Lieb lattice considered in Section \ref{sec:Lieb}. We use the crystallographic setting ${}^{2}4/{}^1m{}^2m_y{}^1m_d$, as listed in Table \ref{tab:nrpiezo}. The only strain component breaking the sublattice mapping (in linear order) is $\varepsilon_{xy}$.

Linear piezomagnetic response is described by a third-rank tensor $\Lambda_{\alpha\beta\gamma}$: $\Delta F=-\Lambda_{\alpha\beta\gamma}H_\alpha\varepsilon_{\beta\gamma}$, so that the induced magnetization is $M_\alpha=\Lambda_{\alpha\beta\gamma}\varepsilon_{\beta\gamma}$. In the nonrelativistic limit considered in this section, we have $\Lambda_{\alpha\beta\gamma}=\hat{L}_\alpha\tilde\Lambda_{\beta\gamma}$, and the only nonzero component for rutile altermagnets is $\tilde\Lambda_{xy}$. Because all three considered fluorides have easy-axis magnetic anisotropy~\cite{erickson1953neutron},  we have $\hat{\mathbf{L}}=\pm\hat z$, and the relevant component of the piezomagnetic tensor is $\Lambda_{zxy}$. Voigt notation ($xy\equiv6$) is often used in the literature \cite{borovik2013}, denoting $\Lambda_{zxy}$ as $\Lambda_{36}$.

Although strain-induced differences in magnetic anisotropy and $g$-factors between the two sublattices can also contribute to $\Lambda_{36}$, we will only consider the nonrelativistic exchange-driven contribution, which is expected to be dominant. 
Strain-induced DMI \cite{Dzyaloshinsky1958,Moriya1960} leads to spin canting and generates transverse (i.e., $\mathbf{M}\perp\mathbf{L}$) piezomagnetic response, which is discussed below in Section~\ref{sec:strain_DMI}. 

We use the Heisenberg model in the form $H=-\sum_{i<j}\tilde J_{ij}\mathbf{S}_i\cdot\mathbf{S}_j$, where $\mathbf{S}_i$ is a quantum spin operator for the ionic site $i$, with $S = 5/2$, 2, and 3/2 for MnF$_2$, FeF$_2$ and CoF$_2$, respectively. We also define $J_{ij}=\tilde J_{ij}S^2$.
The strain-induced sublattice asymmetry of the effective magnetic field stems from the difference in the intrasublattice exchange constants $J_A=\sum_{j\in A}J_{ij}$ (where $i\in A$) and $J_B=\sum_{j\in B}J_{ij}$ (where $i\in B$). Let us denote $\Delta J=(J_A-J_B)/2$.
On the mean-field level, a finite $\Delta J$ is equivalent to an effective magnetic field $L(T)\Delta J/\mu$, where $L(T)$ is the altermagnetic order parameter normalized to 1 at $T=0$, and $\mu$ is the magnetic moment of one ion.
Therefore, the piezomagnetic response is given by:
\begin{equation}\label{lambda36}
\Lambda_{36} = 
\frac{\chi(T) L(T)}{\mu}\frac{\Delta J}{\varepsilon},
\end{equation} 
were, as usual, we assume $\varepsilon_{xy}=\varepsilon_{yx}=\varepsilon/2$, and
$\chi(T)=(\partial M/\partial B)_T$ is the magnetic susceptibility. The first, temperature-dependent factor in Eq. (\ref{lambda36}) is obtained from the mean-field approximation. This factor also appears in the longitudinal exchange-driven magnetoelectric response in magnetoelectric antiferromagnets like Cr$_2$O$_3$~\cite{astrov1960magnetoelectric,folen1961anisotropy,Mostovoy2010,Mu2014}, where sublattice asymmetry is induced by the applied electric field, rather than strain.

\begin{figure}[htb]
\centering
\includegraphics[width=0.85\columnwidth]{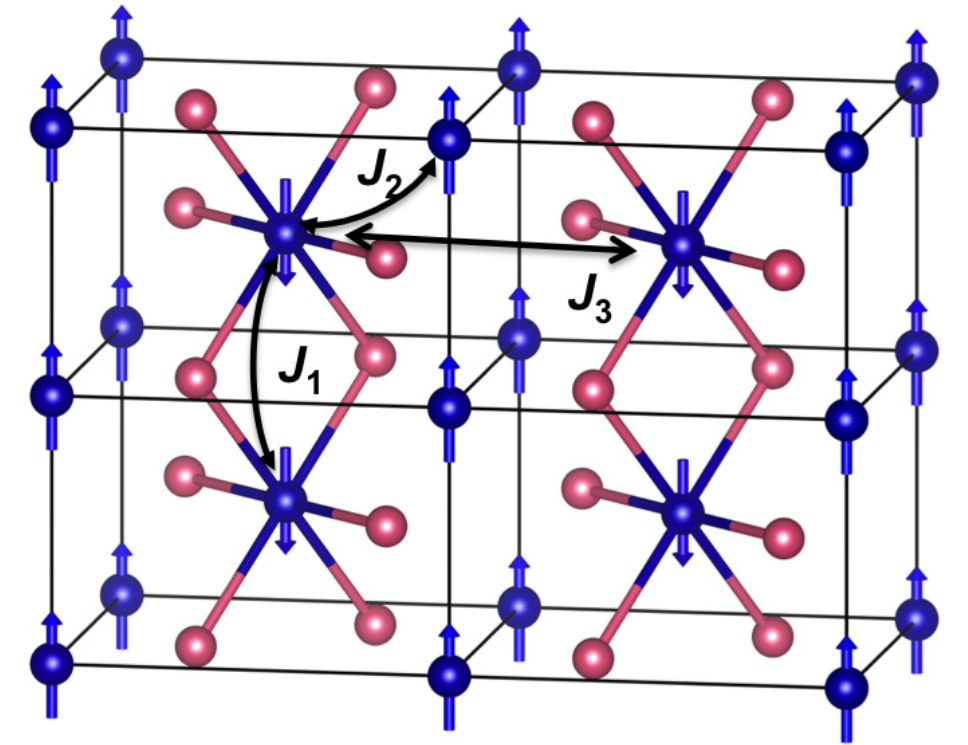}
\caption{1$\times$2$\times$2 supercell of the rutile structure for transition metal fluorides. Blue and red spheres: transition-metal and fluorine atoms, respectively. Blue arrows: spin orientations on the transition-metal ions. Exchange interactions $J_1$, $J_2$, and $J_3$ are shown with black arrows.
}
\label{fig:str}
\end{figure}

The second factor $\Delta J/\varepsilon$ in Eq. (\ref{lambda36}) is obtained from first-principles calculations. For the fluorides, it is sufficient to consider exchange parameters up to third nearest neighbors~\cite{Okazaki,Hutchings,Belorizky,Morano2024}, which are denoted $J_1$, $J_2$, $J_3$ and depicted in Fig.~\ref{fig:str}. The second-neighbor $J_2$ connects transition-metal atoms on different sublattices, while $J_1$ and $J_3$ are intrasublattice couplings. The exchange parameters for the bulk materials obtained from total energy calculations are listed in Table~\ref{exchange} along with the mean-field values of the N\'eel temperature $T_N$. 

The agreement with the exchange parameters obtained from fitting the experimental magnon spectra is good for MnF$_2$ and FeF$_2$. Reference \onlinecite{Das2012} reported considerably larger $J_2$ parameters, which can be attributed to their use of a smaller $U=3$ eV.
For CoF$_2$, the exchange parameters agree with an earlier calculation \cite{Dubrovin2024} but the large value of $J_1$ disagrees with experiment.
We attribute this discrepancy to the deficiency of the single-reference DFT$+U$ description of the partially filled $t_{2g}$ subshell of the Co$^{2+}$ ion. We also noted that the relatively small $J_1$ in MnF$_2$ changes sign if the PBE \cite{PBE} functional is used instead of PBEsol \cite{Perdew}, suggesting the DFT prediction for this parameter is unreliable. However, this difference has almost no effect on the piezomagnetic response.

\begin{table}[htb]
\centering
\caption{Theoretical and experimental exchange parameters $J_n$ (meV), N\'eel temperature ~\cite{Okazaki,Hutchings,Belorizky} $T_N$ (K), the maximum value $\Lambda^{max}_{36}$ ($\mu_{\text{B}}$/f.u.), and the corresponding temperature $T_{max}$ ($K$) for the fluoride altermagnets.}
\begin{tabular}{|l|ccc|cc|c|} 
\hline
     & $J_1$   &   $J_2$  &  $J_3$ & $T^\text{peak}$ & $\Lambda^\text{peak}_{36}$ & $T_\text{N}$  \\
\hline
     MnF$_2$ (theory)   &  0.17 &$-$1.78 & $-$0.09    & 58  & 0.72  &$\ 76.9\ $             \\
     MnF$_2$ \cite{Okazaki}  &   $\ 0.34\ $     & $\ -1.9\ $  &  $\ -\ $  &  $\ -\ $ & $ \ -\ $  &$\ 67.0\ $      \\
    FeF$_2$ (theory)    &   $-$0.07 & $-$2.11 & $-$0.15  &71  & 0.45    & $\ 93.8\ $          \\
    FeF$_2$ \cite{Hutchings}    &  $\ 0.024\ $  & $\ -1.80\ $   &  $\ 0.024\ $   &  $\ -\ $ & $\ -\ $ &  $\ 78.4\ $       \\
    CoF$_2$ (theory)   &  $-$1.18 & $-$1.79 &$-$0.23   &53 & 1.58    & $\ 71.2\ $       \\
    CoF$_2$ \cite{Belorizky}  &   $\ 0.23\ $     & $\ -1.27\ $  &     $\ -\ $  &  $\ -\ $ &  $\ -\ $    &   $\ 38.0\ $        \\ \hline
\end{tabular}
\label{exchange}
\end{table}

When shear strain $\varepsilon_{xy}$ is applied, the intrasublattice $J_1$ and $J_3$ parameters become different for magnetic sublattices A and B. We define $\Delta J_1=(J_1^A-J_1^B)/2$ and $\Delta J_3=(J_3^A-J_3^B)/2$, so that $\Delta J = z_1\Delta J_1 + z_3 \Delta J_3$, where $z_1=2$ and $z_3=4$ are the numbers of first and third nearest neighbors in the rutile crystal.

We use a 
$1 \times 2 \times 2$ supercell shown in Fig.~\ref{fig:str} and four magnetic configurations with either one spin or a whole layer of spins flipped. The total energies of these four configurations, including the applied $\varepsilon_{xy}$ strain, are used to extract $\Delta J_1$ and $\Delta J_3$.
We have checked that $\Delta J_1$ and $\Delta J_3$ obtained in this way depend linearly on the strain. In all three fluorides, we found $\Delta J_3$ is two orders of magnitude smaller than $\Delta J_1$ and, therefore, negligible. The ratios $\Delta J/\varepsilon$ (where $\varepsilon_{xy}=\varepsilon/2$) are 8.6, 7.4, and 27.5 meV for MnF$_2$, FeF$_2$, and CoF$_2$, respectively.

\begin{figure}[htb]
\centering
\includegraphics[width=0.5\textwidth]{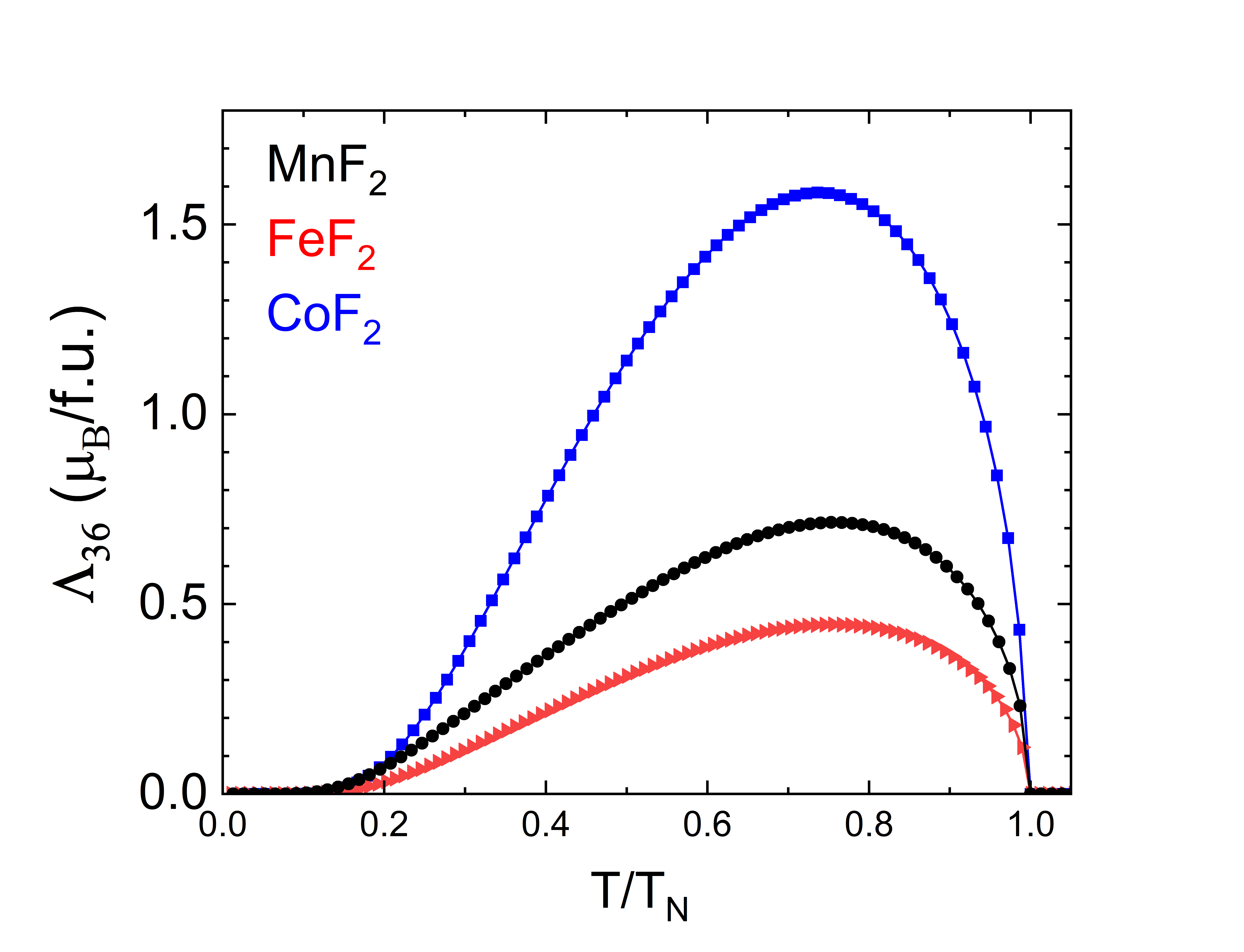}
\caption{Calculated temperature dependence of $\Lambda_{36}$ ($\mu_\text{B}$/f.u.) for the fluoride altermagnets.}
\label{36}
\end{figure} 

The temperature dependence of $\Lambda_{36}$ given by Eq. (\ref{lambda36}) is shown in Fig.~\ref{36}. The response vanishes at $T=0$ because, on the mean-field level, $\chi(T)\to0$ at $T\to0$ in the Heisenberg model.
The maximum value $\Lambda^{max}_{36}$ is reached at  temperature $T_{max}\sim0.75 T_N$ which depends slightly on $S$. The results for $\Lambda^{max}_{36}$ and $T_{max}$ are summarized in Table~\ref{exchange}. 

Among the three compounds, CoF$_2$ exhibits the largest $\Lambda_{36}$, while FeF$_2$ shows the weakest response.
To our knowledge, the only available experimental reference for $\Lambda_{36}$~\cite{borovik2013} reports a value of $8.2 \times 10^{-10}$ Oe$^{-1}$ in CoF$_2$ at 20~K, i.e., $T/T_N\approx0.5$. This value corresponds to 5.6 $\mu_B$/f.u., and our calculated $\Lambda_{36}$ at a similar $T/T_\text{N}$ is 1.1 $\mu_B$/f.u. Given the erroneous sign and magnitude of the calculated $J_1$ for CoF$_2$ that was noted above, it is perhaps surprising that  $\Lambda_{36}$ is in fair agreement with experiment.

\subsection{Nonlinear piezomagnetism in \texorpdfstring{$g$}{g}-wave altermagnets}
\label{sec:CrSb-nonlinear}

\begin{figure}[htb]
    \centering
    \includegraphics[width=0.85\linewidth]{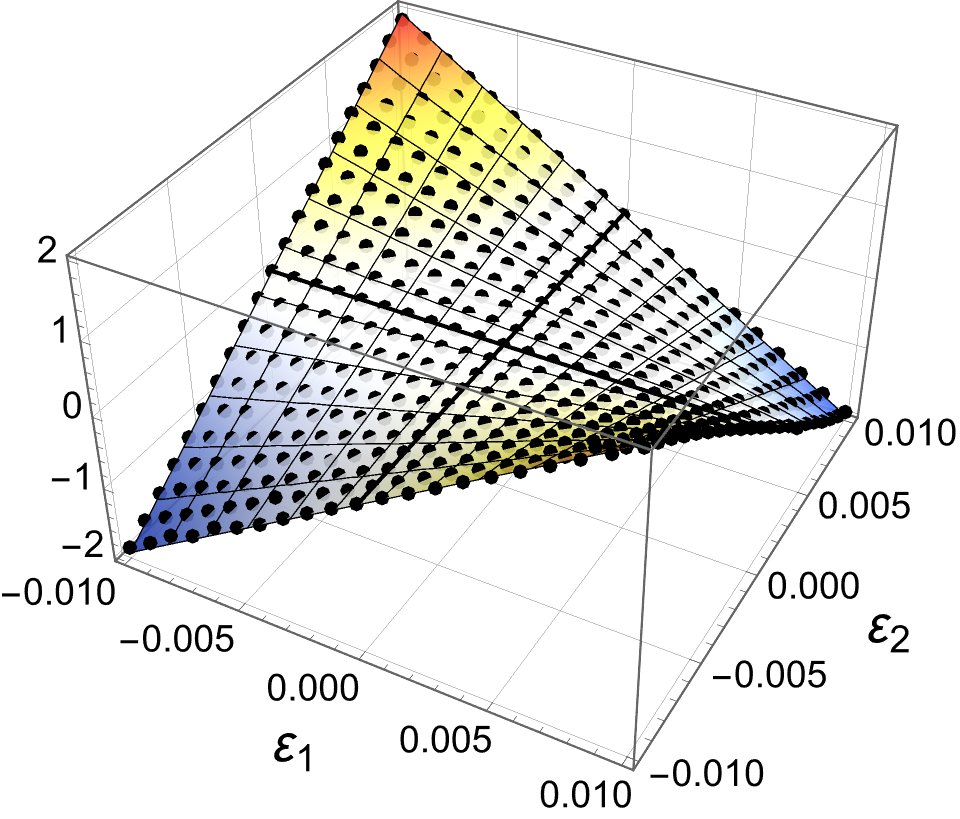}
    \caption{Magnetization in CrSb (units of $\mu_B$ per formula unit) induced by a combination of two shear strain modes, $\varepsilon_{xx}=-\varepsilon_{yy}=\varepsilon_1/2$ and $\varepsilon_{yz}=\varepsilon_{zy}=\varepsilon_2/2$. Spin-orbit coupling is off. The surface shows the fitted function $M(\varepsilon_1,\varepsilon_2)=a\varepsilon_1\varepsilon_2/2$ with $a=-39.4$ $\mu_B$/f.u.}
    \label{fig:CrSb}
\end{figure}

As shown above, in the  nonrelativistic limit the strain-induced magnetization in $g$-wave altermagnets starts with second order in strain. In hexagonal altermagnets with spin point group ${}^26/{}^2m{}^2m{}^1m$, including the widely studied MnTe and CrSb, there is one response parameter, as shown in Table \ref{tab:nrpiezo}:
\begin{equation}
    M=a[(\varepsilon_{xx}-\varepsilon_{yy})\varepsilon_{yz}+2\varepsilon_{xy}\varepsilon_{xz}]
    \label{M-CrSb-quadratic}
\end{equation}
Because CrSb is metallic, this response exists already at zero temperature due to the band-filling effect. To extract the coupling parameter $a$, we use first-principles calculations for a series of strained configurations with small distortions $\varepsilon_{xx}=-\varepsilon_{yy}=\varepsilon_1/2$ and $\varepsilon_{yz}=\varepsilon_{zy}=\varepsilon_2/2$ imposed simultaneously. We used the Fermi smearing method for the Brillouin zone integration with the Fermi temperature of 300~K and a $40\times40\times26$ $k$-point mesh. Figure \ref{fig:CrSb} shows the dependence of the induced magnetization in CrSb as a function of $\varepsilon_1$ and $\varepsilon_2$, which is well fitted by the function $M(\varepsilon_1,\varepsilon_2)=a\varepsilon_1\varepsilon_2/2$ with $a=-39.4$ $\mu_B$/f.u. The sign of the coupling constant $a$ is odd under time reversal.

In Section \ref{rel-symmetry} below we list all bilinear $L$-$M$ coupling terms allowed in the presence of spin-orbit coupling. For the ${}^26/{}^2m{}^2m{}^1m$ SLG with easy-axis anisotropy (i.e., $\mathbf{L}$ along the $z$ axis), we read in Table \ref{tab:relpiezo} that linear piezomagnetic response is purely transverse, coming from the $L_z\left[M_y (\varepsilon_{xx}-\varepsilon_{yy}) + 2 M_x \varepsilon_{xy}\right]$ free-energy invariant. Thus, the nonrelativistic term described by Eq. (\ref{M-CrSb-quadratic}) is expected to be the most important longitudinal strain-induced magnetization response for easy-axis CrSb.

\section{Piezomagnetism due to relativistic effects}
\label{sec:rel-piezo}

\subsection{Symmetry considerations}
\label{rel-symmetry}

Piezomagnetism in magnetic materials is traditionally analyzed within the framework of the magnetic space groups, and the form of the piezomagnetic tensor. 
The antisymmetric Hall conductivity tensor, $\hat{\sigma}_H$ and the magnetization $\mathbf{M}$ transform identically under all unitary and anti-unitary operations. 
Naturally, the structure of the elasto-Hall conductivity tensor \cite{Takahashi2025} follows from the form of the piezomagnetic tensor tabulated for all magnetic space groups \cite{Birss1964,Radaelli2024}. 

In the presence of spin-orbit coupling, some altermagnets become weak ferromagnets \cite{Kluczyk2024,Kotegawa2024,Cheong2024}. 
This weak ferromagnetism can be generally characterized by free energy invariants of the form $F_a = \mathbf{T}_a(\mathbf{L})\cdot\mathbf{M}$, where $\mathbf{T}_a(\mathbf{L})$ is a vector function of the N\'eel vector \cite{McClarty2024,Roig2024a}. 

The most important terms are expected to be linear in $\mathbf{L}$. Such terms can be represented in terms of symmetric or antisymmetric exchange, which usually dominates over higher-order magnetic interactions, such as biquadratic or ring-exchange \cite{Fedorova2015,Szilva2023}. 
If linear terms are absent (as, for example, in MnTe \cite{Gossamer}), the resulting weak magnetization is usually very small. 

In addition, bilinear terms determine the lowest-order harmonics in the angular dependence of the induced magnetization if, for example, the N\'eel order parameter is manipulated by a rotated external field. After  relabeling in terms of the anomalous Hall vector, the same expressions determine the leading harmonics in the angular dependence of the strain-induced anomalous Hall effect (elasto-Hall conductivity \cite{Takahashi2025}).
Therefore, in the following we focus on the $F_a$ invariants that are bilinear in $L$ and $M$.

The allowed invariants depend only on the crystallographic point group and the irreducible representation (irrep) $\Gamma_{\mathbf{L}}$ of the altermagnetic order parameter. This information is equivalently contained in the spin point group. Indeed, the nontrivial spin point group in the form of Eq.~\eqref{eq:sg13} specifies the point group as $\mathbf{H}+ \mathcal{A} \mathbf{H}$. This form automatically identifies the one-dimensional representation $\Gamma_{\mathbf{L}}$ that is even under $\mathbf{H}$ and odd under $\mathcal{A} \mathbf{H}$. Moreover, adding spatial inversion that keeps the sublattices invariant does not forbid any bilinear invariants. Thus, all such invariants can be deduced from the SLG. 

To illustrate the equivalence of the SLG description, as presented here, and the point-group description used in Ref.~\cite{Roig2024a}, we consider two examples.
The $i$-wave SLG ${}^16/{}^1m{}^2m{}^2m$ identifies the point group as $D_{6h}$ and $\Gamma_{\mathbf{L}} = A_{2g}$.
The $d$-wave  SLG ${}^2m_z{}^2m_y{}^1m_x$ identifies the point group as $D_{2h}$ and $\Gamma_{\mathbf{L}} = B_{3g}$.
In our Tables \ref{tab:nrpiezo} and \ref{tab:relpiezo} we do not list equivalent SLGs such as ${}^2m_x{}^2m_y{}^1m_z$ and ${}^2m_x{}^2m_z{}^1m_y$ corresponding to $\Gamma_{\mathbf{L}} = B_{1g}$ and $\Gamma_{\mathbf{L}} = B_{2g}$, respectively, in notations of Ref.~\cite{Roig2024a}.

The allowed bilinear invariants, up to linear order in strain, are listed in Table \ref{tab:relpiezo} for the altermagnetic SLG's. 
Previously, the zero-strain bilinear terms allowed in $F_a$ for all altermagnetic space groups have been tabulated in Ref.~\cite{Roig2024a}.
Our results agree once SLG's are reinterpreted in terms of the point group and its irrep $\Gamma_{\mathbf{L}}$, as in the above examples.

We note a general property: if a certain invariant $f_{\alpha\beta}L_\alpha M_\beta$ is allowed, then so is $f_{\alpha\beta}L_\beta M_\alpha$. If $f_{\alpha\beta}$ is not automatically symmetric or antisymmetric, these invariants are independent.
Each such ``matching'' pair can be equivalently re-expressed in terms of symmetric and antisymmetric invariants, $L_\alpha M_\beta \pm L_\beta M_\alpha$  with coefficients, $f^s_{\alpha\beta}=f_{\alpha\beta}+f_{\beta\alpha}$ and $f^a_{\alpha\beta}=f_{\alpha\beta}-f_{\beta\alpha}$, respectively. 
Antisymmetric invariants correspond to Dzyaloshinskii-Moriya interaction (DMI), and appear in first order in SOC \cite{Roig2024a} for a generic minimal model \cite{Roig2024} of an altermagnet. Strain-induced DMI is further studied below in Section \ref{sec:strain_DMI}. 

\begin{table*}[ht]
\caption{Bilinear $L$-$M$ free-energy invariants in altermagnets in the presence of spin-orbit coupling, to the leading  order in strain. In cases where some $L$-$M$ invariants are allowed at zero strain, only some piezomagnetic invariants are included (see text). We use the following notation: $C_{\alpha\beta}=L_\alpha M_\beta$, and $C^\pm_{\alpha\beta}=C_{\alpha\beta}\pm C_{\beta\alpha}$.}
    \begin{tabular}{|l|l|l|l|}
    \hline
         Type & Spin Laue group & At $\varepsilon_{\alpha\beta}=0$ & Piezomagnetic $L$-$M$ invariants\\
         \hline
         \multirow{4}{*}{$d$-wave} &
         ${}^2m_z{}^2m_y{}^1m_x$ \rule{0pt}{2.3ex} & $C^\pm_{yz}$ & $\varepsilon_{xy}C^\pm_{xz}$, $\varepsilon_{xz}C^\pm_{xy}$, $\varepsilon_{yz}C_{xx}$\\
         & ${}^24/{}^1m$ & $C_{xx}-C_{yy}$, $C^+_{xy}$ & $\varepsilon_{xy}C_{zz}$, $(\varepsilon_{xx}-\varepsilon_{yy})C_{zz}$, $\varepsilon_{xz}C^\pm_{xz}-\varepsilon_{yz}C^\pm_{yz}$, $\varepsilon_{yz}C^\pm_{xz}+\varepsilon_{xz}C^\pm_{yz}$\\
         & ${}^24/{}^1m{}^2m_y{}^1m_d$ & $C^+_{xy}$ & $\varepsilon_{xy}C_{zz}$, $\varepsilon_{yz}C^\pm_{xz}+\varepsilon_{xz}C^\pm_{yz}$\\
         & ${}^22/{}^2m$ & $C^\pm_{xz}$, $C^\pm_{yz}$ &\\
\hline
         \multirow{4}{*}{$g$-wave} & ${}^14/{}^1m{}^2m{}^2m$ \rule{0pt}{2.3ex} & $C^-_{xy}$ & $\varepsilon_{xz}C^\pm_{yz}-\varepsilon_{yz}C^\pm_{xz}$\\
         & ${}^1\bar3{}^2m_y$ & $C^-_{xy}$ & $(\varepsilon_{xx}-\varepsilon_{yy})C^\pm_{yz}+2\varepsilon_{xy}C^\pm_{xz}$, $\varepsilon_{xz}C^\pm_{yz}-\varepsilon_{yz}C^\pm_{xz}$\\
         & ${}^26/{}^2m$ & --- & $(\varepsilon_{xx}-\varepsilon_{yy})C^\pm_{yz}+2\varepsilon_{xy}C^\pm_{xz}$, $(\varepsilon_{xx}-\varepsilon_{yy})C^\pm_{xz}-2\varepsilon_{xy}C^\pm_{yz}$,\\
         &&&$\varepsilon_{yz}(C_{xx}-C_{yy})+\varepsilon_{xz}C^+_{xy}$, $\varepsilon_{xz}(C_{xx}-C_{yy})-\varepsilon_{yz}C^+_{xy}$\\
         & ${}^26/{}^2m{}^2m_y{}^1m_x$ & --- & $\varepsilon_{yz}(C_{xx}-C_{yy})+\varepsilon_{xz}C^+_{xy}$, $(\varepsilon_{xx}-\varepsilon_{yy})C^\pm_{yz} + 2 \varepsilon_{xy}C^\pm_{xz}$\\
\hline
        \multirow{2}{*}{$i$-wave} & ${}^16/{}^1m{}^2m{}^2m$ \rule{0pt}{2.3ex} & $C^-_{xy}$ & $\varepsilon_{xz}C_{yz}-\varepsilon_{yz}C_{xz}$\\
         & ${}^1m{}^1\bar3{}^2m$ & --- &
         $\varepsilon_{xy}C^-_{xy}+\varepsilon_{yz}C^-_{yz}+\varepsilon_{zx}C^-_{zx}$, $(\varepsilon_{yy}-\varepsilon_{zz})C_{xx}+(\varepsilon_{zz}-\varepsilon_{xx})C_{yy}+(\varepsilon_{xx}-\varepsilon_{yy})C_{zz}$\\
\hline
\end{tabular}
\label{tab:relpiezo}
\end{table*}

All altermagnetic $d$-wave SLG's, two of the $g$-wave SLGs (${}^14/{}^1m{}^2m{}^2m$ and ${}^1\bar3{}^2m_y$), and one $i$-wave SLG (${}^16/{}^1m{}^2m{}^2m$) contain bilinear $L$-$M$ terms in the absence of strain. 
However, with the exception of ${}^22/{}^2m$, these SLGs have a special symmetry-enforced axis ($x$ for ${}^2m_z{}^2m_y{}^1m_x$; $z$ for ${}^24/{}^1m$, ${}^24/{}^1m{}^2m_y{}^1m_d$, ${}^14/{}^1m{}^2m{}^2m$, ${}^1\bar3{}^2m_y$, and ${}^16/{}^1m{}^2m{}^2m$) such that the corresponding components of $\mathbf{L}$ and $\mathbf{M}$ do not appear in the bilinear invariants. 
This implies the following two properties satisfied on the bilinear $L$-$M$ level for the unstrained crystal: (1) the magnetization along the special axis vanishes for any orientation of $\mathbf{L}$, and (2) if $\mathbf{L}$ points along the special axis, the magnetization vector vanishes. Because such cases may be of interest in applications, we include in Table \ref{tab:relpiezo} the piezomagnetic terms involving the components of $\mathbf{M}$ or $\mathbf{L}$ along the special axis for the above SLGs. Other kinds of piezomagnetic terms are omitted from Table \ref{tab:relpiezo}. All the component of magnetization are possible in the bilinear invariants for ${}^22/{}^2m$ already at zero strain, and we do not specify allowed bilinear combinations at finite strain for this SLG.
In contrast, the ${}^26/{}^2m$, ${}^26/{}^2m{}^2m_y{}^1m_x$ and ${}^1m{}^1\bar3{}^2m$  SLGs are inconsistent with the bilinear $L$-$M$ terms at zero strain, and we list all such invariants to first order in strain.
Note that the terms with $C_{xx}=L_xM_x$ and $C_{zz}=L_zM_z$ for $d$-wave SLGs in Table \ref{tab:relpiezo} correspond to the isotropic invariants listed in Table \ref{tab:nrpiezo} for the nonrelativistic limit.

\subsection{Strain-induced Dzyaloshinskii-Moriya interaction}\label{sec:strain_DMI}

As noted above, the invariants in Table \ref{tab:relpiezo} containing antisymmetrized $L$-$M$ products $C^-_{\alpha\beta}$ describe strain-induced DMI.
These terms, appearing in first order in SOC, can generate spin canting and a net magnetization. In transition-metal fluorides (SLG: ${}^24/{}^1m{}^2m_y{}^1m_d$), shear strain ($\varepsilon_{yz}$) induces a net magnetization $M_x$ along the $x$ direction, corresponding to the piezomagnetic tensor component $\Lambda_{xyz}$, or $\Lambda_{14}$ in Voigt notation \cite{borovik2013}.

The SLG for MnTe and CrSb is ${}^26/{}^2m{}^2m_y{}^1m_x$.
For MnTe, the order parameter $\mathbf{L}$ is in-plane along $\hat y$. According to Table \ref{tab:relpiezo}, the only strain-induced DMI term corresponds to $\Lambda_{zxx}\approx-\Lambda_{zyy}$, or $\Lambda_{31}\approx-\Lambda_{32}$ in Voigt notation. This equality is approximate, because it excludes higher-order terms beyond bilinear. In CrSb, the order parameter is along the $\hat z$ axis, and the strain-induced DMI corresponds to $\Lambda_{21}=-\Lambda_{22}=\Lambda_{16}$. These equalities for CrSb also follow from the magnetic point group analysis and are, therefore, exact.

To evaluate the piezomagnetic coefficients, we first optimize the internal structural parameters (unless they are fixed by symmetry) under the applied strain. For transition-metal fluorides, we then constrain the local moments on the two sublattices so that their magnetic moments are canted by a small angle $\theta_c$ toward the axis along which the induced magnetization $M_c$ should appear. The total energy $E$ is calculated, with spin-orbit coupling included, as a function of $M_c$; its minimum yields the ground-state energy and the strain-induced magnetization at zero temperature.

Because both DMI and Heisenberg exchange are bilinear operators, the temperature dependence of the DMI-induced magnetization follows $L(T)$, and the canting angle is temperature-independent in the model with fixed exchange and DMI parameters.

Figure \ref{canting14}(a) shows the calculated total energies $E(M_c)$ for a small applied strain $\varepsilon_{yz}=\varepsilon/2$ for the fluorides, with $\varepsilon=0.02$.
In MnF$_2$ the spin canting is below the sensitivity of our calculation, which may be explained by the suppressed orbital moment of the Mn$^{2+}$ ions with the half-filled $3d_\uparrow^5$ configuration. This is in agreement with a very small experimental value for MnF$_2$. A tiny $\Lambda_{14}$ of less than 0.01 $\mu_B$/f.u. can also be inferred from the calculation of Ref.~\onlinecite{Bhowal2024}.  In contrast, strain-induced magnetization is substantial in FeF$_2$ and CoF$_2$.

\begin{figure}[htb]
\centering
\includegraphics*[width=0.45\textwidth]{Figure7.png}
\caption{(a) Total energy as a function of the canted magnetization $M_c=M_x$ for transition-metal fluorides at  $\varepsilon_{yz}=\varepsilon/2$ with $\varepsilon=0.01745$. Solid lines: quadratic fits to the data.
(b-c) Total energy $E(\theta_c)$ (scaled by 0.01 or 0.05; gray symbols) and odd part $\Delta E_b(\theta_c)$ (blue and orange symbols) of the band energy, obtained using the generalized force theorem, as a function of the canting angle $\theta_c$ for (b) MnTe at $\varepsilon=\varepsilon_{xx}-\varepsilon_{yy}=\pm0.04$ and (c) CrSb at $\varepsilon=\varepsilon_{xx}-\varepsilon_{yy}=\pm0.02$. Black and blue (gray and orange) symbols correspond to $\varepsilon>0$ ($\varepsilon<0$). Black line: quartic polynomial fit to $E(\theta_c)$. Blue and orange lines: linear fits to $\Delta E_b(\theta_c)$.}
\label{canting14}
\end{figure} 

For MnTe and CrSb the straightforward calculation of the total energy as a function of the canting angle proved to be computationally unreliable, and we turned to a different strategy. First, we performed self-consistent calculations, without SOC, for a series of constrained canting angles $\theta_c>0$. Then, we used the resulting charge and magnetization densities to evaluate the odd part of $E(\theta_c)$ using the generalized force theorem \cite{Martin2020ElectronicStructure2e}. Specifically, for each angle $\theta_c$ we take the self-consistent constrained magnetization density $\mathbf{m}(\mathbf{r},\theta_c)$ and generate $\mathbf{m}(\mathbf{r},-\theta_c)=\hat C_{2\mathbf{L}}\mathbf{m}(\mathbf{r},\theta_c)$ by applying twofold rotation $C_{2\mathbf{L}}$ around the $\mathbf{L}$ axis at each $\mathbf{r}$. This rotation involves both pseudo-magnetization density and the augmentation occupancies in the PAW method. Finally, we calculate the band energies $E_b(\theta_c)$ and $E_b(-\theta_c)$ corresponding to the frozen charge and magnetization densities $\mathbf{m}(\mathbf{r},\pm\theta_c)$, now without any constraining fields. The odd part $\Delta E_b(\theta_c)= E_b(\theta_c)-E_b(-\theta_c)$ is entirely due to SOC and isolates the angular dependence of strain-induced DMI, which should be linear in $\theta_c$. These calculations for MnTe and CrSb were performed with the strain tensor $\varepsilon_{xx}=-\varepsilon_{yy}=\varepsilon/2$ with $\varepsilon=0.02$ for CrSb and 0.04 for MnTe.

Figures \ref{canting14}(b) and \ref{canting14}(c) show the total energy $E(\theta_c)$ and the odd part $\Delta E_b(\theta_c)$ of the band energy for MnTe and CrSb, respectively, calculated for positive and negative strain. The even part is nearly identical for $\varepsilon\gtrless0$ and is well fitted to an even quartic  polynomial, reflecting the exchange contribution to the energy. The odd part is fitted reasonably well by a linear function with a slope whose sign changes along with that of the strain $\varepsilon$, clearly identifying $\Delta E_b(\theta_c)$ as the strain-induced DMI contribution. The absolute slopes for $\varepsilon\gtrless0$ differ by 9\% in MnTe and 19\% in CrSb, which may reflect either nonlinearity in strain or imperfect convergence. To find the equilibrium canting angle $\theta_{c0}$ and the corresponding piezomagnetic coefficient, we take the averaged slope for $\varepsilon\gtrless0$. This results in $\theta_{c0}=\SI{0.18}{\degree}$ for MnTe and \SI{0.027}{\degree} for CrSb for the chosen values of $\varepsilon$. The linear dependence of the magnetization on $\theta_c$ is taken directly from the first-principles calculation.

Table~\ref{lambda14} lists the piezomagnetic coefficients. For FeF$_2$ the response is appreciable but experimental measurements are not available, to our knowledge. For CoF$_2$ we obtained a sizeable value of 0.9 $\mu_B$/f.u., which is still about one order of magnitude smaller compared to the experimental value at $T/T_N\sim0.5$. As mentioned above, our results for CoF$_2$ should be considered unreliable, given the large disagreement with experiment for the $J_1$ parameter and the limitations of the DFT$+U$ approach.

\begin{table}[htb]
\centering
\caption{Transverse piezomagnetic coefficients ($M_c/\varepsilon$) due to strain-induced DMI (units of $\mu_\text{B}$/f.u.). Experimental values \cite{borovik2013} were measured at 20 K.}
\begin{tabular}{|l|c|c|c|}
\hline
Material & Component & Theory & Experiment \\
\hline
MnF$_2$  & $\Lambda_{14}$ & $\sim0$ & 0.09 \\
FeF$_2$  & $\Lambda_{14}$ & 0.4 & ---\\
CoF$_2$  & $\Lambda_{14}$ & 0.9 & 8.94 \\ 
MnTe     & $\Lambda_{31}$ & 0.38 & --- \\  
CrSb     & $\Lambda_{21}$ & 0.06 & --- \\
\hline
\end{tabular}
\label{lambda14}
\end{table}

The predicted value $\Lambda_{21}=0.06$ $\mu_B$/f.u. for CrSb is smaller, while $\Lambda_{31}=0.38$ $\mu_B$/f.u. for MnTe is similar to FeF$_2$. We note that MnTe already has a tiny magnetization along the $z$ axis at zero strain, on the order of \num{5e-5} $\mu_B$/f.u. \cite{Kluczyk2024}, thanks to a higher-order chiral biquadratic coupling \cite{Wasscher-thesis,McClarty2024,Gossamer}. This means that in-plane shear-strain-induced magnetization exceeds the zero-strain limit in MnTe already at $\varepsilon\sim\num[print-unity-mantissa=false]{1e-4}$.

\section{Strain-induced non-unitary triplet superconductivity}
\label{sec:superconductivity}

In several recent works various manifestations of the altermagnetism in superconductors such as the current-phase relation of a Josephson junction with the altermagnetic normal region have been explored,  \cite{Ouassou2023,Lu2024,Sun2025,Fukaya2025,Zhao2025}.
Superconducting altermagnets are argued to exhibit a specific non-reciprocal critical current, also known as superconducting diode effect, if parity is broken, for example, by an external electric field \cite{Banerjee2024}.

Depending on the parameters, the superconducting order parameter favored by the altermagnet with finite spin-orbit interaction can be singlet, triplet or a mixture of the two that breaks $\mathcal{T}$ \cite{Carvalho2024}.
Here we address the non-relativistic limit of zero SOC where the emergent inversion symmetry inherent in SLGs allows to disentangle singlet and triplet Cooper pair correlations \cite{Maeda2025,Fukaya2025a}. 
As the $\mathcal{T}$ symmetry is broken by the magnetism such symmetry protects the equal spin Cooper pairs, \cite{Mazin2022a}.
These triplet pairs fall into two varieties distinguished by the total angular momentum, $m_s = \pm 1$ along $\mathbf{L}$.
The nontrivial spin symmetry $[C_{2\hat{\mathbf{L}}_\perp}||\mathcal{A}]$ transforms one type of triplet pairs to another.
This ensures that such superconducting state is unitary on average.

The strain deformations listed in the Tab.~\ref{tab:nrpiezo} are odd under $[C_{2\hat{\mathbf{L}}_\perp}||\mathcal{A}]$, and effectively disconnect the two sublattices. In this section we show how such strain makes the superconductivity non-unitary on average. Concomitantly, the strain promotes the superconductivity by increasing the critical temperature $T_c$. Surprisingly, we also find that the magnetization perpendicular to $\mathbf{L}$ controls the relative phase of the two triplet varieties. 

We adopt the phenomenological Ginzburg-Landau description of a coupling of strain and superconductivity.
In many instances the Ginzburg-Landau functional contains $\mathbf{L}$ as an independent degree of freedom.
At zero SOC, such free energy is isotropic in $\mathbf{L}$ reflecting arbitrariness in its direction \cite{McClarty2024}.
The N\'eel vector has to be treated as an independent degree of freedom to study the dynamical coupling of the strain to the magnetization \cite{Steward2023}.

In contrast, herein $\mathbf{L}$ is a fixed background for other degrees of freedom including the superconductor order parameter, $\Delta$, the strain, $\boldsymbol{\varepsilon}$ and magnetization, $\mathbf{M}$ necessarily present at finite strain. 
In particular, electronic spins are subject to the spin-only spin rotation about a fixed $\mathbf{L}$ rather than to the full spin rotation group.

In a collinear altermagnet, parity protects Cooper pairing of electrons with parallel spins. Furthermore, since parity does not exchange the sublattices in an altermagnet, parity-protected pairs are necessarily intrasublattice triplets with the total spin $m_s =\pm 1$ along the N\'eel vector. In the limit of large exchange splitting we then have two sets of parity-protected Cooper pairs: one with spin-up electrons forming pairs residing on sublattice $A$ and another with spin-down electrons on sublattice $B$.

To see how such triplets react to strain and possibly to other perturbations we classify the triplet order parameters according to SLG. As in the more standard cases, the order parameters belonging to the $\ell$-th irrep, $\mathbf{\Gamma}_\ell$ of SLG form sets of $\mathbf{\Gamma}_\ell$ partners, $\boldsymbol{\Delta}^{\mathbf{\Gamma}_\ell} = \{ \Delta_1^{\mathbf{\Gamma}_\ell},\ldots ,\Delta_{\mathrm{dim}\mathbf{\Gamma}_\ell }^{\mathbf{\Gamma}_\ell} \}$.

The Ginzburg-Landau free energy, $F$ contains the contributions of the second and fourth order in the superconducting order parameter denoted as $F^{(2)}$ and $F^{(4)}$, respectively.
Focusing on the a specific irrep $\mathbf{\Gamma}_\ell$, the second order terms read
\begin{align}\label{eq:F2}
    F^{(2)} = (T -T^\ell_{c0}) \sum_{l=1}^{\mathrm{dim}\mathbf{\Gamma}_\ell}|\Delta_l^{\mathbf{\Gamma}_\ell}|^2
    + F^{(2)}_{\mathbf{L}}[\boldsymbol{\Delta}^{\mathbf{\Gamma}_\ell},\mathbf{M},\boldsymbol{\varepsilon}]\, ,
\end{align}
where 
the first term describes the hypothetical superconducting phase transition of an unstrained material at zero magnetization as the temperature $T$ reaches the critical temperature $T^\ell_{c0}$ defined per symmetry channel $\ell$.
It is the only scalar allowed by symmetry at zero strain. 

The second term, $F^{(2)}_{\mathbf{L}}[\boldsymbol{\Delta}^{\mathbf{\Gamma}_\ell},\mathbf{M},\boldsymbol{\varepsilon}]$ 
describes the coupling of the strain and magnetization with the superconducting order parameter at a fixed $\mathbf{L}$.
The coupling of the strain and magnetization has been analyzed in Sec.~\ref{sec:nonrel} and is omitted in Eq.~\eqref{eq:F2}. 
It is fixed by SLG classification of all the degrees of freedom.
It affects the critical temperature, and may cause the superconducting order parameter to become non-unitary as we discuss below.

Microscopically, the superconducting order parameters are found self-consistently from the Bogoliubov-de Gennes (BdG) Hamiltonian, $H_{\mathrm{BdG}}=H_b + H_{p}$ including the band Hamiltonian, $H_b$ and the pairing part, 
\begin{subequations}\label{eq:BdG}
    \begin{align}\label{eq:Hp}
        H_{p} = \sum_{l=1}^{\mathrm{dim}\mathbf{\Gamma}_\ell}  \Delta^{\mathbf{\Gamma}_\ell}_{l} \mathcal{O}^{\mathbf{\Gamma}_\ell}_{l} + h.c.,
    \end{align}
    \begin{align}\label{eq:partners1}
         \mathcal{O}^{\mathbf{\Gamma}_\ell}_{l} = \sum_{\mathbf{k},ij,\sigma\sigma'}  \phi^{\mathbf{\Gamma}_\ell}_{l;\sigma\sigma'}(\mathbf{k},i,j) c^{\dagger}_{\mathbf{k},i \sigma} c^{\dagger}_{-\mathbf{k},j \sigma'},
    \end{align}
\end{subequations}
where $c^\dagger_{\mathbf{k}i \sigma}$ creates an electron at momentum, $\mathbf{k}$, sublattice $i=A,B$ and spin state $\sigma$.
It follows from Eq.~\eqref{eq:BdG} that the classification of the order parameter amounts to the classification of the partner functions $\phi^{\mathbf{\Gamma}_\ell}_{l}$.
For a fixed irrep $\mathbf{\Gamma}_\ell$ we further specify them as
\begin{align}\label{eq:partners2}
    \phi^{\mathbf{\Gamma}_\ell}_{l;\sigma\sigma'}(\mathbf{k},i,j) = [P_l]_{\sigma\sigma'}[\tau_{\mu}]_{ij}\varphi_l(\mathbf{k})\,  ,
\end{align}
where $P_l$ and $\tau_\mu$ specify the spin and lattice dependence of the $l$th partner function.
Depending on $\mathbf{\Gamma}_\ell$ and the partner index, the $P_l$ can be the spin-triplet wave function, $P^t_{m_s}$, $m_s=\pm 1,0$ or spin-singlet wave function, $P^s$.
The Pauli matrices operating in the sublattice space are denoted as $\tau_\mu$, where $\mu=0,x,y,z$, and $\tau_0$ is a unit matrix.
The Pauli matrices are either even or odd upon the exchange of the indices, and this makes it possible to factorize the sublattice dependence of the partner functions \eqref{eq:partners2}.
Hence, Pauli principle implies that if a given partner is odd (even) under combined spin and lattice exchange, the orbital wave function is even (odd), $\varphi_l(\mathbf{k})= \pm\varphi_l(-\mathbf{k})$, respectively. 

For concreteness we focus on the the paradigmatic $d$-wave altermagnets, SLG ${}^24/{}^1m{}^2m_y{}^1m_d$, (see Table~\ref{tab:nrpiezo}).
Within this class we discuss the Lieb lattice, as well the bulk materials with rutile structure. 
Our conclusions are applicable to oxyselenide altermagnets \cite{Junwei2021,Cui2023,KV2Se2O,RbV2Te2O}. 

The site group in this case, $\mathbf{H} = mmm$ ($D_{2h}$), the crystallographic point group, 
$\mathbf{H} + \mathcal{A} \mathbf{H} = 4/mmm$ ($D_{4h}$), where we choose $\mathcal{A}= T_{\mathbf{t}}C_{2x}$, $\mathbf{t}=(1/2,1/2,1/2)$ such that $\mathcal{A}^2 =E$. The conclusions apply to all SLGs with the abelian site group, $\mathbf{H}$. The generalization to non-abelian $\mathbf{H}$ is also possible, and will be given elsewhere. 

Recently, the irreps of spin point groups have been tabulated \cite{Schiff2025}.
It was stressed that in cases where the spin-only group is continuous, such as in collinear altermagnets, an irrep carries the index of the site group irrep and an integer or half-integer label $m_s$ of the irrep of the spin-only group.
As a result, when the spin-only group is represented by nonzero phases the irreps of the full spin group are distinct from the irreps of the parent point group, and, in general, \emph{a priori} unknown. 
We argue that such is the case of the equal-spin triplet Cooper pairs carrying an integer angular momentum, $m_s = \pm 1$.
Hence, we are naturally led to the need to find the SLG irreps carrying integer angular momentum.

In App.~\ref{app:irreps} we implement a modified procedure to identify the irreps of the ${}^24/{}^1m{}^2m_y{}^1m_d$ spin group, which allows us to obtain the representation matrices along with the symmetry-adapted partner functions. In the present context, the latter are interpreted as the Cooper pair wave function.

When the site group, $\mathbf{H}$ is abelian, $[\mathbf{C}_\infty||\mathbf{H}]$ forms the abelian and normal subgroup of the full spin point group. 
In this case the irreps of spin point group follow from the irreps of $[\mathbf{C}_\infty||\mathbf{H}]$ by the method of induction.
For this reason the spin point group irreps carry the subscript index which is the irrep of $\mathbf{H}$.
For the same reason, the irreps carry the total angular momentum, $m_s$ as superscript labeling the single valued irreps of $\mathbf{C}_\infty$.
In our example the irreps of the site group in a standard notation are $A_{g(u)}$, $B_{1g(u)}$, $B_{2g(u)}$ or $B_{3g(u)}$.

To specify the irreps it suffies to state the representation matrices for an element $[C_{2\hat{\mathbf{L}}_{\perp}}||\mathcal{A}]$, and a generic element $g = [C_{\phi\hat{\mathbf{L}}}||h]\in [\mathbf{C}_\infty||\mathbf{H}]$, where $h \in \mathbf{H}$ and $C_{\phi\hat{\mathbf{L}}} \in \mathbf{C}_\infty$ is a rotation around $\mathbf{L}$ by angle $\phi$.

In the present example there are only one- and two-dimensional irreps.
We start with one-dimensional irreps.
In this case, the angular momentum $m_s=0$ and site group irrep subscript completely fixes the numbers representing any element of $[\mathbf{C}_\infty||\mathbf{H}]$ subgroup.
The element $[C_{2\hat{\mathbf{L}}_{\perp}}||\mathcal{A}]$ is represented either by $+1$ or by $-1$ because by construction this element squares to identity.
We introduce the second superscript set to $+$ or $-$ to denote these two alternatives.

With these notations, the eight one-dimensional irreps, ($\beta=A, B_1$),
\begin{align}\label{eq:1Dirrep}
    \mathbf{\Gamma}^{0;\pm}_{\beta_{g(u)}}(g) = \beta_{{g(u)}}(h)\, , \,\,
    \mathbf{\Gamma}^{0;\pm}_{\beta_{g(u)}}([C_{2\hat{\mathbf{L}}_{\perp}}||\mathcal{A}]) = \pm 1\, ,
\end{align}
where specifically the subscripts can be $A_{g}$, $A_{u}$, $B_{1g}$ and $B_{1u}$ which are irreps of $\mathbf{H}$, superscript $0$ means that $\mathbf{C}_\infty$ is represented by zero angular momentum, and finally $\pm$ subscript refers to whether the element $[C_{2\hat{\mathbf{L}}_{\perp}}||\mathcal{A}]$ is represented by $+1$ or $-1$.

An example of the fermion operator transforming trivially, i.e. as $\mathbf{\Gamma}^{0;+}_{A_{g}}$ is the Hamiltonian.
The scalar product $\mathbf{L} \cdot\mathbf{M}$, as well as the functions $f_{\mathrm{P}}(\boldsymbol{\varepsilon})$ in the Tab.~\ref{tab:nrpiezo} change sign under $[C_{2\hat{\mathbf{L}}_{\perp}}||\mathcal{A}]$, and therefore, transform as $\mathbf{\Gamma}^{0;-}_{A_{g}}$.
Since in Eq.~\eqref{eq:1Dirrep} the spin rotations $\mathbf{C}_\infty$ are represented just by a number 1, it defines two irrep of the non-trivial part of the spin group, which is in turn isomorphic to the parent crystallographic point group, $D_{4h}$.
In this way, e.g.
$\mathbf{\Gamma}^{0;+}_{A_{g}}$ is $A_{1g}$ and $\mathbf{\Gamma}^{0;-}_{A_{g}}$ is $B_{2g}$ irreps of $D_{4h}$ \cite{Roig2024}.

The two dimensional irreps are necessary to describe the operators with a non-zero total spin projection $m_s$ on $\mathbf{L}$.
We have, ($\beta=A, B_1$),
\begin{align}\label{eq:2Dirrep}
\mathbf{\Gamma}^{m'_s}_{\beta_{g(u)}}(g)& =
    \begin{bmatrix}
        \beta_{g(u)} (h) e^{-i m'_s \phi} &\! 0 \\ 0 & \!\! \beta_{g(u)} (h)e^{i m'_s \phi} 
    \end{bmatrix},
    \notag \\
\mathbf{\Gamma}^{m_s}_{B_{2g(u)}}(g)& =
    \begin{bmatrix}
        B_{2g(u)} (h) e^{-i m_s \phi} &\! 0 \\ 0 & \!\! B_{3g(u)}(h)e^{i m_s \phi} 
    \end{bmatrix},
    \notag \\
   \mathbf{\Gamma}^{m'_s}_{\beta_{g(u)}}([C_{2\hat{\mathbf{L}}_{\perp}}||\mathcal{A}])  &=\mathbf{\Gamma}^{m_s}_{B_{2g(u)}}([C_{2\hat{\mathbf{L}}_{\perp}}||\mathcal{A}])  = \begin{bmatrix}
        0 & 1 \\ 1 & 0
    \end{bmatrix},
\end{align}
where $m'_s \in \mathbb{N}$, and $m_s \in \mathbb{Z}$.
Both Eqs.~\eqref{eq:1Dirrep} and \eqref{eq:2Dirrep} are in agreement with Ref.~\cite{Schiff2025}.
For instance, the matrices of the irrep $\mathbf{\Gamma}^{m_s}_{B_{2u}}$ is related to the matrices, $D^{(\mu,5)}$ with $\mu =m_s$ by a simple similarity transformation exchanging the two representation partners.

Finally, as found in Ref.~\cite{Schiff2025} by applying the Dimmock test \cite{Dimmock1963}, the anti-unitary operations do not lead to doubling of the representation space.
Physically, this is because the anti-unitary operation $\mathcal{T}[C_{2\hat{\mathbf{L}}_{\perp}}||E]$ leaves the spin orientation unchanged.
It follows that under the $\mathcal{T}[C_{2\hat{\mathbf{L}}_{\perp}}||E]$ operation the partner wave functions for spin singlets and triplets acquire at most a phase factor.
Hence, the representation and co-representation dimensions are the same in the present example.

We are now ready to apply these result to classify the Cooper pair wave functions.
Generally, the equal-spin triplets, $P^t_{\pm 1}$ may form four two-dimensional irreps, 
$\mathbf{\Gamma}^{\pm 1}_{B_{2u}}$, $\mathbf{\Gamma}^{1}_{A_{u}}$ and $\mathbf{\Gamma}^{1}_{B_{1u}}$, Eq.~\eqref{eq:2Dirrep}.
If the spin wave-functions of a Cooper pairs are $P^t_{0}$ or $P^s$, they form one of the four one-dimensional irreps $\mathbf{\Gamma}^{0;\pm}_{A_{g(u)}}$ ($\mathbf{\Gamma}^{0;\pm}_{B_{1g(u)}}$)
when the corresponding $\varphi(\mathbf{k}) \in A_{g(u)} (B_{1g(u)})$, respectively, Eq.~\eqref{eq:1Dirrep}.
Otherwise, they give rise to the two-dimensional irrep $\mathbf{\Gamma}^{0}_{B_{2g(u)}}$.

To reduce the number of possible Cooper pairs here we consider the limit of large exchange splitting $J$.
We have employed the same approximation in Sec.~\ref{sec:Lieb} where the effective model included two bands instead of four.
In this limit, we can limit the consideration to either intra-sublattice equal spin triplets, $P^t_{\pm 1}$ or to the inter-sublattice pairs of opposite spins.
The latter can form anti-parallel spin triplets, $P^t_{0}$ or spin singlets, $P^s$.
Furthermore, in all cases we consider the Cooper pairs formed by nearest or next to nearest neighbor electrons. 

\subsubsection{Cooper pairs on the Lieb lattice}

We first present the possible symmetries of Cooper pairs on a Lieb lattice (see App.~\ref{app:CooperLieb} for details). 
The singlet pairing arising at the level of nearest neighbor hopping is represented by the two operators, 
\begin{subequations}\label{eq:nn_singlets}
\begin{align}\label{eq:nn_singlets_a}
    \phi^{\mathbf{\Gamma}_\ell}_{1;\sigma\sigma'}(\mathbf{k},i,j) = [P^s]_{\sigma\sigma'} [\tau_x]_{ij} \cos\frac{k_x}{2} \cos\frac{k_y}{2}\, ,
\end{align}
\begin{align}\label{eq:nn_singlets_b}
    \phi^{\mathbf{\Gamma}_\ell}_{1;\sigma\sigma'}(\mathbf{k},i,j) = [P^s]_{\sigma\sigma'} [\tau_x]_{ij} 
    \sin\frac{k_x}{2} \sin\frac{k_y}{2}\, 
\end{align}
\end{subequations}
for $\mathbf{\Gamma}_{\ell}=\mathbf{\Gamma}^{0;+}_{A_{1g}}$ and $\mathbf{\Gamma}_{\ell}=\mathbf{\Gamma}^{0;+}_{B_{1g}}$, respectively.  
Equations \eqref{eq:nn_singlets_a} and \eqref{eq:nn_singlets_b} have recently been identified as $s$- and $d$-wave singlet pairing, respectively \cite{Chakraborty2024}.
These Cooper pairs are even in the sublattice interchange.
The coexisting Cooper pairs that are odd in it form triplet states given in Eq.~\eqref{eq:nn_triplets}.
These triplets mixes with the singlets of the same symmetry.
The two-dimensional triplet and singlet Cooper pairs are obtained in Eqs.~\eqref{eq:nn_singlet_B2u} and \eqref{eq:nn_triplets_B2u}, respectively.

All of the above Cooper pairs survive the large $J$ becuase the electrons forming such pairs reside on different sublattices and have opposite spins.
However, since the band dispersion satisfies $E_{\sigma}(\mathbf{k}) \ne E_{-\sigma}(-\mathbf{k})$, such pairs are destroyed by the altermagnetic spin splitting.
The only pairs protected from both the exchange and altermagnetic splitting due to the $E_{\sigma}(\mathbf{k}) = E_{\sigma}(-\mathbf{k})$ symmetry are intra-sublattice parallel spin Cooper pairs discussed next. 

Equal spin triplets arise at the level of next nearest neighbor pairing.
The partner functions of the resulting $\mathbf{\Gamma}^{+1}_{B_{2u}}$ are 
\begin{align}\label{eq:nnn_triplet_B2u}
    \{ \phi^{\mathbf{\Gamma}_\ell}_{1},\phi^{\mathbf{\Gamma}_\ell}_{2}\}
    \!=\!
\left\{ \! (\tau_{0}\!+\!\tau_z) P^t_1 \sin k_y,\! (\tau_{0}\! -\! \tau_z)P^t_{-1} \sin k_x \!\right\} 
\end{align}
with similar expressions for $\mathbf{\Gamma}^{-1}_{B_{2u}}$ irrep.
The Cooper pairs of the type \eqref{eq:nnn_triplet_B2u} can be referred to as $(p_x,p_y)$ triplets.
We note that the spin-triplet Cooper pairs Eq.~\eqref{eq:nnn_triplet_B2u} break the $\mathbf{H}$ symmetry of the lattice.
For instance, the $C_{2y}$ real space rotation changes the sign of the second partner and leaves the sign of the first partner unchanged. 
Such a transformation of the order parameter cannot be undone by a global gauge transformation.
Therefore, even though such order parameter is most protected by the spin symmetry it might not be preferred as breaking of the lattice symmetry by superconductivity could be opposed by the lattice interactions that maintain such symmetry to the first place. Only for the $\mathbf{\Gamma}^{1}_{A_{u}}$ and $\mathbf{\Gamma}^{1}_{B_{1u}}$ irreps the two partners representing Cooper pairs at the two sublattices acquire the same phase under $\mathbf{H}$.
In view of the global gauge invariance only these order parameters preserve $\mathbf{H}$.
In the Lieb lattice model they are forbidden by the horizontal mirror symmetry.
Indeed, the functions transforming as $A_u$ or $B_{1u}$ are antisymmetric under this operation.
It follows that these more robust triplet order parameters might be possible in the bulk structures.
And this is what we show next.

\subsubsection{Cooper pairs in the rutile structure}

At the level of the nearest neighbor hopping the possible spatial symmetries of the Cooper pairs are listed in Tab.~\ref{tab:Rut_nn}.
The conventional Cooper pairing of $\mathbf{\Gamma}^{0;+}_{A_g}$ symmetry is very similar to that on the Lieb lattice, Eq.~\eqref{eq:nn_singlets}.
There are other unconventional singlet and triplet order parameters that can be deduced similar to the Lieb lattice model.

\begin{table}[htb]
\caption{The $\mathbf{H}=D_{2h}$ symmetry classification of the nearest neighbor (nn) hopping amplitudes on the rutile structure of even (top) and odd (bottom) parity.}
\newcolumntype{C}[1]{>{\centering\arraybackslash}p{#1}}
\begin{tabular}{|C{0.6cm}|C{2.9cm}|C{2.2cm}|C{2.2cm}|}
\hline
     nn    &     $A_g$   &   $B_{2g}$  & $B_{3g}$ \\
\cline{2-4}
      even   &  $\cos\frac{k_z}{2} \cos\frac{k_x}{2} \cos\frac{k_y}{2}$  &   $\sin\frac{k_z}{2}\sin\frac{k_x+k_y}{2}$         & $\sin\frac{k_z}{2}\sin\frac{k_x - k_y}{2}$   \\
    &   $\cos\frac{k_z}{2} \sin\frac{k_x}{2} \sin\frac{k_y}{2}$     &              &    \\
\hline
\noalign{\vskip 3pt}
\end{tabular}
\begin{tabular}{|C{0.6cm}|C{2.9cm}|C{2.2cm}|C{2.2cm}|}
\hline
                 nn  &     $B_{1u}$  & $B_{2u}$ &  $B_{3u}$\\
\cline{2-4}
  odd &  $\sin\frac{k_z}{2} \cos\frac{k_x}{2} \cos\frac{k_y}{2}$              & $\cos\frac{k_z}{2} \sin\frac{k_x-k_y}{2}$ & $\cos\frac{kz}{2} \cos\frac{k_x+k_y}{2}$   \\
     &     $\sin\frac{k_z}{2} \sin\frac{k_x}{2} \sin\frac{k_y}{2}$            &    &   \\
\hline
\end{tabular}
\label{tab:Rut_nn}
\end{table}

\begin{table}[htb]
\centering
\caption{The $\mathbf{H}=D_{2h}$ symmetry classification of the next to nearest neighbor (nnn) hopping amplitudes on the rutile structure of even (top) and odd (bottom) parity.}
\newcolumntype{C}[1]{>{\centering\arraybackslash}p{#1}}
\begin{tabular}{|C{0.6cm}|C{3.0cm}|C{2cm}|}
\hline
          nnn &   $A_g$   &  $B_{1g}$  \\
\cline{2-3}
     even  &  $\cos k_x + \cos k_y, \cos k_z$  &   $\cos k_x - \cos k_y$    \\
\hline
\noalign{\vskip 3pt}
\end{tabular}
\begin{tabular}{|C{0.6cm}|C{0.85cm}|C{2cm}|C{2cm}|}
\hline
        nnn  &   $B_{1u}$ & $B_{2u}$  & $B_{3u}$ \\
\cline{2-4}
       odd  & $\sin k_z$   &   $\sin k_x - \sin k_y$     &    $\sin k_x + \sin k_y$           \\
\hline
\end{tabular}
\label{tab:Rut_nnn}
\end{table}

We now turn to triplets of $\mathbf{\Gamma}^{1}_{A_{u}}$ and $\mathbf{\Gamma}^{1}_{B_{1u}}$ symmetry that are special to the bulk altermagnets.
The amplitudes for the nearest neighbor hopping are summarized in Tab.~\ref{tab:Rut_nnn}.
At this hopping range we have no functions transforming as $A_{u}$, and just one function $\sin k_z \in B_{1u}$.
Therefore, the optimal order parameter reads,
\begin{align}\label{eq:nnn_triplet_rutile}
    \{ \phi^{\mathbf{\Gamma}_\ell}_{1},\phi^{\mathbf{\Gamma}_\ell}_{2}\}
    \!=\!
\left\{ \! (\tau_{0}\!+\!\tau_z) P^t_1 \sin k_z,\! (\tau_{0}\! -\! \tau_z)P^t_{-1} \sin k_z \!\right\} 
\end{align}
for $\mathbf{\Gamma}_\ell= \mathbf{\Gamma}^{1}_{B_{1u}}$.
The order parameter \eqref{eq:nnn_triplet_rutile} can be referred to as $p_z$-wave triplet.
It 1) is protected against pair breaking by exchange splitting 2) is immune to the altermagnetic spin splitting 3) does not break the symmetry of the underlying lattice.

The $A_{u}$ pairing results from a more extended Cooper pairs.
The $A_u$ hopping amplitude of a minimal range has a form,  $\varphi(\mathbf{k}) = \sin k_z (\cos k_x - \cos k_y)$.
We argue that the resulting triplet order parameters has three nodal plains, $k_z=0$, $k_x = \pm  k_y$ crossing at one point.
This makes such order parameter less likely to play a role.
Note that in both $\mathbf{\Gamma}^{1}_{A_{u}}$ and $\mathbf{\Gamma}^{1}_{B_{1u}}$,  $k_z=0$ is a nodal plane which makes such superconductivity impossible in a two-dimensional Lieb lattice model as shown above.

In what follows, we focus on $\mathbf{\Gamma}^{1}_{B_{1u}}$ triplet order parameter as the most physically motivated choice.
Such pairing is classified as $p$-wave with the electrons at the adjacent layers forming a Cooper pair. The pairing Hamiltonian \eqref{eq:BdG} takes the form,  
\begin{align}\label{eq:BdG1}
    H_{p} & = \Delta^{\mathbf{\Gamma}^{1}_{B_{1u}}}_1
    \sum_\mathbf{k}  \phi^{B_{1u}}_{\mathbf{k}} c^{\dagger}_{\mathbf{k},A \uparrow} c^{\dagger}_{-\mathbf{k},A \uparrow}
    \notag \\
    & + \Delta^{\mathbf{\Gamma}^{1}_{B_{1u}}}_2
    \sum_\mathbf{k} \phi^{B_{1u}}_{C_{2x}^{-1} \mathbf{k} }c^{\dagger}_{\mathbf{k},B \downarrow} c^{\dagger}_{-\mathbf{k},B \downarrow}  +h.c.\, ,
\end{align}
where $\phi^{B_{1u}}_{\mathbf{k}}$ is a function transforming as $B_{1u}$ under $\mathbf{H}$, and as we argued before the simplest choice is $\phi^{B_{1u}}_{\mathbf{k}} = \sin k_z$.
The two component superconducting order parameter \( \boldsymbol{\Delta}^{\mathbf{\Gamma}^{1}_{B_{1u}}} = 
\{ \Delta^{\mathbf{\Gamma}^{1}_{B_{1u}}}_1,\Delta^{\mathbf{\Gamma}^{1}_{B_{1u}}}_2 \}\in \mathbf{\Gamma}^{1}_{B_{1u}}  \).

\subsection{Effect of strain on equal-spin triplet superconductivity}
To investigate the coupling of \( \boldsymbol{\Delta}^{\mathbf{\Gamma}^{1}_{B_{1u}}} \) to strain and magnetization we invoke the decomposition, 
\begin{align}\label{eq:decomposition}
   [ \mathbf{\Gamma}^{1}_{B_{1u}} ]^* \otimes \mathbf{\Gamma}^{1}_{B_{1u}} = \mathbf{\Gamma}^{0,+}_{A_{g}} \oplus \mathbf{\Gamma}^{0,-}_{A_{g}} \oplus \mathbf{\Gamma}^{2}_{A_{g}}\, .
\end{align}
The bilinear combinations of the order parameters entering the decomposition \eqref{eq:decomposition} are
\begin{align}\label{eq:decomposition1}
    |\Delta^{\mathbf{\Gamma}^{1}_{B_{1u}}}_1|^2 \pm |\Delta^{\mathbf{\Gamma}^{1}_{B_{1u}}}_2|^2  & \in \mathbf{\Gamma}^{0,\pm}_{A_{g}}  \notag \\
    \left\{  \Delta^{B_{1u}}_1[\Delta^{B_{1u}}_2]^* ,[\Delta^{B_{1u}}_1]^* \Delta^{B_{1u}}_2  \right\} & \in \mathbf{\Gamma}^{2}_{A_{g}}\, .
\end{align}

Previously we have classified the magnetization along N\'eel vector, $M_L=\mathbf{M}\cdot\mathbf{L}$ as well as the strain components listed in Tab.~\ref{tab:nrpiezo} as $\mathbf{\Gamma}^{0,-}_{A_{g}}$, see Eq.~\eqref{eq:1Dirrep}. 
Based to Eq.~\eqref{eq:decomposition} they couple to the $\mathbf{\Gamma}^{0,-}_{A_{g}}$ component listed in Eq.~\eqref{eq:decomposition1}.
Accidentally this argument also shows that they couple to each other in agreement with the results of Sec.~\ref{sec:nonrel}.

To classify the magnetization components $\mathbf{M}_{\perp} \perp \mathbf{L}$ we introduce the orthonormal basis for the spin space, $\{\hat{X},\hat{Y},\hat{Z}\}$ such that $\hat{X} \times \hat{Y}= \hat{Z}\parallel \mathbf{L}$.
In these notations,  $\mathbf{M}_{\perp}$ is classified as $\{ M_+, M_-\} \in \mathbf{\Gamma}^{1}_{A_{g}}$ where $M_{\pm}=M_X \pm i M_Y$.
Since $\mathbf{\Gamma}^{1}_{A_{g}}$ is not encountered in the decomposition \eqref{eq:decomposition}, $\mathbf{M}_{\perp}$ does not couple to neither strain nor the superconducting order parameter.

To the second order, however, $\{ M^2_+, M^2_-\} \in \mathbf{\Gamma}^{2}_{A_{g}}$ enter Eq.~\eqref{eq:decomposition}.
It therefore couples to the triplet superconductivity.
In summary the phenomenological free energy can be formulated as 
\begin{align}\label{eq:GL3}
    F^{(2)}_{\mathbf{L}}& = 
    (\kappa_1 \varepsilon_{xy} +  \kappa_2 \mathbf{M}\cdot \mathbf{L})(|\Delta_1^{B_{1u}}|^2-|\Delta_2^{B_{1u}}|^2)  
    \notag \\
   & + \left(\kappa_3  \Delta^{B_{1u}}_1[\Delta^{B_{1u}}_2]^* M_+^2 + c.c. \right) \,.
\end{align}
The first term of Eq.~\eqref{eq:GL3} shows that the strain splits the $T_{c0}^{B_{1u}}$ by an amount $\delta T_c$ proportional to the strain.
This change originate from the change of the density of states for a given band of a bandwidth, $W$.
It can be estimated as $\Delta T_c/T_{c0}^{B_{1u}} \propto \Delta E_\Gamma/W$, where the strain induced splitting at $\Delta E_\Gamma$ has been introduced in Sec.~\ref{sec:Lieb}.

The spin state wave-function of the Cooper pairs is encoded in the $\mathbf{d}$-vector, $(\mathbf{d}_\mathbf{k} \cdot \boldsymbol{\sigma}) i \sigma_y $ \cite{Sigrist1991}.
According to Eq.~\eqref{eq:BdG1} the two types of the Cooper pairs transforming as $\mathbf{\Gamma}^{1}_{B_{1u}}$ can be assigned a sublattice index. 
The two $\mathbf{d}$-vectors read $\mathbf{d}_{\mathbf{k},A} = \phi^{B_{1u}}_{\mathbf{k}}(\hat{X} + i \hat{Y})$ and $\mathbf{d}_{\mathbf{k},B} = \phi^{B_{1u}}_{C_{2x}^{-1}\mathbf{k}}(\hat{X} - i \hat{Y})$   \cite{Brekke2023}.
The two sets of Cooper pairs carry the opposite spin angular momentum, 
$\mathbf{S}_{\mathbf{k},A} = i \mathbf{d}_{\mathbf{k},A} \times [\mathbf{d}_{\mathbf{k},A}]^* =  
|\phi^{B_{1u}}_{\mathbf{k}}|^2 \hat{Z}$ 
and similarly, $\mathbf{S}_{\mathbf{k},b} =
-|\phi^{B_{1u}}_{C_{2x}^{-1}\mathbf{k}}|^2 \hat{Z}$. 
We note that unlike in the conventional relativistic classification, $\mathbf{d}_\mathbf{k}$ is detached from the real space transformations and does not transform as a vector.
Instead, the pairs can be thought of as colored differently for the two-component order parameters or the by the same color for a single-component ones. 

At zero strain and magnetization, $|\Delta_1^{B_{1u}}| = |\Delta_2^{B_{1u}}|$, and the total angular momentum carried by the electrons of the two partners cancel each other out.
The superconductivity hence is on average unitary.
It follows from Eq.~\eqref{eq:GL3}, that certain strain disrupts this balance, $|\Delta_1^{B_{1u}}| \neq |\Delta_2^{B_{1u}}|$ and turns the superconducting state globally non-unitary.

The second term in the free energy, Eq.~\eqref{eq:GL3} shows that the magnetization $\mathbf{M}_\perp \perp\mathbf{L}$ imposes a two-fold variation of the relative phase of the two partner order parameters as $\mathbf{M}_\perp$ rotates. 
Such a magnetization can be induced by a weak field polarizing the itinerant carriers. 
Potentially, this may play a role in phase-sensitive superconducting systems.

\section{Conclusions}
\label{sec:conclusions}

We have analyzed the symmetries and mechanisms responsible for the emergence of the strain-induced magnetization in collinear altermagnets. The symmetry analysis is contained in Table \ref{tab:nrpiezo} listing allowed leading-order responses in the nonrelativistic limit, and in Table \ref{tab:relpiezo} tabulating strain-induced bilinear $L$-$M$ invariants. The nonrelativistic response includes the band-filling effect in metals and the temperature-dependence exchange-driven contribution in all altermagnets, including insulators. The most important relativistic effect is the strain-induced Dzyaloshinskii-Moriya interaction, which can induced the magnetization transverse to the altermagnetic order parameter via spin canting. To illustrate these mechanisms, we have used a simple tight-binding model for the Lieb lattice and first-principles calculations for transition-metal fluorides (MnF$_2$, FeF$_2$, CoF$_2$), CrSb, and MnTe.

We have also examined the hypothetical triplet superconductivity in altermagnets. 
In the absence of strain, such superconductivity is unitary on average, but it becomes non-unitary when piezomagnetically active strain is present. The equal-spin triplets form a two-component superconducting order parameter, with the two components represented by Cooper pairs localized on the two sublattices. The strain that breaks the equivalence of the sublattices simultaneously favors one component of the order parameter over the other, thereby inducing a non-zero total angular momentum carried by the Cooper pairs. 
Such strain also affects triplet superconductivity indirectly through piezomagnetism. 
In addition, the two-component nature of the superconducting order parameter enables control of the relative phase between those components via a magnetic field applied perpendicular to the N\'eel vector.

\emph{Note added:} After the completion of this work, we noticed that the symmetry classification in Table XIV of Ref. \cite{Schiff2025a} overlaps with parts of our Tables I and III. Our results are presented directly in terms of strain invariants, including nonlinear cases and additional relativistic terms in the presence of a symmetry-enforced axis. We also note that the strain-induced nonrelativistic Zeeman-type spin splittings for $d$-wave altermagnets listed in Table II of Ref. \onlinecite{Zhai2025} can be mapped to the corresponding part of our Table \ref{tab:nrpiezo}.

\begin{acknowledgments}
MK acknowledges the financial support from the Israel Science Foundation, Grant No. 2665/20. I.I.M. was supported by the Army Research Office under Cooperative Agreement Number W911NF-22-2-0173.
KDB was supported by the U.S. Department of Energy (DOE) Established Program to Stimulate Competitive Research (EPSCoR) through Grant No. DE-SC0024284. 
S.M. would like to acknowledge the startup fund from the University of South Carolina. 
This work utilized the Holland Computing Center of the University of Nebraska, which receives support from the UNL Office of Research and Economic Development and from the Nebraska Research Initiative.
This work also used the Expanse supercomputer at the San Diego Supercomputer Center through allocation PHY230093 from the Advanced Cyberinfrastructure Coordination Ecosystem: Services \& Support (ACCESS) program, which is supported by National Science Foundation Grants
No. 2138259, No. 2138286, No. 2138307, No. 2137603, and No. 2138296.
\end{acknowledgments}

\appendix
\section{Irreducible representations of the spin point group}
\label{app:irreps}
In this appendix we derive the irreps of the spin point groups including those transforming non-trivially under the spin-only group elements forming $\mathbf{C}_{\infty}$. We focus on unitary operations forming a unitary group $\mathcal{G}_u= \mathbf{C}_{\infty} \times \mathbf{R}_s$.
According to Eq.~\eqref{eq:sg13} and the discussion therein,
\begin{align}\label{eq:CP13}
    \mathcal{G}_u = [\mathbf{C}_\infty||\mathbf{H}]+[C_{2\hat{\mathbf{L}}_{\perp}}||\mathcal{A}][\mathbf{C}_\infty||\mathbf{H}]\, ,
\end{align}
In the cases of interest we have been able to choose the lattice operation, $\mathcal{A}$ to satisfy $\mathcal{A}^2 = E$.
As Cooper pairs have an integer spin this entails $[C_{2\hat{\mathbf{L}}_{\perp}}||\mathcal{A}]^2 = E $, where we set $[E||E] =E$.
We conclude that unitary spin symmetries are a semidirect product, $\mathcal{G}_u = [\mathbf{C}_\infty||\mathbf{H}] \rtimes \mathbf{G}_2$, where $\mathbf{G}_2 = \{E, [C_{2\hat{\mathbf{L}}_\perp}||\mathcal{A}]\}$.
The group $[\mathbf{C}_\infty||\mathbf{H}]$ is a normal subgroup of $\mathcal{G}_u$ of index 2.
A generic element $g$ of $[\mathbf{C}_\infty||\mathbf{H}]$ has a form, $g=[C_{\phi\hat{\mathbf{L}}}||h]$, where 
$C_{\phi\hat{\mathbf{L}}}$ is a spin rotation by an angle $\phi$ around $\mathbf{L}$, and $h \in \mathbf{H}$.

We follow the standard algorithm for constructing the irreps of $\mathcal{G}_u$ out of the irreps of its normal subgroup \cite{Bradley1972}.
For clarity, we illustrate each step for the paradigmatic $d$-wave rutile altermagnet with spin group ${}^24/{}^1m{}^2m_y{}^1m_d$ and crystallographic point group $4/mmm$ ($D_{4h}$) (see Table~\ref{tab:nrpiezo}).
In this case, the site group $\mathbf{H} = mmm$ ($D_{2h}$) is abelian.
We set $\mathcal{A}= T_{\mathbf{t}}C_{2x}$, where $\mathbf{t}=(1/2,1/2,1/2)$, so that $\mathcal{A}^2 =E$. 

The irreps of $[\mathbf{C}_\infty||\mathbf{H}]$ are $\mathbf{D}^{m_s}_{\alpha_{g(u)}} =D^s_{m_s} \times \alpha_{{g(u)}}$, where 
$D^s_{m_s}$ are irreps of $\mathbf{C}_\infty$ labeled by the total angular momentum component along $\hat{L}$, $m_s$.
$\alpha_{g(u)}$ are the irreps of $\mathbf{H}$.
For the rutile structure, $\alpha = A, B_{i}$, $i=1,2,3$.

The next step of the algorithm is to break the irreps $\mathbf{D}$ into orbits.
The orbits of length one are $\{ \mathbf{D}^{0}_{A_{g(u)}} \}$ and $ \{ \mathbf{D}^{0}_{B_{1g(u)}} \} $.
The remaining orbits are of length two.
These include ($m_s>0$)
$\{ \mathbf{D}^{0}_{B_{2g(u)}},\mathbf{D}^{0}_{B_{3g(u)}} \}$, 
$ \{ \mathbf{D}^{\pm m_s}_{B_{2g(u)}},\mathbf{D}^{ \mp m_s}_{B_{3g(u)}} \}$, and
$\{ \mathbf{D}^{m_s}_{\beta_{g(u)}},\mathbf{D}^{-m_s}_{\beta_{g(u)}} \}$ for $\beta=A, B_1$ throughout the section.

The irreps of $\mathcal{G}_u$ denoted as $\mathbf{\Gamma}$ are obtained by induction of the small irreps of the little group for each orbit.
For the orbits of length one the little group is $\mathcal{G}_u$ itself, and hence no induction is needed.

The resulting irreps are obtained by combining the irrep of the given orbit with the irreps of $\mathbf{G}_2$.
The two trivial irreps of the latter assign $\pm 1$ to the element $r$.
Correspondingly, each orbit of length one produces  two one-dimensional irreps of $\mathcal{G}_u$ distinguished by the $\pm$ superscript.
They are specified by the representation of $g = [C_{\phi\hat{L}}||h]\in [\mathbf{C}_\infty||\mathbf{H}]$ and the representation of the element $r$.
In summary, we obtain the one-dimensional irreps listed in Eq.~\eqref{eq:1Dirrep}.

Each orbit of length two gives rise to one two-dimensional irrep of $\mathcal{G}_u$.
These irreps can be induced from any one of the orbit representative.
Therefore, we have the following irreps: 
$\mathbf{\Gamma}^{0}_{B_{2g(u)}} = \mathbf{D}^{0}_{B_{2g(u)}}\!\! \uparrow\!\mathcal{G}_u$,
$\mathbf{\Gamma}^{\pm m_s}_{B_{2g(u)}} = \mathbf{D}^{\pm m_s}_{B_{2g(u)}} \!\! \uparrow\! \mathcal{G}_u$,
and 
$\mathbf{\Gamma}^{m_s}_{\beta_{g(u)}} = \mathbf{D}^{m_s}_{\beta_{g(u)}}\!\! \uparrow\! \mathcal{G}_u$.
The resulting two-dimensional irreps are fully specified by Eq.~\eqref{eq:2Dirrep}.

The generalization to other SLG with abelian site group $\mathbf{H}$, such as ${}^2m{}^2m{}^1m$ describing the $d$-wave FeSb$_2$ altermagnetic candidate \cite{Mazin2021,Attias2024}, is straightforward. 
The cases with non-abelian $\mathbf{H}$ can be dealt with by the method of Ref. \onlinecite{Bradley1972} since 
the index of a normal subgroup in all cases is $2$, which is a prime number.
We expect some of the triplet Cooper pair wave-functions to transform as four-dimensional irreps for $g$-wave altermagnets, and as six-dimensional irreps for $i$-wave superconductors. 

\subsection{Nonrelativistic classification of Cooper pairs on the Lieb lattice}
\label{app:CooperLieb}

Here we apply the results of Sec.~\ref{sec:superconductivity} and App.~\ref{app:irreps} to the Lieb lattice shown in Fig.~\ref{fig:Lieb} in the ${}^24/{}^1m{}^2m_d{}^1m_y$ setting.
We restrict our consideration to the nearest neighbor (nn) and next-nearest neighbor (nnn) pairing.
The orbital wave-functions are classified with respect to the site group $\mathbf{H}=D_{2h}$.

\begin{table}[htb]
\centering
\caption{The $\mathbf{H} =D_{2h}$ classification of the hopping amplitude up to next-nearest neighbors on the Lieb lattice.}
\begin{tabular}{|c|c|c|c|c|}
\hline
                    & $A_g$   &   $B_{1g}$  & $B_{2u}$ &  $B_{3u}$\\
\hline
      nn  &   $\cos\frac{k_x}{2} \cos\frac{k_y}{2}$  & $\sin\frac{k_x}{2} \sin\frac{k_y}{2}$              & $\cos\frac{k_x}{2} \sin\frac{k_y}{2}$ & $\sin\frac{k_x}{2} \cos\frac{k_y}{2}$   \\
      nnn &   $\cos k_x, \cos k_y$     &               & $\sin k_y$   & $\sin k_x$  \\
\hline
\end{tabular}
\label{tab:LiebClass}
\end{table}

Below we separately classify the nn and nnn Cooper pairs in the large $J$ limit.
Some of the Cooper pairs are destroyed in this limit, and some are not.
However, all the nn Cooper pairs are suppressed by the altermagnetic spin splitting due to the broken time reversal symmetry. 
Such suppression is illustrated by the splitting of the Fermi contours shown in Fig.~\ref{fig:LiebPockets}.
We will see that some of the nnn Cooper pairs survive both the large $J$ and altermagnetic splitting. 
Such spin triplet pairs are formed by electrons residing on the same Fermi contour.  
For completeness we present both the nn and nnn Cooper pair classification. 

The general form of the Cooper pair amplitude is defined by Eq.~\eqref{eq:partners2}.
We therefore have to classify the the momentum dependent part $\varphi_l(\mathbf{k})$ for nn and nnn pairing. 

\subsubsection{The nn Cooper pairs}
We first classify the nn Cooper pairs.
The momentum dependent part $\varphi_l(\mathbf{k})$ for nn are presented in the first line of Tab.~\ref{tab:LiebClass}.
Combining these functions with all the possible spin- and lattice- dependent functions as given in Eq.~\eqref{eq:partners2}, and complying with the Pauli principle results in all possible Cooper pair wave functions summarized in Tab.~\ref{tab:LiebClassNN}.
In particular, as indicated by the last column of Tab.~\ref{tab:LiebClassNN} the singlet pairing $P^s$ is represented by two operators given by Eq.~\eqref{eq:nn_singlets}. 
These singlets survive the large $J$ limit.

\begin{table}[htb]
\centering
\caption{Classification of the nearest neighbor Cooper pairing according to the irreps of the spin point group. The symbols $\checkmark$ and $\xmark$ denote irreps surviving or not surviving at large exchange splitting $J$.
All the nn Cooper pairs are suppressed by the altermagnetic spin splitting.}
\begin{tabular}{|c|c|c|c|}
\hline
    nn                & $\{ P^t_1,P^t_{-1} \}$ $\xmark$   &  $P^t_{0}$ $\checkmark$ &  $P^s$ $\checkmark$\\
\hline
      $\tau_x$ &   $\mathbf{\Gamma}^{+ 1}_{B_{2u}}, \mathbf{\Gamma}^{- 1}_{B_{2u}}$  & $\mathbf{\Gamma}^{ 0}_{B_{2u}}$              & $\mathbf{\Gamma}^{0;+ 1}_{A_{g}}, \mathbf{\Gamma}^{0;+1}_{B_{1g}}$   \\
      $\tau_y$ &   $\mathbf{\Gamma}^{+ 1}_{A_{g}}, \mathbf{\Gamma}^{+1}_{B_{1g}}$     &    $\mathbf{\Gamma}^{0;+}_{A_{g}}, \mathbf{\Gamma}^{0;+}_{B_{1g}}$        & $\mathbf{\Gamma}^{ 0}_{B_{2u}}$ \\
\hline
\end{tabular}
\label{tab:LiebClassNN}
\end{table}

Similarly, zero-spin triplet pairs $P^t_0$ comply with the Pauli principle by being antisymmetric with respect to the sublattice index, and have the same momentum dependence:

\begin{subequations}\label{eq:nn_triplets}
\begin{align}\label{eq:nn_triplets_a}
    \phi^{\mathbf{\Gamma}_\ell}_{1;\sigma\sigma'}(\mathbf{k},i,j) = [P^t_0]_{\sigma\sigma'} [\tau_y]_{ij} \cos\frac{k_x}{2} \cos\frac{k_y}{2}\, ,
\end{align}
\begin{align}\label{eq:nn_triplets_b}
    \phi^{\mathbf{\Gamma}_\ell}_{1;\sigma\sigma'}(\mathbf{k},i,j) = [P^t_0]_{\sigma\sigma'} [\tau_y]_{ij} 
    \sin\frac{k_x}{2} \sin\frac{k_y}{2}\, 
\end{align}
\end{subequations}
for $\mathbf{\Gamma}_{\ell}=\mathbf{\Gamma}^{0;+}_{A_{g}}$ and $\mathbf{\Gamma}_{\ell}=\mathbf{\Gamma}^{0;+}_{B_{1g}}$, respectively. 
These irreps carry superscript $+$ as they are even under the $(C_{2\mathbf{L}_{\perp}}||\mathcal{A})$ operation. 
Indeed, the spin part of the wave function $|\uparrow \downarrow \rangle +|\downarrow \uparrow \rangle $ acquires the total Berry phase $\pi$ and changes the sign under the two-fold rotation in the spin space, $[C_{2\hat{\mathbf{L}}_{\perp}}||E]$, and the orbital part is odd under the sublattice exchange affected by the 
$[E||\mathcal{A}]$ operation. 

The singlet and triplet Cooper pairs \eqref{eq:nn_singlets}, \eqref{eq:nn_triplets} could be equally important as both survive the large $J$ limit.
In contrast, nn equal-spin triplets $P^t_{\pm1}$ are suppressed at large $J$. 
Hence we only present the remaining zero-spin triplets forming a two-component order parameter of the $\mathbf{\Gamma}^0_{B_{2u}}$ symmetry,
\begin{align}\label{eq:nn_triplets_B2u}
    & \left\{ \phi^{\mathbf{\Gamma}_\ell}_{1;\sigma\sigma'}(\mathbf{k},i,j), \phi^{\mathbf{\Gamma}_\ell}_{2;\sigma\sigma'}(\mathbf{k},i,j) \right\}   = 
    [P^t_0]_{\sigma \sigma'} [\tau_x]_{ij} 
    \notag \\
    &\times 
\left\{ \cos\frac{k_x}{2} \sin\frac{k_y}{2}, \sin\frac{k_x}{2} \cos\frac{k_y}{2}
    \right\}
    \, ,(\mathbf{\Gamma}_\ell = \mathbf{\Gamma}^0_{B_{2u}})
\end{align}
identified in \cite{Chakraborty2024} as a mixed spin, $p$-wave triplet Cooper pair wave-function.
Note that it coexists with the two-component singlet Cooper pair wave-function 

\begin{align}\label{eq:nn_singlet_B2u}
    & \left\{ \phi^{\mathbf{\Gamma}_\ell}_{1;\sigma\sigma'}(\mathbf{k},i,j), \phi^{\mathbf{\Gamma}_\ell}_{2;\sigma\sigma'}(\mathbf{k},i,j) \right\}   = 
    [P^s]_{\sigma \sigma'} [\tau_y]_{ij} 
    \notag \\
    &\times 
\left\{ \cos\frac{k_x}{2} \sin\frac{k_y}{2}, \sin\frac{k_x}{2} \cos\frac{k_y}{2}
    \right\}
    \, ,(\mathbf{\Gamma}_\ell = \mathbf{\Gamma}^0_{B_{2u}})
\end{align}
having the same momentum dependence and belonging to the same $\mathbf{\Gamma}^0_{B_{2u}}$ spin group symmetry. 
\subsubsection{The nnn Cooper pairs}
The momentum dependence of hopping amplitudes is given in the last row of Tab.~\ref{tab:LiebClass}. 
As the nnn Cooper pairs on the Lieb lattice are intra-sublattice, they are clearly symmetric with respect to the sublattice interchange. 
As in the case of nn, combining these functions with all the possible spin- and lattice- dependent functions as given in Eq.~\eqref{eq:partners2}, results in all possible Cooper pair wave functions summarized in Tab.~\ref{tab:LiebClassNNN}.

\begin{table}[htb]
\centering
\caption{Classification of the next-nearest neighbor Cooper pairing according to the irreps of the spin point group. $\checkmark$ and $\xmark$ denote irreps surviving the large exchange splitting $J$.
The nnn parallel spin triplets are not suppressed by exchange and altermagnetic spin splitting.}
\begin{tabular}{|c|c|c|c|}
\hline
    nnn                & $\{ P^t_1,P^t_{-1} \}$ $\checkmark$   &  $P^t_{0}$ $\xmark$ &  $P^s$ $\xmark$\\
\hline
      $\tau_{0,z}$ &   $\mathbf{\Gamma}^{+ 1}_{B_{2u}}, \mathbf{\Gamma}^{- 1}_{B_{2u}}$  & $\mathbf{\Gamma}^{ 0}_{B_{2u}}$              & $\mathbf{\Gamma}^{0;+ 1}_{A_{g}}$   \\
\hline
\end{tabular}
\label{tab:LiebClassNNN}
\end{table}

In contrast to the nn pairing only the equal-spin triplets are robust against exchange $J$.
The corresponding Cooper pair wave function is given by Eq.~\eqref{eq:nnn_triplet_B2u}.
In addition thanks to the dispersion symmetry, 
Due to the spectral symmetry $E_\sigma(\mathbf{k}) = E_\sigma(-\mathbf{k})$ these triplet Cooper pairs are immune to the pair breaking imposed by altermagnetic spin splitting. 
This makes these triplet Cooper pairs are the most robust among all the Cooper pairs we have considered.

\subsection{Classification of the hopping amplitudes in the rutile structure}
\label{sec:rutile_pairs}

For the rutile structure we use the setting ${}^24/{}^1m{}^2m_y{}^1m_d$ which differs from that used for the Lieb lattice by a rotation by $\pi/4$.
Again, we restrict the analysis to the case of the large $J$ limit.
The spin-singlet Cooper pairs surviving in this limit are formed by electrons at different sublattices, while the spin-triplet pairs are intrasublattice.
It is sufficient to classify the nn and nnn Cooper pairs. 
The nn hopping amplitudes of even (odd) parity are classified in Tab.~\ref{tab:Rut_nn}.

\section{Computational details}
\begin{table}[htb]
\centering
\caption{Calculated structure information for MF$_2$ with $U$ and $J$ values from constrained occupation method. Experimental structural information \cite{Stout} are shown in bracket. The local moments on transition metal ions from neutron scattering experiments for are also listed in the brackets.}
\begin{tabular}{|c|c|c|c|}
\hline
                    & MnF$_2$   &   FeF$_2$  & CoF$_2$\\
\hline
      $U$ (eV)          &   5.98          & 5.90              & 5.01    \\
      $J$  (eV)         &   0.98          & 0.85             & 0.91      \\
\hline
      $a$ (\AA)         &       4.885 (4.873)         & 4.689 (4.697)  &  4.676 (4.695)\\
      $c$ (\AA)         &       3.299 (3.310)         & 3.281 (3.309)  &  3.171 (3.110)\\
      $u$                     &         0.305 (0.310)        & 0.299 (0.305)   & 0.305 (0.308)\\
      M-F$_1$ (\AA)      & 2.11  &1.99 & 2.02 \\
      M-F$_2$ (\AA)     & 2.13 &2.11  & 2.04\\
      $\mu_{M}$ ($\mu_B$)  &  4.72 (4.6 \cite{Yamani})            & 3.81 (3.75 \cite{Almeida})       &  2.80 (2.57 \cite{Chatterji})    \\  \hline
\end{tabular}
\label{UJ}
\end{table}

Kohn-Sham equations were solved using projector augmented wave (PAW) method \cite{Blochl} as implemented in the Vienna ab initio simulation package (VASP) \cite{Kresse,Kresse2}. 
For transition metal fluorides, we employed the PBEsol exchange-correlation functional \cite{Perdew} and pseudopotentials with $2p$ orbitals on the transition-metal atoms treated as core.
The energy cutoff for the plane wave expansion was set to 520 eV, and a 6$\times$6$\times$8 Brillouin zone mesh was used to relax the  tetragonal unit cell.
We adopted a Gaussian smearing of 0.02 eV for Brillouin zone integration for relaxation calculations, and tetrahedron method with Bl\"ochl corrections in the total energy calculations. Hellmann-Feynman forces were  converged to 0.005 eV/\AA. To extract the exchange parameters in fluorides, we extended to an orthorhombic 1$\times$2$\times$2 supercell to include more neighbor shells, and fitted the total energy of eight different spin configurations to Heisenberg Model with exchange interactions up to the third nearest neighbor shell. Spin-orbit coupling was included for the constraint field calculations for strain-induced canting moments.  

For MnTe and CrSb, the $2p$ orbitals for Cr and Sb atoms were treated as valence electrons, and we used the PBE functional \cite{PBE}. In the calculations of strain-induced DMI energies using the generalized force theorem, we used the Brillouin zone mesh of 22$\times$22$\times$15 and the plane wave expansion cutoff to 600 eV.

To take into account electronic correlation on transition-metal ions in fluorides and MnTe, we adopted the DFT+$U$ method \cite{Liechtenstein}. The values of $U$ and $J$ were calculated using the constrained occupation method using the FLEUR code~\cite{fleurCode}, and they are listed in Table \ref{UJ} along with the structural parameters. For MnTe we used $U=4.8$ and $J=0.8$ eV \cite{Mu2019,Mazin2023,Belashchenko2025}.


\begin{thebibliography}{96}%
\makeatletter
\providecommand \@ifxundefined [1]{%
 \@ifx{#1\undefined}
}%
\providecommand \@ifnum [1]{%
 \ifnum #1\expandafter \@firstoftwo
 \else \expandafter \@secondoftwo
 \fi
}%
\providecommand \@ifx [1]{%
 \ifx #1\expandafter \@firstoftwo
 \else \expandafter \@secondoftwo
 \fi
}%
\providecommand \natexlab [1]{#1}%
\providecommand \enquote  [1]{``#1''}%
\providecommand \bibnamefont  [1]{#1}%
\providecommand \bibfnamefont [1]{#1}%
\providecommand \citenamefont [1]{#1}%
\providecommand \href@noop [0]{\@secondoftwo}%
\providecommand \href [0]{\begingroup \@sanitize@url \@href}%
\providecommand \@href[1]{\@@startlink{#1}\@@href}%
\providecommand \@@href[1]{\endgroup#1\@@endlink}%
\providecommand \@sanitize@url [0]{\catcode `\\12\catcode `\$12\catcode `\&12\catcode `\#12\catcode `\^12\catcode `\_12\catcode `\%12\relax}%
\providecommand \@@startlink[1]{}%
\providecommand \@@endlink[0]{}%
\providecommand \url  [0]{\begingroup\@sanitize@url \@url }%
\providecommand \@url [1]{\endgroup\@href {#1}{\urlprefix }}%
\providecommand \urlprefix  [0]{URL }%
\providecommand \Eprint [0]{\href }%
\providecommand \doibase [0]{https://doi.org/}%
\providecommand \selectlanguage [0]{\@gobble}%
\providecommand \bibinfo  [0]{\@secondoftwo}%
\providecommand \bibfield  [0]{\@secondoftwo}%
\providecommand \translation [1]{[#1]}%
\providecommand \BibitemOpen [0]{}%
\providecommand \bibitemStop [0]{}%
\providecommand \bibitemNoStop [0]{.\EOS\space}%
\providecommand \EOS [0]{\spacefactor3000\relax}%
\providecommand \BibitemShut  [1]{\csname bibitem#1\endcsname}%
\let\auto@bib@innerbib\@empty
\bibitem [{\citenamefont {\ifmmode~\check{S}\else \v{S}\fi{}mejkal}\ \emph {et~al.}(2022{\natexlab{a}})\citenamefont {\ifmmode~\check{S}\else \v{S}\fi{}mejkal}, \citenamefont {Sinova},\ and\ \citenamefont {Jungwirth}}]{Smejkal2022a}%
  \BibitemOpen
  \bibfield  {author} {\bibinfo {author} {\bibfnamefont {L.}~\bibnamefont {\ifmmode~\check{S}\else \v{S}\fi{}mejkal}}, \bibinfo {author} {\bibfnamefont {J.}~\bibnamefont {Sinova}},\ and\ \bibinfo {author} {\bibfnamefont {T.}~\bibnamefont {Jungwirth}},\ }\bibfield  {title} {\bibinfo {title} {Emerging research landscape of altermagnetism},\ }\href {https://doi.org/10.1103/PhysRevX.12.040501} {\bibfield  {journal} {\bibinfo  {journal} {Phys. Rev. X}\ }\textbf {\bibinfo {volume} {12}},\ \bibinfo {pages} {040501} (\bibinfo {year} {2022}{\natexlab{a}})}\BibitemShut {NoStop}%
\bibitem [{\citenamefont {Mazin}(2022)}]{Mazin2022}%
  \BibitemOpen
  \bibfield  {author} {\bibinfo {author} {\bibfnamefont {I.}~\bibnamefont {Mazin}} (\bibinfo {collaboration} {The PRX Editors}),\ }\bibfield  {title} {\bibinfo {title} {Editorial: Altermagnetism---a new punch line of fundamental magnetism},\ }\href {https://doi.org/10.1103/PhysRevX.12.040002} {\bibfield  {journal} {\bibinfo  {journal} {Phys. Rev. X}\ }\textbf {\bibinfo {volume} {12}},\ \bibinfo {pages} {040002} (\bibinfo {year} {2022})}\BibitemShut {NoStop}%
\bibitem [{\citenamefont {Bai}\ \emph {et~al.}(2024)\citenamefont {Bai}, \citenamefont {Feng}, \citenamefont {Liu}, \citenamefont {Šmejkal}, \citenamefont {Mokrousov},\ and\ \citenamefont {Yao}}]{Bai2024}%
  \BibitemOpen
  \bibfield  {author} {\bibinfo {author} {\bibfnamefont {L.}~\bibnamefont {Bai}}, \bibinfo {author} {\bibfnamefont {W.}~\bibnamefont {Feng}}, \bibinfo {author} {\bibfnamefont {S.}~\bibnamefont {Liu}}, \bibinfo {author} {\bibfnamefont {L.}~\bibnamefont {Šmejkal}}, \bibinfo {author} {\bibfnamefont {Y.}~\bibnamefont {Mokrousov}},\ and\ \bibinfo {author} {\bibfnamefont {Y.}~\bibnamefont {Yao}},\ }\bibfield  {title} {\bibinfo {title} {Altermagnetism: Exploring new frontiers in magnetism and spintronics},\ }\href {https://doi.org/https://doi.org/10.1002/adfm.202409327} {\bibfield  {journal} {\bibinfo  {journal} {Advanced Functional Materials}\ }\textbf {\bibinfo {volume} {34}},\ \bibinfo {pages} {2409327} (\bibinfo {year} {2024})}\BibitemShut {NoStop}%
\bibitem [{\citenamefont {Tamang}\ \emph {et~al.}(2025)\citenamefont {Tamang}, \citenamefont {Gurung}, \citenamefont {Rai}, \citenamefont {Brahimi},\ and\ \citenamefont {Lounis}}]{Tamang2024}%
  \BibitemOpen
  \bibfield  {author} {\bibinfo {author} {\bibfnamefont {R.}~\bibnamefont {Tamang}}, \bibinfo {author} {\bibfnamefont {S.}~\bibnamefont {Gurung}}, \bibinfo {author} {\bibfnamefont {D.~P.}\ \bibnamefont {Rai}}, \bibinfo {author} {\bibfnamefont {S.}~\bibnamefont {Brahimi}},\ and\ \bibinfo {author} {\bibfnamefont {S.}~\bibnamefont {Lounis}},\ }\bibfield  {title} {\bibinfo {title} {Altermagnetism and altermagnets: A brief review},\ }\bibfield  {journal} {\bibinfo  {journal} {Magnetism}\ }\textbf {\bibinfo {volume} {5}},\ \href {https://doi.org/10.3390/magnetism5030017} {10.3390/magnetism5030017} (\bibinfo {year} {2025})\BibitemShut {NoStop}%
\bibitem [{\citenamefont {Ma}\ \emph {et~al.}(2021)\citenamefont {Ma}, \citenamefont {Hu}, \citenamefont {Li}, \citenamefont {Liu}, \citenamefont {Yao}, \citenamefont {Jia},\ and\ \citenamefont {Liu}}]{Junwei2021}%
  \BibitemOpen
  \bibfield  {author} {\bibinfo {author} {\bibfnamefont {H.-Y.}\ \bibnamefont {Ma}}, \bibinfo {author} {\bibfnamefont {M.}~\bibnamefont {Hu}}, \bibinfo {author} {\bibfnamefont {N.}~\bibnamefont {Li}}, \bibinfo {author} {\bibfnamefont {J.}~\bibnamefont {Liu}}, \bibinfo {author} {\bibfnamefont {W.}~\bibnamefont {Yao}}, \bibinfo {author} {\bibfnamefont {J.-F.}\ \bibnamefont {Jia}},\ and\ \bibinfo {author} {\bibfnamefont {J.}~\bibnamefont {Liu}},\ }\bibfield  {title} {\bibinfo {title} {Multifunctional antiferromagnetic materials with giant piezomagnetism and noncollinear spin current},\ }\href {https://doi.org/10.1038/s41467-021-23127-7} {\bibfield  {journal} {\bibinfo  {journal} {Nat. Commun.}\ }\textbf {\bibinfo {volume} {12}},\ \bibinfo {pages} {2846} (\bibinfo {year} {2021})}\BibitemShut {NoStop}%
\bibitem [{\citenamefont {Aoyama}\ and\ \citenamefont {Ohgushi}(2024)}]{Aoyama2024}%
  \BibitemOpen
  \bibfield  {author} {\bibinfo {author} {\bibfnamefont {T.}~\bibnamefont {Aoyama}}\ and\ \bibinfo {author} {\bibfnamefont {K.}~\bibnamefont {Ohgushi}},\ }\bibfield  {title} {\bibinfo {title} {Piezomagnetic properties in altermagnetic {MnTe}},\ }\href {https://doi.org/10.1103/PhysRevMaterials.8.L041402} {\bibfield  {journal} {\bibinfo  {journal} {Phys. Rev. Mater.}\ }\textbf {\bibinfo {volume} {8}},\ \bibinfo {pages} {L041402} (\bibinfo {year} {2024})}\BibitemShut {NoStop}%
\bibitem [{\citenamefont {Chakraborty}\ \emph {et~al.}(2024)\citenamefont {Chakraborty}, \citenamefont {Gonz\'alez~Hern\'andez}, \citenamefont {\ifmmode~\check{S}\else \v{S}\fi{}mejkal},\ and\ \citenamefont {Sinova}}]{Chakraborty2024a}%
  \BibitemOpen
  \bibfield  {author} {\bibinfo {author} {\bibfnamefont {A.}~\bibnamefont {Chakraborty}}, \bibinfo {author} {\bibfnamefont {R.}~\bibnamefont {Gonz\'alez~Hern\'andez}}, \bibinfo {author} {\bibfnamefont {L.}~\bibnamefont {\ifmmode~\check{S}\else \v{S}\fi{}mejkal}},\ and\ \bibinfo {author} {\bibfnamefont {J.}~\bibnamefont {Sinova}},\ }\bibfield  {title} {\bibinfo {title} {Strain-induced phase transition from antiferromagnet to altermagnet},\ }\href {https://doi.org/10.1103/PhysRevB.109.144421} {\bibfield  {journal} {\bibinfo  {journal} {Phys. Rev. B}\ }\textbf {\bibinfo {volume} {109}},\ \bibinfo {pages} {144421} (\bibinfo {year} {2024})}\BibitemShut {NoStop}%
\bibitem [{\citenamefont {Belashchenko}(2025)}]{Belashchenko2025}%
  \BibitemOpen
  \bibfield  {author} {\bibinfo {author} {\bibfnamefont {K.~D.}\ \bibnamefont {Belashchenko}},\ }\bibfield  {title} {\bibinfo {title} {Giant strain-induced spin splitting effect in {MnTe}, a $g$-wave altermagnetic semiconductor},\ }\href {https://doi.org/10.1103/PhysRevLett.134.086701} {\bibfield  {journal} {\bibinfo  {journal} {Phys. Rev. Lett.}\ }\textbf {\bibinfo {volume} {134}},\ \bibinfo {pages} {086701} (\bibinfo {year} {2025})}\BibitemShut {NoStop}%
\bibitem [{\citenamefont {Takahashi}\ \emph {et~al.}(2025)\citenamefont {Takahashi}, \citenamefont {Steward}, \citenamefont {Ogata}, \citenamefont {Fernandes},\ and\ \citenamefont {Schmalian}}]{Takahashi2025}%
  \BibitemOpen
  \bibfield  {author} {\bibinfo {author} {\bibfnamefont {K.}~\bibnamefont {Takahashi}}, \bibinfo {author} {\bibfnamefont {C.~R.~W.}\ \bibnamefont {Steward}}, \bibinfo {author} {\bibfnamefont {M.}~\bibnamefont {Ogata}}, \bibinfo {author} {\bibfnamefont {R.~M.}\ \bibnamefont {Fernandes}},\ and\ \bibinfo {author} {\bibfnamefont {J.}~\bibnamefont {Schmalian}},\ }\bibfield  {title} {\bibinfo {title} {{Elasto-Hall} conductivity and the anomalous {Hall} effect in altermagnets},\ }\href {https://doi.org/10.1103/PhysRevB.111.184408} {\bibfield  {journal} {\bibinfo  {journal} {Phys. Rev. B}\ }\textbf {\bibinfo {volume} {111}},\ \bibinfo {pages} {184408} (\bibinfo {year} {2025})}\BibitemShut {NoStop}%
\bibitem [{\citenamefont {Karetta}\ \emph {et~al.}(2025)\citenamefont {Karetta}, \citenamefont {Verbeek}, \citenamefont {Jaeschke-Ubiergo}, \citenamefont {Šmejkal},\ and\ \citenamefont {Sinova}}]{Karetta2025}%
  \BibitemOpen
  \bibfield  {author} {\bibinfo {author} {\bibfnamefont {B.}~\bibnamefont {Karetta}}, \bibinfo {author} {\bibfnamefont {X.~H.}\ \bibnamefont {Verbeek}}, \bibinfo {author} {\bibfnamefont {R.}~\bibnamefont {Jaeschke-Ubiergo}}, \bibinfo {author} {\bibfnamefont {L.}~\bibnamefont {Šmejkal}},\ and\ \bibinfo {author} {\bibfnamefont {J.}~\bibnamefont {Sinova}},\ }\href {https://arxiv.org/abs/2505.21293} {\bibinfo {title} {Strain controlled $g$- to $d$-wave transition in altermagnetic {CrSb}}} (\bibinfo {year} {2025}),\ \Eprint {https://arxiv.org/abs/2505.21293} {arXiv:2505.21293} \BibitemShut {NoStop}%
\bibitem [{\citenamefont {Birss}(1964)}]{Birss1964}%
  \BibitemOpen
  \bibfield  {author} {\bibinfo {author} {\bibfnamefont {R.}~\bibnamefont {Birss}},\ }\href@noop {} {\emph {\bibinfo {title} {Symmetry and Magnetism}}},\ Selected topics in solid state physics\ (\bibinfo  {publisher} {North-Holland Publishing Company},\ \bibinfo {year} {1964})\BibitemShut {NoStop}%
\bibitem [{\citenamefont {\ifmmode~\check{S}\else \v{S}\fi{}mejkal}\ \emph {et~al.}(2022{\natexlab{b}})\citenamefont {\ifmmode~\check{S}\else \v{S}\fi{}mejkal}, \citenamefont {Sinova},\ and\ \citenamefont {Jungwirth}}]{Smejkal2022}%
  \BibitemOpen
  \bibfield  {author} {\bibinfo {author} {\bibfnamefont {L.}~\bibnamefont {\ifmmode~\check{S}\else \v{S}\fi{}mejkal}}, \bibinfo {author} {\bibfnamefont {J.}~\bibnamefont {Sinova}},\ and\ \bibinfo {author} {\bibfnamefont {T.}~\bibnamefont {Jungwirth}},\ }\bibfield  {title} {\bibinfo {title} {Beyond conventional ferromagnetism and antiferromagnetism: A phase with nonrelativistic spin and crystal rotation symmetry},\ }\href {https://doi.org/10.1103/PhysRevX.12.031042} {\bibfield  {journal} {\bibinfo  {journal} {Phys. Rev. X}\ }\textbf {\bibinfo {volume} {12}},\ \bibinfo {pages} {031042} (\bibinfo {year} {2022}{\natexlab{b}})}\BibitemShut {NoStop}%
\bibitem [{\citenamefont {Litvin}\ and\ \citenamefont {Opechowski}(1974)}]{Litvin1974}%
  \BibitemOpen
  \bibfield  {author} {\bibinfo {author} {\bibfnamefont {D.}~\bibnamefont {Litvin}}\ and\ \bibinfo {author} {\bibfnamefont {W.}~\bibnamefont {Opechowski}},\ }\bibfield  {title} {\bibinfo {title} {Spin groups},\ }\href {https://doi.org/https://doi.org/10.1016/0031-8914(74)90157-8} {\bibfield  {journal} {\bibinfo  {journal} {Physica}\ }\textbf {\bibinfo {volume} {76}},\ \bibinfo {pages} {538} (\bibinfo {year} {1974})}\BibitemShut {NoStop}%
\bibitem [{\citenamefont {Litvin}(1977)}]{Litvin:a14103}%
  \BibitemOpen
  \bibfield  {author} {\bibinfo {author} {\bibfnamefont {D.~B.}\ \bibnamefont {Litvin}},\ }\bibfield  {title} {\bibinfo {title} {{Spin point groups}},\ }\href {https://doi.org/10.1107/S0567739477000709} {\bibfield  {journal} {\bibinfo  {journal} {Acta Crystallogr. Sect. A}\ }\textbf {\bibinfo {volume} {33}},\ \bibinfo {pages} {279} (\bibinfo {year} {1977})}\BibitemShut {NoStop}%
\bibitem [{\citenamefont {Mazin}(2025)}]{Mazin2022a}%
  \BibitemOpen
  \bibfield  {author} {\bibinfo {author} {\bibfnamefont {I.~I.}\ \bibnamefont {Mazin}},\ }\bibfield  {title} {\bibinfo {title} {Notes on altermagnetism and superconductivity},\ }\href {https://doi.org/10.1007/s43673-025-00158-6} {\bibfield  {journal} {\bibinfo  {journal} {AAPPS Bulletin}\ }\textbf {\bibinfo {volume} {35}},\ \bibinfo {pages} {18} (\bibinfo {year} {2025})}\BibitemShut {NoStop}%
\bibitem [{\citenamefont {Bose}\ \emph {et~al.}(2024)\citenamefont {Bose}, \citenamefont {Vadnais},\ and\ \citenamefont {Paramekanti}}]{Bose2024}%
  \BibitemOpen
  \bibfield  {author} {\bibinfo {author} {\bibfnamefont {A.}~\bibnamefont {Bose}}, \bibinfo {author} {\bibfnamefont {S.}~\bibnamefont {Vadnais}},\ and\ \bibinfo {author} {\bibfnamefont {A.}~\bibnamefont {Paramekanti}},\ }\bibfield  {title} {\bibinfo {title} {Altermagnetism and superconductivity in a multiorbital {$t$-$J$} model},\ }\href {https://doi.org/10.1103/PhysRevB.110.205120} {\bibfield  {journal} {\bibinfo  {journal} {Phys. Rev. B}\ }\textbf {\bibinfo {volume} {110}},\ \bibinfo {pages} {205120} (\bibinfo {year} {2024})}\BibitemShut {NoStop}%
\bibitem [{\citenamefont {Sumita}\ \emph {et~al.}(2023)\citenamefont {Sumita}, \citenamefont {Naka},\ and\ \citenamefont {Seo}}]{Sumita2023}%
  \BibitemOpen
  \bibfield  {author} {\bibinfo {author} {\bibfnamefont {S.}~\bibnamefont {Sumita}}, \bibinfo {author} {\bibfnamefont {M.}~\bibnamefont {Naka}},\ and\ \bibinfo {author} {\bibfnamefont {H.}~\bibnamefont {Seo}},\ }\bibfield  {title} {\bibinfo {title} {{Fulde-Ferrell-Larkin-Ovchinnikov} state induced by antiferromagnetic order in $\ensuremath{\kappa}$-type organic conductors},\ }\href {https://doi.org/10.1103/PhysRevResearch.5.043171} {\bibfield  {journal} {\bibinfo  {journal} {Phys. Rev. Res.}\ }\textbf {\bibinfo {volume} {5}},\ \bibinfo {pages} {043171} (\bibinfo {year} {2023})}\BibitemShut {NoStop}%
\bibitem [{\citenamefont {Chakraborty}\ and\ \citenamefont {Black-Schaffer}(2025)}]{Chakraborty2024}%
  \BibitemOpen
  \bibfield  {author} {\bibinfo {author} {\bibfnamefont {D.}~\bibnamefont {Chakraborty}}\ and\ \bibinfo {author} {\bibfnamefont {A.~M.}\ \bibnamefont {Black-Schaffer}},\ }\bibfield  {title} {\bibinfo {title} {Constraints on superconducting pairing in altermagnets},\ }\href {https://doi.org/10.1103/zylh-rqxl} {\bibfield  {journal} {\bibinfo  {journal} {Phys. Rev. B}\ }\textbf {\bibinfo {volume} {112}},\ \bibinfo {pages} {014516} (\bibinfo {year} {2025})}\BibitemShut {NoStop}%
\bibitem [{\citenamefont {de~Carvalho}\ and\ \citenamefont {Freire}(2024)}]{Carvalho2024}%
  \BibitemOpen
  \bibfield  {author} {\bibinfo {author} {\bibfnamefont {V.~S.}\ \bibnamefont {de~Carvalho}}\ and\ \bibinfo {author} {\bibfnamefont {H.}~\bibnamefont {Freire}},\ }\bibfield  {title} {\bibinfo {title} {Unconventional superconductivity in altermagnets with spin-orbit coupling},\ }\href {https://doi.org/10.1103/PhysRevB.110.L220503} {\bibfield  {journal} {\bibinfo  {journal} {Phys. Rev. B}\ }\textbf {\bibinfo {volume} {110}},\ \bibinfo {pages} {L220503} (\bibinfo {year} {2024})}\BibitemShut {NoStop}%
\bibitem [{\citenamefont {Brekke}\ \emph {et~al.}(2023)\citenamefont {Brekke}, \citenamefont {Brataas},\ and\ \citenamefont {Sudb\o{}}}]{Brekke2023}%
  \BibitemOpen
  \bibfield  {author} {\bibinfo {author} {\bibfnamefont {B.}~\bibnamefont {Brekke}}, \bibinfo {author} {\bibfnamefont {A.}~\bibnamefont {Brataas}},\ and\ \bibinfo {author} {\bibfnamefont {A.}~\bibnamefont {Sudb\o{}}},\ }\bibfield  {title} {\bibinfo {title} {Two-dimensional altermagnets: Superconductivity in a minimal microscopic model},\ }\href {https://doi.org/10.1103/PhysRevB.108.224421} {\bibfield  {journal} {\bibinfo  {journal} {Phys. Rev. B}\ }\textbf {\bibinfo {volume} {108}},\ \bibinfo {pages} {224421} (\bibinfo {year} {2023})}\BibitemShut {NoStop}%
\bibitem [{\citenamefont {Roig}\ \emph {et~al.}(2024)\citenamefont {Roig}, \citenamefont {Kreisel}, \citenamefont {Yu}, \citenamefont {Andersen},\ and\ \citenamefont {Agterberg}}]{Roig2024}%
  \BibitemOpen
  \bibfield  {author} {\bibinfo {author} {\bibfnamefont {M.}~\bibnamefont {Roig}}, \bibinfo {author} {\bibfnamefont {A.}~\bibnamefont {Kreisel}}, \bibinfo {author} {\bibfnamefont {Y.}~\bibnamefont {Yu}}, \bibinfo {author} {\bibfnamefont {B.~M.}\ \bibnamefont {Andersen}},\ and\ \bibinfo {author} {\bibfnamefont {D.~F.}\ \bibnamefont {Agterberg}},\ }\bibfield  {title} {\bibinfo {title} {Minimal models for altermagnetism},\ }\href {https://doi.org/10.1103/PhysRevB.110.144412} {\bibfield  {journal} {\bibinfo  {journal} {Phys. Rev. B}\ }\textbf {\bibinfo {volume} {110}},\ \bibinfo {pages} {144412} (\bibinfo {year} {2024})}\BibitemShut {NoStop}%
\bibitem [{\citenamefont {Parshukov}\ \emph {et~al.}(2025)\citenamefont {Parshukov}, \citenamefont {Wiedmann},\ and\ \citenamefont {Schnyder}}]{Parshukov2024}%
  \BibitemOpen
  \bibfield  {author} {\bibinfo {author} {\bibfnamefont {K.}~\bibnamefont {Parshukov}}, \bibinfo {author} {\bibfnamefont {R.}~\bibnamefont {Wiedmann}},\ and\ \bibinfo {author} {\bibfnamefont {A.~P.}\ \bibnamefont {Schnyder}},\ }\bibfield  {title} {\bibinfo {title} {Topological crossings in two-dimensional altermagnets: Symmetry classification and topological responses},\ }\href {https://doi.org/10.1103/PhysRevB.111.224406} {\bibfield  {journal} {\bibinfo  {journal} {Phys. Rev. B}\ }\textbf {\bibinfo {volume} {111}},\ \bibinfo {pages} {224406} (\bibinfo {year} {2025})}\BibitemShut {NoStop}%
\bibitem [{\citenamefont {Attias}\ \emph {et~al.}(2024)\citenamefont {Attias}, \citenamefont {Levchenko},\ and\ \citenamefont {Khodas}}]{Attias2024}%
  \BibitemOpen
  \bibfield  {author} {\bibinfo {author} {\bibfnamefont {L.}~\bibnamefont {Attias}}, \bibinfo {author} {\bibfnamefont {A.}~\bibnamefont {Levchenko}},\ and\ \bibinfo {author} {\bibfnamefont {M.}~\bibnamefont {Khodas}},\ }\bibfield  {title} {\bibinfo {title} {Intrinsic anomalous {Hall} effect in altermagnets},\ }\href {https://doi.org/10.1103/PhysRevB.110.094425} {\bibfield  {journal} {\bibinfo  {journal} {Phys. Rev. B}\ }\textbf {\bibinfo {volume} {110}},\ \bibinfo {pages} {094425} (\bibinfo {year} {2024})}\BibitemShut {NoStop}%
\bibitem [{\citenamefont {Rao}\ \emph {et~al.}(2024)\citenamefont {Rao}, \citenamefont {Mook},\ and\ \citenamefont {Knolle}}]{Rao2024}%
  \BibitemOpen
  \bibfield  {author} {\bibinfo {author} {\bibfnamefont {P.}~\bibnamefont {Rao}}, \bibinfo {author} {\bibfnamefont {A.}~\bibnamefont {Mook}},\ and\ \bibinfo {author} {\bibfnamefont {J.}~\bibnamefont {Knolle}},\ }\bibfield  {title} {\bibinfo {title} {Tunable band topology and optical conductivity in altermagnets},\ }\href {https://doi.org/10.1103/PhysRevB.110.024425} {\bibfield  {journal} {\bibinfo  {journal} {Phys. Rev. B}\ }\textbf {\bibinfo {volume} {110}},\ \bibinfo {pages} {024425} (\bibinfo {year} {2024})}\BibitemShut {NoStop}%
\bibitem [{\citenamefont {Antonenko}\ \emph {et~al.}(2025)\citenamefont {Antonenko}, \citenamefont {Fernandes},\ and\ \citenamefont {Venderbos}}]{Antonenko2025}%
  \BibitemOpen
  \bibfield  {author} {\bibinfo {author} {\bibfnamefont {D.~S.}\ \bibnamefont {Antonenko}}, \bibinfo {author} {\bibfnamefont {R.~M.}\ \bibnamefont {Fernandes}},\ and\ \bibinfo {author} {\bibfnamefont {J.~W.~F.}\ \bibnamefont {Venderbos}},\ }\bibfield  {title} {\bibinfo {title} {Mirror {Chern} bands and {Weyl} nodal loops in altermagnets},\ }\href {https://doi.org/10.1103/PhysRevLett.134.096703} {\bibfield  {journal} {\bibinfo  {journal} {Phys. Rev. Lett.}\ }\textbf {\bibinfo {volume} {134}},\ \bibinfo {pages} {096703} (\bibinfo {year} {2025})}\BibitemShut {NoStop}%
\bibitem [{\citenamefont {Vila}\ \emph {et~al.}(2025)\citenamefont {Vila}, \citenamefont {Sunko},\ and\ \citenamefont {Moore}}]{Vila2024}%
  \BibitemOpen
  \bibfield  {author} {\bibinfo {author} {\bibfnamefont {M.}~\bibnamefont {Vila}}, \bibinfo {author} {\bibfnamefont {V.}~\bibnamefont {Sunko}},\ and\ \bibinfo {author} {\bibfnamefont {J.~E.}\ \bibnamefont {Moore}},\ }\bibfield  {title} {\bibinfo {title} {Orbital-spin locking and its optical signatures in altermagnets},\ }\href {https://doi.org/10.1103/bzzy-ngcs} {\bibfield  {journal} {\bibinfo  {journal} {Phys. Rev. B}\ }\textbf {\bibinfo {volume} {112}},\ \bibinfo {pages} {L020401} (\bibinfo {year} {2025})}\BibitemShut {NoStop}%
\bibitem [{\citenamefont {Gondolf}\ \emph {et~al.}(2025)\citenamefont {Gondolf}, \citenamefont {Kreisel}, \citenamefont {Roig}, \citenamefont {Yu}, \citenamefont {Agterberg},\ and\ \citenamefont {Andersen}}]{Gondolf2025}%
  \BibitemOpen
  \bibfield  {author} {\bibinfo {author} {\bibfnamefont {J.}~\bibnamefont {Gondolf}}, \bibinfo {author} {\bibfnamefont {A.}~\bibnamefont {Kreisel}}, \bibinfo {author} {\bibfnamefont {M.}~\bibnamefont {Roig}}, \bibinfo {author} {\bibfnamefont {Y.}~\bibnamefont {Yu}}, \bibinfo {author} {\bibfnamefont {D.~F.}\ \bibnamefont {Agterberg}},\ and\ \bibinfo {author} {\bibfnamefont {B.~M.}\ \bibnamefont {Andersen}},\ }\bibfield  {title} {\bibinfo {title} {Local signatures of altermagnetism},\ }\href {https://doi.org/10.1103/PhysRevB.111.174436} {\bibfield  {journal} {\bibinfo  {journal} {Phys. Rev. B}\ }\textbf {\bibinfo {volume} {111}},\ \bibinfo {pages} {174436} (\bibinfo {year} {2025})}\BibitemShut {NoStop}%
\bibitem [{\citenamefont {Mu}(2014)}]{Mu-thesis}%
  \BibitemOpen
  \bibfield  {author} {\bibinfo {author} {\bibfnamefont {S.}~\bibnamefont {Mu}},\ }\emph {\bibinfo {title} {First-principles study of magnetoelectric materials}},\ \href {https://www.proquest.com/docview/1641140930} {Ph.D. thesis},\ \bibinfo  {school} {University of Nebraska-Lincoln} (\bibinfo {year} {2014})\BibitemShut {NoStop}%
\bibitem [{\citenamefont {Chen}\ \emph {et~al.}(2024)\citenamefont {Chen}, \citenamefont {Ren}, \citenamefont {Zhu}, \citenamefont {Yu}, \citenamefont {Zhang}, \citenamefont {Liu}, \citenamefont {Li}, \citenamefont {Liu}, \citenamefont {Li},\ and\ \citenamefont {Liu}}]{Chen2024}%
  \BibitemOpen
  \bibfield  {author} {\bibinfo {author} {\bibfnamefont {X.}~\bibnamefont {Chen}}, \bibinfo {author} {\bibfnamefont {J.}~\bibnamefont {Ren}}, \bibinfo {author} {\bibfnamefont {Y.}~\bibnamefont {Zhu}}, \bibinfo {author} {\bibfnamefont {Y.}~\bibnamefont {Yu}}, \bibinfo {author} {\bibfnamefont {A.}~\bibnamefont {Zhang}}, \bibinfo {author} {\bibfnamefont {P.}~\bibnamefont {Liu}}, \bibinfo {author} {\bibfnamefont {J.}~\bibnamefont {Li}}, \bibinfo {author} {\bibfnamefont {Y.}~\bibnamefont {Liu}}, \bibinfo {author} {\bibfnamefont {C.}~\bibnamefont {Li}},\ and\ \bibinfo {author} {\bibfnamefont {Q.}~\bibnamefont {Liu}},\ }\bibfield  {title} {\bibinfo {title} {Enumeration and representation theory of spin space groups},\ }\href {https://doi.org/10.1103/PhysRevX.14.031038} {\bibfield  {journal} {\bibinfo  {journal} {Phys. Rev. X}\ }\textbf {\bibinfo {volume} {14}},\ \bibinfo {pages} {031038} (\bibinfo {year} {2024})}\BibitemShut {NoStop}%
\bibitem [{\citenamefont {Jiang}\ \emph {et~al.}(2024)\citenamefont {Jiang}, \citenamefont {Song}, \citenamefont {Zhu}, \citenamefont {Fang}, \citenamefont {Weng}, \citenamefont {Liu}, \citenamefont {Yang},\ and\ \citenamefont {Fang}}]{Jiang2024}%
  \BibitemOpen
  \bibfield  {author} {\bibinfo {author} {\bibfnamefont {Y.}~\bibnamefont {Jiang}}, \bibinfo {author} {\bibfnamefont {Z.}~\bibnamefont {Song}}, \bibinfo {author} {\bibfnamefont {T.}~\bibnamefont {Zhu}}, \bibinfo {author} {\bibfnamefont {Z.}~\bibnamefont {Fang}}, \bibinfo {author} {\bibfnamefont {H.}~\bibnamefont {Weng}}, \bibinfo {author} {\bibfnamefont {Z.-X.}\ \bibnamefont {Liu}}, \bibinfo {author} {\bibfnamefont {J.}~\bibnamefont {Yang}},\ and\ \bibinfo {author} {\bibfnamefont {C.}~\bibnamefont {Fang}},\ }\bibfield  {title} {\bibinfo {title} {Enumeration of spin-space groups: Toward a complete description of symmetries of magnetic orders},\ }\href {https://doi.org/10.1103/PhysRevX.14.031039} {\bibfield  {journal} {\bibinfo  {journal} {Phys. Rev. X}\ }\textbf {\bibinfo {volume} {14}},\ \bibinfo {pages} {031039} (\bibinfo {year} {2024})}\BibitemShut {NoStop}%
\bibitem [{\citenamefont {Xiao}\ \emph {et~al.}(2024)\citenamefont {Xiao}, \citenamefont {Zhao}, \citenamefont {Li}, \citenamefont {Shindou},\ and\ \citenamefont {Song}}]{Xiao2024}%
  \BibitemOpen
  \bibfield  {author} {\bibinfo {author} {\bibfnamefont {Z.}~\bibnamefont {Xiao}}, \bibinfo {author} {\bibfnamefont {J.}~\bibnamefont {Zhao}}, \bibinfo {author} {\bibfnamefont {Y.}~\bibnamefont {Li}}, \bibinfo {author} {\bibfnamefont {R.}~\bibnamefont {Shindou}},\ and\ \bibinfo {author} {\bibfnamefont {Z.-D.}\ \bibnamefont {Song}},\ }\bibfield  {title} {\bibinfo {title} {Spin space groups: Full classification and applications},\ }\href {https://doi.org/10.1103/PhysRevX.14.031037} {\bibfield  {journal} {\bibinfo  {journal} {Phys. Rev. X}\ }\textbf {\bibinfo {volume} {14}},\ \bibinfo {pages} {031037} (\bibinfo {year} {2024})}\BibitemShut {NoStop}%
\bibitem [{\citenamefont {Lifshitz}(2005)}]{Lifshitz2005}%
  \BibitemOpen
  \bibfield  {author} {\bibinfo {author} {\bibfnamefont {R.}~\bibnamefont {Lifshitz}},\ }\bibfield  {title} {\bibinfo {title} {Magnetic point groups and space groups},\ }\href {https://doi.org/10.1016/B0-12-369401-9/00748-8} {\bibfield  {journal} {\bibinfo  {journal} {Encyclopedia of Condensed Matter Physics, Ed. F. Bassani, G.L. Liedl, and P. Wyder,}\ }\textbf {\bibinfo {volume} {3}},\ \bibinfo {pages} {219} (\bibinfo {year} {2005})}\BibitemShut {NoStop}%
\bibitem [{\citenamefont {Heesch}(1930)}]{Heesch1930}%
  \BibitemOpen
  \bibfield  {author} {\bibinfo {author} {\bibfnamefont {H.}~\bibnamefont {Heesch}},\ }\bibfield  {title} {\bibinfo {title} {{\"U}ber die vierdimensionalen {Gruppen} des dreidimensionalen {Raumes}},\ }\href {https://doi.org/doi:10.1524/zkri.1930.73.1.325} {\bibfield  {journal} {\bibinfo  {journal} {Z. Kristallogr. Cryst. Mater.}\ }\textbf {\bibinfo {volume} {73}},\ \bibinfo {pages} {325} (\bibinfo {year} {1930})}\BibitemShut {NoStop}%
\bibitem [{\citenamefont {Shubnikov}(1951)}]{Shubnikov1951}%
  \BibitemOpen
  \bibfield  {author} {\bibinfo {author} {\bibfnamefont {A.~V.}\ \bibnamefont {Shubnikov}},\ }\href@noop {} {\emph {\bibinfo {title} {Symmetry and Antisymmetry of Finite Figures}}}\ (\bibinfo  {publisher} {Izd. AN SSSR},\ \bibinfo {address} {Moscow},\ \bibinfo {year} {1951})\ \bibinfo {note} {[{\it Colored Symmetry}, edited by W. T. Holser (MacMillan, New York, 1964)]}\BibitemShut {NoStop}%
\bibitem [{\citenamefont {McClarty}\ and\ \citenamefont {Rau}(2024)}]{McClarty2024}%
  \BibitemOpen
  \bibfield  {author} {\bibinfo {author} {\bibfnamefont {P.~A.}\ \bibnamefont {McClarty}}\ and\ \bibinfo {author} {\bibfnamefont {J.~G.}\ \bibnamefont {Rau}},\ }\bibfield  {title} {\bibinfo {title} {Landau theory of altermagnetism},\ }\href {https://doi.org/10.1103/PhysRevLett.132.176702} {\bibfield  {journal} {\bibinfo  {journal} {Phys. Rev. Lett.}\ }\textbf {\bibinfo {volume} {132}},\ \bibinfo {pages} {176702} (\bibinfo {year} {2024})}\BibitemShut {NoStop}%
\bibitem [{\citenamefont {C\^onsoli}\ \emph {et~al.}(2021)\citenamefont {C\^onsoli}, \citenamefont {Fornoville},\ and\ \citenamefont {Vojta}}]{Consoli2021}%
  \BibitemOpen
  \bibfield  {author} {\bibinfo {author} {\bibfnamefont {P.~M.}\ \bibnamefont {C\^onsoli}}, \bibinfo {author} {\bibfnamefont {M.}~\bibnamefont {Fornoville}},\ and\ \bibinfo {author} {\bibfnamefont {M.}~\bibnamefont {Vojta}},\ }\bibfield  {title} {\bibinfo {title} {Fluctuation-induced ferrimagnetism in sublattice-imbalanced antiferromagnets with application to {${\mathrm{SrCu}}_{2}$(${\mathrm{BO}}_{3}{)}_{2}$} under pressure},\ }\href {https://doi.org/10.1103/PhysRevB.104.064422} {\bibfield  {journal} {\bibinfo  {journal} {Phys. Rev. B}\ }\textbf {\bibinfo {volume} {104}},\ \bibinfo {pages} {064422} (\bibinfo {year} {2021})}\BibitemShut {NoStop}%
\bibitem [{\citenamefont {Yershov}\ \emph {et~al.}(2024)\citenamefont {Yershov}, \citenamefont {Kravchuk}, \citenamefont {Daghofer},\ and\ \citenamefont {van~den Brink}}]{Yershov2024}%
  \BibitemOpen
  \bibfield  {author} {\bibinfo {author} {\bibfnamefont {K.~V.}\ \bibnamefont {Yershov}}, \bibinfo {author} {\bibfnamefont {V.~P.}\ \bibnamefont {Kravchuk}}, \bibinfo {author} {\bibfnamefont {M.}~\bibnamefont {Daghofer}},\ and\ \bibinfo {author} {\bibfnamefont {J.}~\bibnamefont {van~den Brink}},\ }\bibfield  {title} {\bibinfo {title} {Fluctuation-induced piezomagnetism in local moment altermagnets},\ }\href {https://doi.org/10.1103/PhysRevB.110.144421} {\bibfield  {journal} {\bibinfo  {journal} {Phys. Rev. B}\ }\textbf {\bibinfo {volume} {110}},\ \bibinfo {pages} {144421} (\bibinfo {year} {2024})}\BibitemShut {NoStop}%
\bibitem [{\citenamefont {Mostovoy}\ \emph {et~al.}(2010)\citenamefont {Mostovoy}, \citenamefont {Scaramucci}, \citenamefont {Spaldin},\ and\ \citenamefont {Delaney}}]{Mostovoy2010}%
  \BibitemOpen
  \bibfield  {author} {\bibinfo {author} {\bibfnamefont {M.}~\bibnamefont {Mostovoy}}, \bibinfo {author} {\bibfnamefont {A.}~\bibnamefont {Scaramucci}}, \bibinfo {author} {\bibfnamefont {N.~A.}\ \bibnamefont {Spaldin}},\ and\ \bibinfo {author} {\bibfnamefont {K.~T.}\ \bibnamefont {Delaney}},\ }\bibfield  {title} {\bibinfo {title} {Temperature-dependent magnetoelectric effect from first principles},\ }\href {https://doi.org/10.1103/PhysRevLett.105.087202} {\bibfield  {journal} {\bibinfo  {journal} {Phys. Rev. Lett.}\ }\textbf {\bibinfo {volume} {105}},\ \bibinfo {pages} {087202} (\bibinfo {year} {2010})}\BibitemShut {NoStop}%
\bibitem [{\citenamefont {Mu}\ \emph {et~al.}(2014)\citenamefont {Mu}, \citenamefont {Wysocki},\ and\ \citenamefont {Belashchenko}}]{Mu2014}%
  \BibitemOpen
  \bibfield  {author} {\bibinfo {author} {\bibfnamefont {S.}~\bibnamefont {Mu}}, \bibinfo {author} {\bibfnamefont {A.~L.}\ \bibnamefont {Wysocki}},\ and\ \bibinfo {author} {\bibfnamefont {K.~D.}\ \bibnamefont {Belashchenko}},\ }\bibfield  {title} {\bibinfo {title} {First-principles microscopic model of exchange-driven magnetoelectric response with application to {${\text{Cr}}_{2}$${\text{O}}_{3}$}},\ }\href {https://doi.org/10.1103/PhysRevB.89.174413} {\bibfield  {journal} {\bibinfo  {journal} {Phys. Rev. B}\ }\textbf {\bibinfo {volume} {89}},\ \bibinfo {pages} {174413} (\bibinfo {year} {2014})}\BibitemShut {NoStop}%
\bibitem [{\citenamefont {Zeng}\ and\ \citenamefont {Zhao}(2024)}]{Zeng2024}%
  \BibitemOpen
  \bibfield  {author} {\bibinfo {author} {\bibfnamefont {S.}~\bibnamefont {Zeng}}\ and\ \bibinfo {author} {\bibfnamefont {Y.-J.}\ \bibnamefont {Zhao}},\ }\bibfield  {title} {\bibinfo {title} {Description of two-dimensional altermagnetism: Categorization using spin group theory},\ }\href {https://doi.org/10.1103/PhysRevB.110.054406} {\bibfield  {journal} {\bibinfo  {journal} {Phys. Rev. B}\ }\textbf {\bibinfo {volume} {110}},\ \bibinfo {pages} {054406} (\bibinfo {year} {2024})}\BibitemShut {NoStop}%
\bibitem [{\citenamefont {Cui}\ \emph {et~al.}(2023)\citenamefont {Cui}, \citenamefont {Zhu}, \citenamefont {Yao}, \citenamefont {Cui},\ and\ \citenamefont {Yang}}]{Cui2023}%
  \BibitemOpen
  \bibfield  {author} {\bibinfo {author} {\bibfnamefont {Q.}~\bibnamefont {Cui}}, \bibinfo {author} {\bibfnamefont {Y.}~\bibnamefont {Zhu}}, \bibinfo {author} {\bibfnamefont {X.}~\bibnamefont {Yao}}, \bibinfo {author} {\bibfnamefont {P.}~\bibnamefont {Cui}},\ and\ \bibinfo {author} {\bibfnamefont {H.}~\bibnamefont {Yang}},\ }\bibfield  {title} {\bibinfo {title} {Giant spin-{Hall} and tunneling magnetoresistance effects based on a two-dimensional nonrelativistic antiferromagnetic metal},\ }\href {https://doi.org/10.1103/PhysRevB.108.024410} {\bibfield  {journal} {\bibinfo  {journal} {Phys. Rev. B}\ }\textbf {\bibinfo {volume} {108}},\ \bibinfo {pages} {024410} (\bibinfo {year} {2023})}\BibitemShut {NoStop}%
\bibitem [{\citenamefont {Jiang}\ \emph {et~al.}(2025)\citenamefont {Jiang}, \citenamefont {Hu}, \citenamefont {Bai}, \citenamefont {Song}, \citenamefont {Mu}, \citenamefont {Qu}, \citenamefont {Li}, \citenamefont {Zhu}, \citenamefont {Pi}, \citenamefont {Wei}, \citenamefont {Sun}, \citenamefont {Huang}, \citenamefont {Zheng}, \citenamefont {Peng}, \citenamefont {He}, \citenamefont {Li}, \citenamefont {Luo}, \citenamefont {Li}, \citenamefont {Chen}, \citenamefont {Li}, \citenamefont {Weng},\ and\ \citenamefont {Qian}}]{KV2Se2O}%
  \BibitemOpen
  \bibfield  {author} {\bibinfo {author} {\bibfnamefont {B.}~\bibnamefont {Jiang}}, \bibinfo {author} {\bibfnamefont {M.}~\bibnamefont {Hu}}, \bibinfo {author} {\bibfnamefont {J.}~\bibnamefont {Bai}}, \bibinfo {author} {\bibfnamefont {Z.}~\bibnamefont {Song}}, \bibinfo {author} {\bibfnamefont {C.}~\bibnamefont {Mu}}, \bibinfo {author} {\bibfnamefont {G.}~\bibnamefont {Qu}}, \bibinfo {author} {\bibfnamefont {W.}~\bibnamefont {Li}}, \bibinfo {author} {\bibfnamefont {W.}~\bibnamefont {Zhu}}, \bibinfo {author} {\bibfnamefont {H.}~\bibnamefont {Pi}}, \bibinfo {author} {\bibfnamefont {Z.}~\bibnamefont {Wei}}, \bibinfo {author} {\bibfnamefont {Y.-J.}\ \bibnamefont {Sun}}, \bibinfo {author} {\bibfnamefont {Y.}~\bibnamefont {Huang}}, \bibinfo {author} {\bibfnamefont {X.}~\bibnamefont {Zheng}}, \bibinfo {author} {\bibfnamefont {Y.}~\bibnamefont {Peng}}, \bibinfo {author} {\bibfnamefont {L.}~\bibnamefont {He}}, \bibinfo {author} {\bibfnamefont {S.}~\bibnamefont {Li}}, \bibinfo {author} {\bibfnamefont {J.}~\bibnamefont
  {Luo}}, \bibinfo {author} {\bibfnamefont {Z.}~\bibnamefont {Li}}, \bibinfo {author} {\bibfnamefont {G.}~\bibnamefont {Chen}}, \bibinfo {author} {\bibfnamefont {H.}~\bibnamefont {Li}}, \bibinfo {author} {\bibfnamefont {H.}~\bibnamefont {Weng}},\ and\ \bibinfo {author} {\bibfnamefont {T.}~\bibnamefont {Qian}},\ }\bibfield  {title} {\bibinfo {title} {A metallic room-temperature $d$-wave altermagnet},\ }\href {https://doi.org/10.1038/s41567-025-02822-y} {\bibfield  {journal} {\bibinfo  {journal} {Nat. Phys.}\ }\textbf {\bibinfo {volume} {21}},\ \bibinfo {pages} {754} (\bibinfo {year} {2025})}\BibitemShut {NoStop}%
\bibitem [{\citenamefont {Zhang}\ \emph {et~al.}(2025)\citenamefont {Zhang}, \citenamefont {Cheng}, \citenamefont {Yin}, \citenamefont {Liu}, \citenamefont {Deng}, \citenamefont {Qiao}, \citenamefont {Shi}, \citenamefont {Zhang}, \citenamefont {Lin}, \citenamefont {Liu}, \citenamefont {Ye}, \citenamefont {Huang}, \citenamefont {Meng}, \citenamefont {Zhang}, \citenamefont {Okuda}, \citenamefont {Shimada}, \citenamefont {Cui}, \citenamefont {Zhao}, \citenamefont {Cao}, \citenamefont {Qiao}, \citenamefont {Liu},\ and\ \citenamefont {Chen}}]{RbV2Te2O}%
  \BibitemOpen
  \bibfield  {author} {\bibinfo {author} {\bibfnamefont {F.}~\bibnamefont {Zhang}}, \bibinfo {author} {\bibfnamefont {X.}~\bibnamefont {Cheng}}, \bibinfo {author} {\bibfnamefont {Z.}~\bibnamefont {Yin}}, \bibinfo {author} {\bibfnamefont {C.}~\bibnamefont {Liu}}, \bibinfo {author} {\bibfnamefont {L.}~\bibnamefont {Deng}}, \bibinfo {author} {\bibfnamefont {Y.}~\bibnamefont {Qiao}}, \bibinfo {author} {\bibfnamefont {Z.}~\bibnamefont {Shi}}, \bibinfo {author} {\bibfnamefont {S.}~\bibnamefont {Zhang}}, \bibinfo {author} {\bibfnamefont {J.}~\bibnamefont {Lin}}, \bibinfo {author} {\bibfnamefont {Z.}~\bibnamefont {Liu}}, \bibinfo {author} {\bibfnamefont {M.}~\bibnamefont {Ye}}, \bibinfo {author} {\bibfnamefont {Y.}~\bibnamefont {Huang}}, \bibinfo {author} {\bibfnamefont {X.}~\bibnamefont {Meng}}, \bibinfo {author} {\bibfnamefont {C.}~\bibnamefont {Zhang}}, \bibinfo {author} {\bibfnamefont {T.}~\bibnamefont {Okuda}}, \bibinfo {author} {\bibfnamefont {K.}~\bibnamefont {Shimada}}, \bibinfo {author} {\bibfnamefont
  {S.}~\bibnamefont {Cui}}, \bibinfo {author} {\bibfnamefont {Y.}~\bibnamefont {Zhao}}, \bibinfo {author} {\bibfnamefont {G.-H.}\ \bibnamefont {Cao}}, \bibinfo {author} {\bibfnamefont {S.}~\bibnamefont {Qiao}}, \bibinfo {author} {\bibfnamefont {J.}~\bibnamefont {Liu}},\ and\ \bibinfo {author} {\bibfnamefont {C.}~\bibnamefont {Chen}},\ }\bibfield  {title} {\bibinfo {title} {Crystal-symmetry-paired spin-valley locking in a layered room-temperature metallic altermagnet candidate},\ }\href {https://doi.org/10.1038/s41567-025-02864-2} {\bibfield  {journal} {\bibinfo  {journal} {Nat. Phys.}\ }\textbf {\bibinfo {volume} {21}},\ \bibinfo {pages} {760} (\bibinfo {year} {2025})}\BibitemShut {NoStop}%
\bibitem [{\citenamefont {Yuan}\ \emph {et~al.}(2024)\citenamefont {Yuan}, \citenamefont {Georgescu},\ and\ \citenamefont {Rondinelli}}]{Yuan2024}%
  \BibitemOpen
  \bibfield  {author} {\bibinfo {author} {\bibfnamefont {L.-D.}\ \bibnamefont {Yuan}}, \bibinfo {author} {\bibfnamefont {A.~B.}\ \bibnamefont {Georgescu}},\ and\ \bibinfo {author} {\bibfnamefont {J.~M.}\ \bibnamefont {Rondinelli}},\ }\bibfield  {title} {\bibinfo {title} {Nonrelativistic spin splitting at the {Brillouin} zone center in compensated magnets},\ }\href {https://doi.org/10.1103/PhysRevLett.133.216701} {\bibfield  {journal} {\bibinfo  {journal} {Phys. Rev. Lett.}\ }\textbf {\bibinfo {volume} {133}},\ \bibinfo {pages} {216701} (\bibinfo {year} {2024})}\BibitemShut {NoStop}%
\bibitem [{\citenamefont {Erickson}(1953)}]{erickson1953neutron}%
  \BibitemOpen
  \bibfield  {author} {\bibinfo {author} {\bibfnamefont {R.~A.}\ \bibnamefont {Erickson}},\ }\bibfield  {title} {\bibinfo {title} {Neutron diffraction studies of antiferromagnetism in manganous fluoride and some isomorphous compounds},\ }\href {https://journals.aps.org/pr/abstract/10.1103/PhysRev.90.779} {\bibfield  {journal} {\bibinfo  {journal} {Phys. Rev.}\ }\textbf {\bibinfo {volume} {90}},\ \bibinfo {pages} {779} (\bibinfo {year} {1953})}\BibitemShut {NoStop}%
\bibitem [{\citenamefont {Borovik-Romanov}\ \emph {et~al.}(2013)\citenamefont {Borovik-Romanov}, \citenamefont {Grimmer},\ and\ \citenamefont {Kenzelmann}}]{borovik2013}%
  \BibitemOpen
  \bibfield  {author} {\bibinfo {author} {\bibfnamefont {A.~S.}\ \bibnamefont {Borovik-Romanov}}, \bibinfo {author} {\bibfnamefont {H.}~\bibnamefont {Grimmer}},\ and\ \bibinfo {author} {\bibfnamefont {M.}~\bibnamefont {Kenzelmann}},\ }\bibinfo {title} {Magnetic properties},\ in\ \href {https://doi.org/10.1107/97809553602060000904} {\emph {\bibinfo {booktitle} {International Tables for Crystallography}}}\ (\bibinfo  {publisher} {Wiley},\ \bibinfo {year} {2013})\ Chap.\ \bibinfo {chapter} {1.5}, pp.\ \bibinfo {pages} {106--152}\BibitemShut {NoStop}%
\bibitem [{\citenamefont {Dzyaloshinsky}(1958)}]{Dzyaloshinsky1958}%
  \BibitemOpen
  \bibfield  {author} {\bibinfo {author} {\bibfnamefont {I.}~\bibnamefont {Dzyaloshinsky}},\ }\bibfield  {title} {\bibinfo {title} {A thermodynamic theory of “weak” ferromagnetism of antiferromagnetics},\ }\href {https://doi.org/https://doi.org/10.1016/0022-3697(58)90076-3} {\bibfield  {journal} {\bibinfo  {journal} {Journal of Physics and Chemistry of Solids}\ }\textbf {\bibinfo {volume} {4}},\ \bibinfo {pages} {241} (\bibinfo {year} {1958})}\BibitemShut {NoStop}%
\bibitem [{\citenamefont {Moriya}(1960)}]{Moriya1960}%
  \BibitemOpen
  \bibfield  {author} {\bibinfo {author} {\bibfnamefont {T.}~\bibnamefont {Moriya}},\ }\bibfield  {title} {\bibinfo {title} {Anisotropic superexchange interaction and weak ferromagnetism},\ }\href {https://doi.org/10.1103/PhysRev.120.91} {\bibfield  {journal} {\bibinfo  {journal} {Phys. Rev.}\ }\textbf {\bibinfo {volume} {120}},\ \bibinfo {pages} {91} (\bibinfo {year} {1960})}\BibitemShut {NoStop}%
\bibitem [{\citenamefont {Astrov}(1960)}]{astrov1960magnetoelectric}%
  \BibitemOpen
  \bibfield  {author} {\bibinfo {author} {\bibfnamefont {D.}~\bibnamefont {Astrov}},\ }\bibfield  {title} {\bibinfo {title} {The magnetoelectric effect in antiferromagnetics},\ }\href {http://www.jetp.ras.ru/cgi-bin/dn/e_011_03_0708.pdf} {\bibfield  {journal} {\bibinfo  {journal} {Sov. Phys. JETP}\ }\textbf {\bibinfo {volume} {11}},\ \bibinfo {pages} {708} (\bibinfo {year} {1960})}\BibitemShut {NoStop}%
\bibitem [{\citenamefont {Folen}\ \emph {et~al.}(1961)\citenamefont {Folen}, \citenamefont {Rado},\ and\ \citenamefont {Stalder}}]{folen1961anisotropy}%
  \BibitemOpen
  \bibfield  {author} {\bibinfo {author} {\bibfnamefont {V.}~\bibnamefont {Folen}}, \bibinfo {author} {\bibfnamefont {G.}~\bibnamefont {Rado}},\ and\ \bibinfo {author} {\bibfnamefont {E.}~\bibnamefont {Stalder}},\ }\bibfield  {title} {\bibinfo {title} {Anisotropy of the magnetoelectric effect in {Cr$_2$O$_3$}},\ }\href {https://journals.aps.org/prl/abstract/10.1103/PhysRevLett.6.607} {\bibfield  {journal} {\bibinfo  {journal} {Phys. Rev. Lett.}\ }\textbf {\bibinfo {volume} {6}},\ \bibinfo {pages} {607} (\bibinfo {year} {1961})}\BibitemShut {NoStop}%
\bibitem [{\citenamefont {Okazaki}\ \emph {et~al.}(1964)\citenamefont {Okazaki}, \citenamefont {Turberfield},\ and\ \citenamefont {Stevenson}}]{Okazaki}%
  \BibitemOpen
  \bibfield  {author} {\bibinfo {author} {\bibfnamefont {A.}~\bibnamefont {Okazaki}}, \bibinfo {author} {\bibfnamefont {K.}~\bibnamefont {Turberfield}},\ and\ \bibinfo {author} {\bibfnamefont {R.~W.}\ \bibnamefont {Stevenson}},\ }\bibfield  {title} {\bibinfo {title} {Neutron inelastic scattering measurements of antiferromagnetic excitations in {MnF$_2$} at 4.2 {K} and at temperatures up to the {N\'eel} point},\ }\href {https://doi.org/10.1016/0031-9163(64)90774-7} {\bibfield  {journal} {\bibinfo  {journal} {Phys. Lett.}\ }\textbf {\bibinfo {volume} {8}} (\bibinfo {year} {1964})}\BibitemShut {NoStop}%
\bibitem [{\citenamefont {Hutchings}\ \emph {et~al.}(1970)\citenamefont {Hutchings}, \citenamefont {Rainford},\ and\ \citenamefont {Guggenheim}}]{Hutchings}%
  \BibitemOpen
  \bibfield  {author} {\bibinfo {author} {\bibfnamefont {M.}~\bibnamefont {Hutchings}}, \bibinfo {author} {\bibfnamefont {B.}~\bibnamefont {Rainford}},\ and\ \bibinfo {author} {\bibfnamefont {H.}~\bibnamefont {Guggenheim}},\ }\bibfield  {title} {\bibinfo {title} {Spin waves in antiferromagnetic {FeF$_2$}},\ }\href {https://iopscience.iop.org/article/10.1088/0022-3719/3/2/013} {\bibfield  {journal} {\bibinfo  {journal} {J. Phys. C: Solid State Phys.}\ }\textbf {\bibinfo {volume} {3}},\ \bibinfo {pages} {307} (\bibinfo {year} {1970})}\BibitemShut {NoStop}%
\bibitem [{\citenamefont {Belorizky}\ \emph {et~al.}(1969)\citenamefont {Belorizky}, \citenamefont {Ng},\ and\ \citenamefont {Phillips}}]{Belorizky}%
  \BibitemOpen
  \bibfield  {author} {\bibinfo {author} {\bibfnamefont {E.}~\bibnamefont {Belorizky}}, \bibinfo {author} {\bibfnamefont {S.}~\bibnamefont {Ng}},\ and\ \bibinfo {author} {\bibfnamefont {T.}~\bibnamefont {Phillips}},\ }\bibfield  {title} {\bibinfo {title} {Determination of exchange interactions between coupled {Co$^{2+}$} ions in {MgF$_2$} by far-infrared spectroscopy},\ }\href {https://doi.org/10.1103/PhysRev.181.467} {\bibfield  {journal} {\bibinfo  {journal} {Phys. Rev.}\ }\textbf {\bibinfo {volume} {181}},\ \bibinfo {pages} {467} (\bibinfo {year} {1969})}\BibitemShut {NoStop}%
\bibitem [{\citenamefont {Morano}\ \emph {et~al.}(2024)\citenamefont {Morano}, \citenamefont {Maesen}, \citenamefont {Nikitin}, \citenamefont {Lass}, \citenamefont {Mazzone},\ and\ \citenamefont {Zaharko}}]{Morano2024}%
  \BibitemOpen
  \bibfield  {author} {\bibinfo {author} {\bibfnamefont {V.~C.}\ \bibnamefont {Morano}}, \bibinfo {author} {\bibfnamefont {Z.}~\bibnamefont {Maesen}}, \bibinfo {author} {\bibfnamefont {S.~E.}\ \bibnamefont {Nikitin}}, \bibinfo {author} {\bibfnamefont {J.}~\bibnamefont {Lass}}, \bibinfo {author} {\bibfnamefont {D.~G.}\ \bibnamefont {Mazzone}},\ and\ \bibinfo {author} {\bibfnamefont {O.}~\bibnamefont {Zaharko}},\ }\href {https://arxiv.org/abs/2412.03545} {\bibinfo {title} {Absence of altermagnetic magnon band splitting in {MnF$_2$}}} (\bibinfo {year} {2024}),\ \Eprint {https://arxiv.org/abs/2412.03545} {arXiv:2412.03545} \BibitemShut {NoStop}%
\bibitem [{\citenamefont {Das}\ \emph {et~al.}(2012)\citenamefont {Das}, \citenamefont {Kanungo},\ and\ \citenamefont {Saha-Dasgupta}}]{Das2012}%
  \BibitemOpen
  \bibfield  {author} {\bibinfo {author} {\bibfnamefont {H.}~\bibnamefont {Das}}, \bibinfo {author} {\bibfnamefont {S.}~\bibnamefont {Kanungo}},\ and\ \bibinfo {author} {\bibfnamefont {T.}~\bibnamefont {Saha-Dasgupta}},\ }\bibfield  {title} {\bibinfo {title} {First-principles study of magnetoelastic effect in the difluoride compounds {$M\mathrm{F}_{2}$} ({$M=\mathrm{Mn}$}, $\mathrm{Fe}$, $\mathrm{Co}$, $\mathrm{Ni}$)},\ }\href {https://doi.org/10.1103/PhysRevB.86.054422} {\bibfield  {journal} {\bibinfo  {journal} {Phys. Rev. B}\ }\textbf {\bibinfo {volume} {86}},\ \bibinfo {pages} {054422} (\bibinfo {year} {2012})}\BibitemShut {NoStop}%
\bibitem [{\citenamefont {Dubrovin}\ \emph {et~al.}(2024)\citenamefont {Dubrovin}, \citenamefont {Tellez-Mora}, \citenamefont {Garcia-Castro}, \citenamefont {Siverin}, \citenamefont {Novikova}, \citenamefont {Boldyrev}, \citenamefont {Mashkovich}, \citenamefont {Romero},\ and\ \citenamefont {Pisarev}}]{Dubrovin2024}%
  \BibitemOpen
  \bibfield  {author} {\bibinfo {author} {\bibfnamefont {R.~M.}\ \bibnamefont {Dubrovin}}, \bibinfo {author} {\bibfnamefont {A.}~\bibnamefont {Tellez-Mora}}, \bibinfo {author} {\bibfnamefont {A.~C.}\ \bibnamefont {Garcia-Castro}}, \bibinfo {author} {\bibfnamefont {N.~V.}\ \bibnamefont {Siverin}}, \bibinfo {author} {\bibfnamefont {N.~N.}\ \bibnamefont {Novikova}}, \bibinfo {author} {\bibfnamefont {K.~N.}\ \bibnamefont {Boldyrev}}, \bibinfo {author} {\bibfnamefont {E.~A.}\ \bibnamefont {Mashkovich}}, \bibinfo {author} {\bibfnamefont {A.~H.}\ \bibnamefont {Romero}},\ and\ \bibinfo {author} {\bibfnamefont {R.~V.}\ \bibnamefont {Pisarev}},\ }\bibfield  {title} {\bibinfo {title} {Polar phonons and magnetic excitations in the antiferromagnet {${\mathrm{CoF}}_{2}$}},\ }\href {https://doi.org/10.1103/PhysRevB.109.224312} {\bibfield  {journal} {\bibinfo  {journal} {Phys. Rev. B}\ }\textbf {\bibinfo {volume} {109}},\ \bibinfo {pages} {224312} (\bibinfo {year} {2024})}\BibitemShut {NoStop}%
\bibitem [{\citenamefont {Perdew}\ \emph {et~al.}(1996)\citenamefont {Perdew}, \citenamefont {Burke},\ and\ \citenamefont {Ernzerhof}}]{PBE}%
  \BibitemOpen
  \bibfield  {author} {\bibinfo {author} {\bibfnamefont {J.~P.}\ \bibnamefont {Perdew}}, \bibinfo {author} {\bibfnamefont {K.}~\bibnamefont {Burke}},\ and\ \bibinfo {author} {\bibfnamefont {M.}~\bibnamefont {Ernzerhof}},\ }\bibfield  {title} {\bibinfo {title} {Generalized gradient approximation made simple},\ }\href {https://doi.org/10.1103/PhysRevLett.77.3865} {\bibfield  {journal} {\bibinfo  {journal} {Phys. Rev. Lett.}\ }\textbf {\bibinfo {volume} {77}},\ \bibinfo {pages} {3865} (\bibinfo {year} {1996})}\BibitemShut {NoStop}%
\bibitem [{\citenamefont {Perdew}\ \emph {et~al.}(2008)\citenamefont {Perdew}, \citenamefont {Ruzsinszky}, \citenamefont {Csonka}, \citenamefont {Vydrov}, \citenamefont {Scuseria}, \citenamefont {Constantin}, \citenamefont {Zhou},\ and\ \citenamefont {Burke}}]{Perdew}%
  \BibitemOpen
  \bibfield  {author} {\bibinfo {author} {\bibfnamefont {J.~P.}\ \bibnamefont {Perdew}}, \bibinfo {author} {\bibfnamefont {A.}~\bibnamefont {Ruzsinszky}}, \bibinfo {author} {\bibfnamefont {G.~I.}\ \bibnamefont {Csonka}}, \bibinfo {author} {\bibfnamefont {O.~A.}\ \bibnamefont {Vydrov}}, \bibinfo {author} {\bibfnamefont {G.~E.}\ \bibnamefont {Scuseria}}, \bibinfo {author} {\bibfnamefont {L.~A.}\ \bibnamefont {Constantin}}, \bibinfo {author} {\bibfnamefont {X.}~\bibnamefont {Zhou}},\ and\ \bibinfo {author} {\bibfnamefont {K.}~\bibnamefont {Burke}},\ }\bibfield  {title} {\bibinfo {title} {Restoring the density-gradient expansion for exchange in solids and surfaces},\ }\href {https://doi.org/10.1103/PhysRevLett.100.136406} {\bibfield  {journal} {\bibinfo  {journal} {Phys. Rev. Lett.}\ }\textbf {\bibinfo {volume} {100}},\ \bibinfo {pages} {136406} (\bibinfo {year} {2008})}\BibitemShut {NoStop}%
\bibitem [{\citenamefont {Radaelli}(2024)}]{Radaelli2024}%
  \BibitemOpen
  \bibfield  {author} {\bibinfo {author} {\bibfnamefont {P.~G.}\ \bibnamefont {Radaelli}},\ }\bibfield  {title} {\bibinfo {title} {Tensorial approach to altermagnetism},\ }\href {https://doi.org/10.1103/PhysRevB.110.214428} {\bibfield  {journal} {\bibinfo  {journal} {Phys. Rev. B}\ }\textbf {\bibinfo {volume} {110}},\ \bibinfo {pages} {214428} (\bibinfo {year} {2024})}\BibitemShut {NoStop}%
\bibitem [{\citenamefont {Kluczyk}\ \emph {et~al.}(2024)\citenamefont {Kluczyk}, \citenamefont {Gas}, \citenamefont {Grzybowski}, \citenamefont {Skupi\'nski}, \citenamefont {Borysiewicz}, \citenamefont {F\k{a}s}, \citenamefont {Suffczy\'nski}, \citenamefont {Domagala}, \citenamefont {Grasza}, \citenamefont {Mycielski}, \citenamefont {Baj}, \citenamefont {Ahn}, \citenamefont {V\'yborn\'y}, \citenamefont {Sawicki},\ and\ \citenamefont {Gryglas-Borysiewicz}}]{Kluczyk2024}%
  \BibitemOpen
  \bibfield  {author} {\bibinfo {author} {\bibfnamefont {K.~P.}\ \bibnamefont {Kluczyk}}, \bibinfo {author} {\bibfnamefont {K.}~\bibnamefont {Gas}}, \bibinfo {author} {\bibfnamefont {M.~J.}\ \bibnamefont {Grzybowski}}, \bibinfo {author} {\bibfnamefont {P.}~\bibnamefont {Skupi\'nski}}, \bibinfo {author} {\bibfnamefont {M.~A.}\ \bibnamefont {Borysiewicz}}, \bibinfo {author} {\bibfnamefont {T.}~\bibnamefont {F\k{a}s}}, \bibinfo {author} {\bibfnamefont {J.}~\bibnamefont {Suffczy\'nski}}, \bibinfo {author} {\bibfnamefont {J.~Z.}\ \bibnamefont {Domagala}}, \bibinfo {author} {\bibfnamefont {K.}~\bibnamefont {Grasza}}, \bibinfo {author} {\bibfnamefont {A.}~\bibnamefont {Mycielski}}, \bibinfo {author} {\bibfnamefont {M.}~\bibnamefont {Baj}}, \bibinfo {author} {\bibfnamefont {K.~H.}\ \bibnamefont {Ahn}}, \bibinfo {author} {\bibfnamefont {K.}~\bibnamefont {V\'yborn\'y}}, \bibinfo {author} {\bibfnamefont {M.}~\bibnamefont {Sawicki}},\ and\ \bibinfo {author} {\bibfnamefont {M.}~\bibnamefont {Gryglas-Borysiewicz}},\
  }\bibfield  {title} {\bibinfo {title} {Coexistence of anomalous {Hall} effect and weak magnetization in a nominally collinear antiferromagnet {MnTe}},\ }\href {https://doi.org/10.1103/PhysRevB.110.155201} {\bibfield  {journal} {\bibinfo  {journal} {Phys. Rev. B}\ }\textbf {\bibinfo {volume} {110}},\ \bibinfo {pages} {155201} (\bibinfo {year} {2024})}\BibitemShut {NoStop}%
\bibitem [{\citenamefont {Kotegawa}\ \emph {et~al.}(2024)\citenamefont {Kotegawa}, \citenamefont {Nakamura}, \citenamefont {Huyen}, \citenamefont {Arai}, \citenamefont {Tou}, \citenamefont {Sugawara}, \citenamefont {Hayashi}, \citenamefont {Takeda}, \citenamefont {Tabata}, \citenamefont {Kaneko}, \citenamefont {Kodama},\ and\ \citenamefont {Suzuki}}]{Kotegawa2024}%
  \BibitemOpen
  \bibfield  {author} {\bibinfo {author} {\bibfnamefont {H.}~\bibnamefont {Kotegawa}}, \bibinfo {author} {\bibfnamefont {A.}~\bibnamefont {Nakamura}}, \bibinfo {author} {\bibfnamefont {V.~T.~N.}\ \bibnamefont {Huyen}}, \bibinfo {author} {\bibfnamefont {Y.}~\bibnamefont {Arai}}, \bibinfo {author} {\bibfnamefont {H.}~\bibnamefont {Tou}}, \bibinfo {author} {\bibfnamefont {H.}~\bibnamefont {Sugawara}}, \bibinfo {author} {\bibfnamefont {J.}~\bibnamefont {Hayashi}}, \bibinfo {author} {\bibfnamefont {K.}~\bibnamefont {Takeda}}, \bibinfo {author} {\bibfnamefont {C.}~\bibnamefont {Tabata}}, \bibinfo {author} {\bibfnamefont {K.}~\bibnamefont {Kaneko}}, \bibinfo {author} {\bibfnamefont {K.}~\bibnamefont {Kodama}},\ and\ \bibinfo {author} {\bibfnamefont {M.-T.}\ \bibnamefont {Suzuki}},\ }\bibfield  {title} {\bibinfo {title} {Large spontaneous {Hall} effect with flexible domain control in the antiferromagnetic material {TaMnP}},\ }\href {https://doi.org/10.1103/PhysRevB.110.214417} {\bibfield  {journal} {\bibinfo
  {journal} {Phys. Rev. B}\ }\textbf {\bibinfo {volume} {110}},\ \bibinfo {pages} {214417} (\bibinfo {year} {2024})}\BibitemShut {NoStop}%
\bibitem [{\citenamefont {Cheong}\ and\ \citenamefont {Huang}(2024)}]{Cheong2024}%
  \BibitemOpen
  \bibfield  {author} {\bibinfo {author} {\bibfnamefont {S.-W.}\ \bibnamefont {Cheong}}\ and\ \bibinfo {author} {\bibfnamefont {F.-T.}\ \bibnamefont {Huang}},\ }\bibfield  {title} {\bibinfo {title} {Altermagnetism with non-collinear spins},\ }\href {https://doi.org/10.1038/s41535-024-00626-6} {\bibfield  {journal} {\bibinfo  {journal} {npj Quantum Mater.}\ }\textbf {\bibinfo {volume} {9}},\ \bibinfo {pages} {13} (\bibinfo {year} {2024})}\BibitemShut {NoStop}%
\bibitem [{\citenamefont {Roig}\ \emph {et~al.}(2025)\citenamefont {Roig}, \citenamefont {Yu}, \citenamefont {Ekman}, \citenamefont {Kreisel}, \citenamefont {Andersen},\ and\ \citenamefont {Agterberg}}]{Roig2024a}%
  \BibitemOpen
  \bibfield  {author} {\bibinfo {author} {\bibfnamefont {M.}~\bibnamefont {Roig}}, \bibinfo {author} {\bibfnamefont {Y.}~\bibnamefont {Yu}}, \bibinfo {author} {\bibfnamefont {R.~C.}\ \bibnamefont {Ekman}}, \bibinfo {author} {\bibfnamefont {A.}~\bibnamefont {Kreisel}}, \bibinfo {author} {\bibfnamefont {B.~M.}\ \bibnamefont {Andersen}},\ and\ \bibinfo {author} {\bibfnamefont {D.~F.}\ \bibnamefont {Agterberg}},\ }\bibfield  {title} {\bibinfo {title} {Quasisymmetry-constrained spin ferromagnetism in altermagnets},\ }\href {https://doi.org/10.1103/839n-rckn} {\bibfield  {journal} {\bibinfo  {journal} {Phys. Rev. Lett.}\ }\textbf {\bibinfo {volume} {135}},\ \bibinfo {pages} {016703} (\bibinfo {year} {2025})}\BibitemShut {NoStop}%
\bibitem [{\citenamefont {Fedorova}\ \emph {et~al.}(2015)\citenamefont {Fedorova}, \citenamefont {Ederer}, \citenamefont {Spaldin},\ and\ \citenamefont {Scaramucci}}]{Fedorova2015}%
  \BibitemOpen
  \bibfield  {author} {\bibinfo {author} {\bibfnamefont {N.~S.}\ \bibnamefont {Fedorova}}, \bibinfo {author} {\bibfnamefont {C.}~\bibnamefont {Ederer}}, \bibinfo {author} {\bibfnamefont {N.~A.}\ \bibnamefont {Spaldin}},\ and\ \bibinfo {author} {\bibfnamefont {A.}~\bibnamefont {Scaramucci}},\ }\bibfield  {title} {\bibinfo {title} {Biquadratic and ring exchange interactions in orthorhombic perovskite manganites},\ }\href {https://doi.org/10.1103/PhysRevB.91.165122} {\bibfield  {journal} {\bibinfo  {journal} {Phys. Rev. B}\ }\textbf {\bibinfo {volume} {91}},\ \bibinfo {pages} {165122} (\bibinfo {year} {2015})}\BibitemShut {NoStop}%
\bibitem [{\citenamefont {Szilva}\ \emph {et~al.}(2023)\citenamefont {Szilva}, \citenamefont {Kvashnin}, \citenamefont {Stepanov}, \citenamefont {Nordstr\"om}, \citenamefont {Eriksson}, \citenamefont {Lichtenstein},\ and\ \citenamefont {Katsnelson}}]{Szilva2023}%
  \BibitemOpen
  \bibfield  {author} {\bibinfo {author} {\bibfnamefont {A.}~\bibnamefont {Szilva}}, \bibinfo {author} {\bibfnamefont {Y.}~\bibnamefont {Kvashnin}}, \bibinfo {author} {\bibfnamefont {E.~A.}\ \bibnamefont {Stepanov}}, \bibinfo {author} {\bibfnamefont {L.}~\bibnamefont {Nordstr\"om}}, \bibinfo {author} {\bibfnamefont {O.}~\bibnamefont {Eriksson}}, \bibinfo {author} {\bibfnamefont {A.~I.}\ \bibnamefont {Lichtenstein}},\ and\ \bibinfo {author} {\bibfnamefont {M.~I.}\ \bibnamefont {Katsnelson}},\ }\bibfield  {title} {\bibinfo {title} {Quantitative theory of magnetic interactions in solids},\ }\href {https://doi.org/10.1103/RevModPhys.95.035004} {\bibfield  {journal} {\bibinfo  {journal} {Rev. Mod. Phys.}\ }\textbf {\bibinfo {volume} {95}},\ \bibinfo {pages} {035004} (\bibinfo {year} {2023})}\BibitemShut {NoStop}%
\bibitem [{\citenamefont {Mazin}\ and\ \citenamefont {Belashchenko}(2024)}]{Gossamer}%
  \BibitemOpen
  \bibfield  {author} {\bibinfo {author} {\bibfnamefont {I.~I.}\ \bibnamefont {Mazin}}\ and\ \bibinfo {author} {\bibfnamefont {K.~D.}\ \bibnamefont {Belashchenko}},\ }\bibfield  {title} {\bibinfo {title} {Origin of the gossamer ferromagnetism in {MnTe}},\ }\href {https://doi.org/10.1103/PhysRevB.110.214436} {\bibfield  {journal} {\bibinfo  {journal} {Phys. Rev. B}\ }\textbf {\bibinfo {volume} {110}},\ \bibinfo {pages} {214436} (\bibinfo {year} {2024})}\BibitemShut {NoStop}%
\bibitem [{\citenamefont {Bhowal}\ and\ \citenamefont {Spaldin}(2024)}]{Bhowal2024}%
  \BibitemOpen
  \bibfield  {author} {\bibinfo {author} {\bibfnamefont {S.}~\bibnamefont {Bhowal}}\ and\ \bibinfo {author} {\bibfnamefont {N.~A.}\ \bibnamefont {Spaldin}},\ }\bibfield  {title} {\bibinfo {title} {Ferroically ordered magnetic octupoles in $d$-wave altermagnets},\ }\href {https://doi.org/10.1103/PhysRevX.14.011019} {\bibfield  {journal} {\bibinfo  {journal} {Phys. Rev. X}\ }\textbf {\bibinfo {volume} {14}},\ \bibinfo {pages} {011019} (\bibinfo {year} {2024})}\BibitemShut {NoStop}%
\bibitem [{\citenamefont {Martin}(2020)}]{Martin2020ElectronicStructure2e}%
  \BibitemOpen
  \bibfield  {author} {\bibinfo {author} {\bibfnamefont {R.~M.}\ \bibnamefont {Martin}},\ }\href {https://doi.org/10.1017/9781108555586} {\emph {\bibinfo {title} {Electronic Structure: Basic Theory and Practical Methods}}},\ \bibinfo {edition} {2nd}\ ed.\ (\bibinfo  {publisher} {Cambridge University Press},\ \bibinfo {year} {2020})\BibitemShut {NoStop}%
\bibitem [{\citenamefont {Wasscher}(1969)}]{Wasscher-thesis}%
  \BibitemOpen
  \bibfield  {author} {\bibinfo {author} {\bibfnamefont {J.}~\bibnamefont {Wasscher}},\ }\emph {\bibinfo {title} {Electrical transport phenomena in {{MnTe}}, an antiferromagnetic semiconductor}},\ \href {https://doi.org/10.6100/IR43336} {Ph.D. thesis},\ \bibinfo  {school} {Technische Hogeschool Eindhoven} (\bibinfo {year} {1969})\BibitemShut {NoStop}%
\bibitem [{\citenamefont {Ouassou}\ \emph {et~al.}(2023)\citenamefont {Ouassou}, \citenamefont {Brataas},\ and\ \citenamefont {Linder}}]{Ouassou2023}%
  \BibitemOpen
  \bibfield  {author} {\bibinfo {author} {\bibfnamefont {J.~A.}\ \bibnamefont {Ouassou}}, \bibinfo {author} {\bibfnamefont {A.}~\bibnamefont {Brataas}},\ and\ \bibinfo {author} {\bibfnamefont {J.}~\bibnamefont {Linder}},\ }\bibfield  {title} {\bibinfo {title} {dc {Josephson} effect in altermagnets},\ }\href {https://doi.org/10.1103/PhysRevLett.131.076003} {\bibfield  {journal} {\bibinfo  {journal} {Phys. Rev. Lett.}\ }\textbf {\bibinfo {volume} {131}},\ \bibinfo {pages} {076003} (\bibinfo {year} {2023})}\BibitemShut {NoStop}%
\bibitem [{\citenamefont {Lu}\ \emph {et~al.}(2024)\citenamefont {Lu}, \citenamefont {Maeda}, \citenamefont {Ito}, \citenamefont {Yada},\ and\ \citenamefont {Tanaka}}]{Lu2024}%
  \BibitemOpen
  \bibfield  {author} {\bibinfo {author} {\bibfnamefont {B.}~\bibnamefont {Lu}}, \bibinfo {author} {\bibfnamefont {K.}~\bibnamefont {Maeda}}, \bibinfo {author} {\bibfnamefont {H.}~\bibnamefont {Ito}}, \bibinfo {author} {\bibfnamefont {K.}~\bibnamefont {Yada}},\ and\ \bibinfo {author} {\bibfnamefont {Y.}~\bibnamefont {Tanaka}},\ }\bibfield  {title} {\bibinfo {title} {$\ensuremath{\varphi}$ {Josephson} junction induced by altermagnetism},\ }\href {https://doi.org/10.1103/PhysRevLett.133.226002} {\bibfield  {journal} {\bibinfo  {journal} {Phys. Rev. Lett.}\ }\textbf {\bibinfo {volume} {133}},\ \bibinfo {pages} {226002} (\bibinfo {year} {2024})}\BibitemShut {NoStop}%
\bibitem [{\citenamefont {Sun}\ \emph {et~al.}(2025)\citenamefont {Sun}, \citenamefont {Zhang}, \citenamefont {Li},\ and\ \citenamefont {Trauzettel}}]{Sun2025}%
  \BibitemOpen
  \bibfield  {author} {\bibinfo {author} {\bibfnamefont {H.-P.}\ \bibnamefont {Sun}}, \bibinfo {author} {\bibfnamefont {S.-B.}\ \bibnamefont {Zhang}}, \bibinfo {author} {\bibfnamefont {C.-A.}\ \bibnamefont {Li}},\ and\ \bibinfo {author} {\bibfnamefont {B.}~\bibnamefont {Trauzettel}},\ }\bibfield  {title} {\bibinfo {title} {Tunable second harmonic in altermagnetic {Josephson} junctions},\ }\href {https://doi.org/10.1103/PhysRevB.111.165406} {\bibfield  {journal} {\bibinfo  {journal} {Phys. Rev. B}\ }\textbf {\bibinfo {volume} {111}},\ \bibinfo {pages} {165406} (\bibinfo {year} {2025})}\BibitemShut {NoStop}%
\bibitem [{\citenamefont {Fukaya}\ \emph {et~al.}(2025{\natexlab{a}})\citenamefont {Fukaya}, \citenamefont {Maeda}, \citenamefont {Yada}, \citenamefont {Cayao}, \citenamefont {Tanaka},\ and\ \citenamefont {Lu}}]{Fukaya2025}%
  \BibitemOpen
  \bibfield  {author} {\bibinfo {author} {\bibfnamefont {Y.}~\bibnamefont {Fukaya}}, \bibinfo {author} {\bibfnamefont {K.}~\bibnamefont {Maeda}}, \bibinfo {author} {\bibfnamefont {K.}~\bibnamefont {Yada}}, \bibinfo {author} {\bibfnamefont {J.}~\bibnamefont {Cayao}}, \bibinfo {author} {\bibfnamefont {Y.}~\bibnamefont {Tanaka}},\ and\ \bibinfo {author} {\bibfnamefont {B.}~\bibnamefont {Lu}},\ }\bibfield  {title} {\bibinfo {title} {Josephson effect and odd-frequency pairing in superconducting junctions with unconventional magnets},\ }\href {https://doi.org/10.1103/PhysRevB.111.064502} {\bibfield  {journal} {\bibinfo  {journal} {Phys. Rev. B}\ }\textbf {\bibinfo {volume} {111}},\ \bibinfo {pages} {064502} (\bibinfo {year} {2025}{\natexlab{a}})}\BibitemShut {NoStop}%
\bibitem [{\citenamefont {Zhao}\ \emph {et~al.}(2025)\citenamefont {Zhao}, \citenamefont {Fukaya}, \citenamefont {Burset}, \citenamefont {Cayao}, \citenamefont {Tanaka},\ and\ \citenamefont {Lu}}]{Zhao2025}%
  \BibitemOpen
  \bibfield  {author} {\bibinfo {author} {\bibfnamefont {W.}~\bibnamefont {Zhao}}, \bibinfo {author} {\bibfnamefont {Y.}~\bibnamefont {Fukaya}}, \bibinfo {author} {\bibfnamefont {P.}~\bibnamefont {Burset}}, \bibinfo {author} {\bibfnamefont {J.}~\bibnamefont {Cayao}}, \bibinfo {author} {\bibfnamefont {Y.}~\bibnamefont {Tanaka}},\ and\ \bibinfo {author} {\bibfnamefont {B.}~\bibnamefont {Lu}},\ }\bibfield  {title} {\bibinfo {title} {Orientation-dependent transport in junctions formed by $d$-wave altermagnets and $d$-wave superconductors},\ }\href {https://doi.org/10.1103/PhysRevB.111.184515} {\bibfield  {journal} {\bibinfo  {journal} {Phys. Rev. B}\ }\textbf {\bibinfo {volume} {111}},\ \bibinfo {pages} {184515} (\bibinfo {year} {2025})}\BibitemShut {NoStop}%
\bibitem [{\citenamefont {Banerjee}\ and\ \citenamefont {Scheurer}(2024)}]{Banerjee2024}%
  \BibitemOpen
  \bibfield  {author} {\bibinfo {author} {\bibfnamefont {S.}~\bibnamefont {Banerjee}}\ and\ \bibinfo {author} {\bibfnamefont {M.~S.}\ \bibnamefont {Scheurer}},\ }\bibfield  {title} {\bibinfo {title} {Altermagnetic superconducting diode effect},\ }\href {https://doi.org/10.1103/PhysRevB.110.024503} {\bibfield  {journal} {\bibinfo  {journal} {Phys. Rev. B}\ }\textbf {\bibinfo {volume} {110}},\ \bibinfo {pages} {024503} (\bibinfo {year} {2024})}\BibitemShut {NoStop}%
\bibitem [{\citenamefont {Maeda}\ \emph {et~al.}(2025)\citenamefont {Maeda}, \citenamefont {Fukaya}, \citenamefont {Yada}, \citenamefont {Lu}, \citenamefont {Tanaka},\ and\ \citenamefont {Cayao}}]{Maeda2025}%
  \BibitemOpen
  \bibfield  {author} {\bibinfo {author} {\bibfnamefont {K.}~\bibnamefont {Maeda}}, \bibinfo {author} {\bibfnamefont {Y.}~\bibnamefont {Fukaya}}, \bibinfo {author} {\bibfnamefont {K.}~\bibnamefont {Yada}}, \bibinfo {author} {\bibfnamefont {B.}~\bibnamefont {Lu}}, \bibinfo {author} {\bibfnamefont {Y.}~\bibnamefont {Tanaka}},\ and\ \bibinfo {author} {\bibfnamefont {J.}~\bibnamefont {Cayao}},\ }\bibfield  {title} {\bibinfo {title} {Classification of pair symmetries in superconductors with unconventional magnetism},\ }\href {https://doi.org/10.1103/PhysRevB.111.144508} {\bibfield  {journal} {\bibinfo  {journal} {Phys. Rev. B}\ }\textbf {\bibinfo {volume} {111}},\ \bibinfo {pages} {144508} (\bibinfo {year} {2025})}\BibitemShut {NoStop}%
\bibitem [{\citenamefont {Fukaya}\ \emph {et~al.}(2025{\natexlab{b}})\citenamefont {Fukaya}, \citenamefont {Lu}, \citenamefont {Yada}, \citenamefont {Tanaka},\ and\ \citenamefont {Cayao}}]{Fukaya2025a}%
  \BibitemOpen
  \bibfield  {author} {\bibinfo {author} {\bibfnamefont {Y.}~\bibnamefont {Fukaya}}, \bibinfo {author} {\bibfnamefont {B.}~\bibnamefont {Lu}}, \bibinfo {author} {\bibfnamefont {K.}~\bibnamefont {Yada}}, \bibinfo {author} {\bibfnamefont {Y.}~\bibnamefont {Tanaka}},\ and\ \bibinfo {author} {\bibfnamefont {J.}~\bibnamefont {Cayao}},\ }\bibfield  {title} {\bibinfo {title} {Superconducting phenomena in systems with unconventional magnets},\ }\href {https://doi.org/10.1088/1361-648X/adf1cf} {\bibfield  {journal} {\bibinfo  {journal} {Journal of Physics: Condensed Matter}\ }\textbf {\bibinfo {volume} {37}},\ \bibinfo {pages} {313003} (\bibinfo {year} {2025}{\natexlab{b}})}\BibitemShut {NoStop}%
\bibitem [{\citenamefont {Steward}\ \emph {et~al.}(2023)\citenamefont {Steward}, \citenamefont {Fernandes},\ and\ \citenamefont {Schmalian}}]{Steward2023}%
  \BibitemOpen
  \bibfield  {author} {\bibinfo {author} {\bibfnamefont {C.~R.~W.}\ \bibnamefont {Steward}}, \bibinfo {author} {\bibfnamefont {R.~M.}\ \bibnamefont {Fernandes}},\ and\ \bibinfo {author} {\bibfnamefont {J.}~\bibnamefont {Schmalian}},\ }\bibfield  {title} {\bibinfo {title} {Dynamic paramagnon-polarons in altermagnets},\ }\href {https://doi.org/10.1103/PhysRevB.108.144418} {\bibfield  {journal} {\bibinfo  {journal} {Phys. Rev. B}\ }\textbf {\bibinfo {volume} {108}},\ \bibinfo {pages} {144418} (\bibinfo {year} {2023})}\BibitemShut {NoStop}%
\bibitem [{\citenamefont {Schiff}\ \emph {et~al.}(2025{\natexlab{a}})\citenamefont {Schiff}, \citenamefont {Corticelli}, \citenamefont {Guerreiro}, \citenamefont {Romhányi},\ and\ \citenamefont {McClarty}}]{Schiff2025}%
  \BibitemOpen
  \bibfield  {author} {\bibinfo {author} {\bibfnamefont {H.}~\bibnamefont {Schiff}}, \bibinfo {author} {\bibfnamefont {A.}~\bibnamefont {Corticelli}}, \bibinfo {author} {\bibfnamefont {A.}~\bibnamefont {Guerreiro}}, \bibinfo {author} {\bibfnamefont {J.}~\bibnamefont {Romhányi}},\ and\ \bibinfo {author} {\bibfnamefont {P.}~\bibnamefont {McClarty}},\ }\bibfield  {title} {\bibinfo {title} {{The crystallographic spin point groups and their representations}},\ }\href {https://doi.org/10.21468/SciPostPhys.18.3.109} {\bibfield  {journal} {\bibinfo  {journal} {SciPost Phys.}\ }\textbf {\bibinfo {volume} {18}},\ \bibinfo {pages} {109} (\bibinfo {year} {2025}{\natexlab{a}})}\BibitemShut {NoStop}%
\bibitem [{\citenamefont {Dimmock}(1963)}]{Dimmock1963}%
  \BibitemOpen
  \bibfield  {author} {\bibinfo {author} {\bibfnamefont {J.~O.}\ \bibnamefont {Dimmock}},\ }\bibfield  {title} {\bibinfo {title} {Representation theory for nonunitary groups},\ }\href {https://doi.org/10.1063/1.1703905} {\bibfield  {journal} {\bibinfo  {journal} {Journal of Mathematical Physics}\ }\textbf {\bibinfo {volume} {4}},\ \bibinfo {pages} {1307} (\bibinfo {year} {1963})}\BibitemShut {NoStop}%
\bibitem [{\citenamefont {Sigrist}\ and\ \citenamefont {Ueda}(1991)}]{Sigrist1991}%
  \BibitemOpen
  \bibfield  {author} {\bibinfo {author} {\bibfnamefont {M.}~\bibnamefont {Sigrist}}\ and\ \bibinfo {author} {\bibfnamefont {K.}~\bibnamefont {Ueda}},\ }\bibfield  {title} {\bibinfo {title} {Phenomenological theory of unconventional superconductivity},\ }\href {https://doi.org/10.1103/RevModPhys.63.239} {\bibfield  {journal} {\bibinfo  {journal} {Rev. Mod. Phys.}\ }\textbf {\bibinfo {volume} {63}},\ \bibinfo {pages} {239} (\bibinfo {year} {1991})}\BibitemShut {NoStop}%
\bibitem [{\citenamefont {Schiff}\ \emph {et~al.}(2025{\natexlab{b}})\citenamefont {Schiff}, \citenamefont {McClarty}, \citenamefont {Rau},\ and\ \citenamefont {Romh\'anyi}}]{Schiff2025a}%
  \BibitemOpen
  \bibfield  {author} {\bibinfo {author} {\bibfnamefont {H.}~\bibnamefont {Schiff}}, \bibinfo {author} {\bibfnamefont {P.}~\bibnamefont {McClarty}}, \bibinfo {author} {\bibfnamefont {J.~G.}\ \bibnamefont {Rau}},\ and\ \bibinfo {author} {\bibfnamefont {J.}~\bibnamefont {Romh\'anyi}},\ }\bibfield  {title} {\bibinfo {title} {Collinear altermagnets and their {Landau} theories},\ }\href {https://doi.org/10.1103/q44z-ynbr} {\bibfield  {journal} {\bibinfo  {journal} {Phys. Rev. Res.}\ }\textbf {\bibinfo {volume} {7}},\ \bibinfo {pages} {033301} (\bibinfo {year} {2025}{\natexlab{b}})}\BibitemShut {NoStop}%
\bibitem [{\citenamefont {Zhai}\ \emph {et~al.}(2025)\citenamefont {Zhai}, \citenamefont {Yu}, \citenamefont {Lv}, \citenamefont {Zhang},\ and\ \citenamefont {Zhao}}]{Zhai2025}%
  \BibitemOpen
  \bibfield  {author} {\bibinfo {author} {\bibfnamefont {Y.}~\bibnamefont {Zhai}}, \bibinfo {author} {\bibfnamefont {L.}~\bibnamefont {Yu}}, \bibinfo {author} {\bibfnamefont {J.}~\bibnamefont {Lv}}, \bibinfo {author} {\bibfnamefont {W.}~\bibnamefont {Zhang}},\ and\ \bibinfo {author} {\bibfnamefont {H.~J.}\ \bibnamefont {Zhao}},\ }\bibfield  {title} {\bibinfo {title} {Zeeman-type spin splittings in strained $d$-wave altermagnets},\ }\href {https://doi.org/10.1103/7q2d-jcqg} {\bibfield  {journal} {\bibinfo  {journal} {Phys. Rev. B}\ }\textbf {\bibinfo {volume} {112}},\ \bibinfo {pages} {174411} (\bibinfo {year} {2025})}\BibitemShut {NoStop}%
\bibitem [{\citenamefont {Bradley}\ and\ \citenamefont {Cracknell}(1972)}]{Bradley1972}%
  \BibitemOpen
  \bibfield  {author} {\bibinfo {author} {\bibfnamefont {C.~J.}\ \bibnamefont {Bradley}}\ and\ \bibinfo {author} {\bibfnamefont {A.~P.}\ \bibnamefont {Cracknell}},\ }\href@noop {} {\emph {\bibinfo {title} {The Mathematical Theory of Symmetry in Solids: Representation Theory for Point Groups and Space Groups}}}\ (\bibinfo  {publisher} {Clarendon Press},\ \bibinfo {year} {1972})\BibitemShut {NoStop}%
\bibitem [{\citenamefont {Mazin}\ \emph {et~al.}(2021)\citenamefont {Mazin}, \citenamefont {Koepernik}, \citenamefont {Johannes}, \citenamefont {Gonz{\'a}lez-Hern{\'a}ndez},\ and\ \citenamefont {{\v S}mejkal}}]{Mazin2021}%
  \BibitemOpen
  \bibfield  {author} {\bibinfo {author} {\bibfnamefont {I.~I.}\ \bibnamefont {Mazin}}, \bibinfo {author} {\bibfnamefont {K.}~\bibnamefont {Koepernik}}, \bibinfo {author} {\bibfnamefont {M.~D.}\ \bibnamefont {Johannes}}, \bibinfo {author} {\bibfnamefont {R.}~\bibnamefont {Gonz{\'a}lez-Hern{\'a}ndez}},\ and\ \bibinfo {author} {\bibfnamefont {L.}~\bibnamefont {{\v S}mejkal}},\ }\bibfield  {title} {\bibinfo {title} {Prediction of unconventional magnetism in doped {FeSb}$_2$},\ }\href {https://doi.org/10.1073/pnas.2108924118} {\bibfield  {journal} {\bibinfo  {journal} {Proc. Natl. Acad. Sci. USA}\ }\textbf {\bibinfo {volume} {118}},\ \bibinfo {pages} {e2108924118} (\bibinfo {year} {2021})}\BibitemShut {NoStop}%
\bibitem [{\citenamefont {Stout}\ and\ \citenamefont {Reed}(1954)}]{Stout}%
  \BibitemOpen
  \bibfield  {author} {\bibinfo {author} {\bibfnamefont {J.}~\bibnamefont {Stout}}\ and\ \bibinfo {author} {\bibfnamefont {S.~A.}\ \bibnamefont {Reed}},\ }\bibfield  {title} {\bibinfo {title} {The crystal structure of {MnF$_2$}, {FeF$_2$}, {CoF$_2$}, {NiF$_2$} and {ZnF$_2$}},\ }\href {https://pubs.acs.org/doi/10.1021/ja01650a005} {\bibfield  {journal} {\bibinfo  {journal} {J. Am. Chem. Soc.}\ }\textbf {\bibinfo {volume} {76}},\ \bibinfo {pages} {5279} (\bibinfo {year} {1954})}\BibitemShut {NoStop}%
\bibitem [{\citenamefont {Yamani}\ \emph {et~al.}(2010)\citenamefont {Yamani}, \citenamefont {Tun},\ and\ \citenamefont {Ryan}}]{Yamani}%
  \BibitemOpen
  \bibfield  {author} {\bibinfo {author} {\bibfnamefont {Z.}~\bibnamefont {Yamani}}, \bibinfo {author} {\bibfnamefont {Z.}~\bibnamefont {Tun}},\ and\ \bibinfo {author} {\bibfnamefont {D.}~\bibnamefont {Ryan}},\ }\bibfield  {title} {\bibinfo {title} {Neutron scattering study of the classical antiferromagnet {MnF$_2$}: a perfect hands-on neutron scattering teaching course},\ }\href {https://doi.org/10.1139/P10-081} {\bibfield  {journal} {\bibinfo  {journal} {Can. J. Phys.}\ }\textbf {\bibinfo {volume} {88}},\ \bibinfo {pages} {771} (\bibinfo {year} {2010})}\BibitemShut {NoStop}%
\bibitem [{\citenamefont {de~Almeida}(1988)}]{Almeida}%
  \BibitemOpen
  \bibfield  {author} {\bibinfo {author} {\bibfnamefont {M.}~\bibnamefont {de~Almeida}},\ }\bibfield  {title} {\bibinfo {title} {Magnetisation density and covalency in ferrous fluoride},\ }\href {https://iopscience.iop.org/article/10.1088/0022-3719/21/6/017} {\bibfield  {journal} {\bibinfo  {journal} {J. Phys. C: Solid State Phys.}\ }\textbf {\bibinfo {volume} {21}},\ \bibinfo {pages} {1111} (\bibinfo {year} {1988})}\BibitemShut {NoStop}%
\bibitem [{\citenamefont {Thomson}\ \emph {et~al.}(2014)\citenamefont {Thomson}, \citenamefont {Chatterji},\ and\ \citenamefont {Carpenter}}]{Chatterji}%
  \BibitemOpen
  \bibfield  {author} {\bibinfo {author} {\bibfnamefont {R.}~\bibnamefont {Thomson}}, \bibinfo {author} {\bibfnamefont {T.}~\bibnamefont {Chatterji}},\ and\ \bibinfo {author} {\bibfnamefont {M.}~\bibnamefont {Carpenter}},\ }\bibfield  {title} {\bibinfo {title} {{CoF$_2$}: a model system for magnetoelastic coupling and elastic softening mechanisms associated with paramagnetic $\leftrightarrow$ antiferromagnetic phase transitions},\ }\href {https://iopscience.iop.org/article/10.1088/0953-8984/26/14/146001} {\bibfield  {journal} {\bibinfo  {journal} {J. Phys.: Condens. Matter}\ }\textbf {\bibinfo {volume} {26}},\ \bibinfo {pages} {146001} (\bibinfo {year} {2014})}\BibitemShut {NoStop}%
\bibitem [{\citenamefont {Bl{\"o}chl}(1994)}]{Blochl}%
  \BibitemOpen
  \bibfield  {author} {\bibinfo {author} {\bibfnamefont {P.~E.}\ \bibnamefont {Bl{\"o}chl}},\ }\bibfield  {title} {\bibinfo {title} {Projector augmented-wave method},\ }\href {https://doi.org/10.1103/PhysRevB.50.17953} {\bibfield  {journal} {\bibinfo  {journal} {Phys. Rev. B}\ }\textbf {\bibinfo {volume} {50}},\ \bibinfo {pages} {17953} (\bibinfo {year} {1994})}\BibitemShut {NoStop}%
\bibitem [{\citenamefont {Kresse}\ and\ \citenamefont {Hafner}(1993)}]{Kresse}%
  \BibitemOpen
  \bibfield  {author} {\bibinfo {author} {\bibfnamefont {G.}~\bibnamefont {Kresse}}\ and\ \bibinfo {author} {\bibfnamefont {J.}~\bibnamefont {Hafner}},\ }\bibfield  {title} {\bibinfo {title} {Ab initio molecular dynamics for open-shell transition metals},\ }\href {https://doi.org/10.1103/PhysRevB.48.13115} {\bibfield  {journal} {\bibinfo  {journal} {Phys. Rev. B}\ }\textbf {\bibinfo {volume} {48}},\ \bibinfo {pages} {13115} (\bibinfo {year} {1993})}\BibitemShut {NoStop}%
\bibitem [{\citenamefont {Kresse}\ and\ \citenamefont {Furthm{\"u}ller}(1996)}]{Kresse2}%
  \BibitemOpen
  \bibfield  {author} {\bibinfo {author} {\bibfnamefont {G.}~\bibnamefont {Kresse}}\ and\ \bibinfo {author} {\bibfnamefont {J.}~\bibnamefont {Furthm{\"u}ller}},\ }\bibfield  {title} {\bibinfo {title} {Efficient iterative schemes for ab initio total-energy calculations using a plane-wave basis set},\ }\href {https://doi.org/10.1103/PhysRevB.54.11169} {\bibfield  {journal} {\bibinfo  {journal} {Phys. Rev. B}\ }\textbf {\bibinfo {volume} {54}},\ \bibinfo {pages} {11169} (\bibinfo {year} {1996})}\BibitemShut {NoStop}%
\bibitem [{\citenamefont {Liechtenstein}\ \emph {et~al.}(1995)\citenamefont {Liechtenstein}, \citenamefont {Anisimov},\ and\ \citenamefont {Zaanen}}]{Liechtenstein}%
  \BibitemOpen
  \bibfield  {author} {\bibinfo {author} {\bibfnamefont {A.}~\bibnamefont {Liechtenstein}}, \bibinfo {author} {\bibfnamefont {V.~I.}\ \bibnamefont {Anisimov}},\ and\ \bibinfo {author} {\bibfnamefont {J.}~\bibnamefont {Zaanen}},\ }\bibfield  {title} {\bibinfo {title} {Density-functional theory and strong interactions: Orbital ordering in {Mott-Hubbard} insulators},\ }\href {https://doi.org/10.1103/PhysRevB.52.R5467} {\bibfield  {journal} {\bibinfo  {journal} {Phys. Rev. B}\ }\textbf {\bibinfo {volume} {52}},\ \bibinfo {pages} {R5467} (\bibinfo {year} {1995})}\BibitemShut {NoStop}%
\bibitem [{\citenamefont {Wortmann}\ \emph {et~al.}(2023)\citenamefont {Wortmann}, \citenamefont {Michalicek}, \citenamefont {Baadji}, \citenamefont {Betzinger}, \citenamefont {Bihlmayer}, \citenamefont {Br\"oder}, \citenamefont {Burnus}, \citenamefont {Enkovaara}, \citenamefont {Freimuth}, \citenamefont {Friedrich}, \citenamefont {Gerhorst}, \citenamefont {Granberg~Cauchi}, \citenamefont {Grytsiuk}, \citenamefont {Hanke}, \citenamefont {Hanke}, \citenamefont {Heide}, \citenamefont {Heinze}, \citenamefont {Hilgers}, \citenamefont {Janssen}, \citenamefont {Kl\"uppelberg}, \citenamefont {Kovacik}, \citenamefont {Kurz}, \citenamefont {Lezaic}, \citenamefont {Madsen}, \citenamefont {Mokrousov}, \citenamefont {Neukirchen}, \citenamefont {Redies}, \citenamefont {Rost}, \citenamefont {Schlipf}, \citenamefont {Schindlmayr}, \citenamefont {Winkelmann},\ and\ \citenamefont {Bl\"ugel}}]{fleurCode}%
  \BibitemOpen
  \bibfield  {author} {\bibinfo {author} {\bibfnamefont {D.}~\bibnamefont {Wortmann}}, \bibinfo {author} {\bibfnamefont {G.}~\bibnamefont {Michalicek}}, \bibinfo {author} {\bibfnamefont {N.}~\bibnamefont {Baadji}}, \bibinfo {author} {\bibfnamefont {M.}~\bibnamefont {Betzinger}}, \bibinfo {author} {\bibfnamefont {G.}~\bibnamefont {Bihlmayer}}, \bibinfo {author} {\bibfnamefont {J.}~\bibnamefont {Br\"oder}}, \bibinfo {author} {\bibfnamefont {T.}~\bibnamefont {Burnus}}, \bibinfo {author} {\bibfnamefont {J.}~\bibnamefont {Enkovaara}}, \bibinfo {author} {\bibfnamefont {F.}~\bibnamefont {Freimuth}}, \bibinfo {author} {\bibfnamefont {C.}~\bibnamefont {Friedrich}}, \bibinfo {author} {\bibfnamefont {C.-R.}\ \bibnamefont {Gerhorst}}, \bibinfo {author} {\bibfnamefont {S.}~\bibnamefont {Granberg~Cauchi}}, \bibinfo {author} {\bibfnamefont {U.}~\bibnamefont {Grytsiuk}}, \bibinfo {author} {\bibfnamefont {A.}~\bibnamefont {Hanke}}, \bibinfo {author} {\bibfnamefont {J.-P.}\ \bibnamefont {Hanke}}, \bibinfo {author}
  {\bibfnamefont {M.}~\bibnamefont {Heide}}, \bibinfo {author} {\bibfnamefont {S.}~\bibnamefont {Heinze}}, \bibinfo {author} {\bibfnamefont {R.}~\bibnamefont {Hilgers}}, \bibinfo {author} {\bibfnamefont {H.}~\bibnamefont {Janssen}}, \bibinfo {author} {\bibfnamefont {D.~A.}\ \bibnamefont {Kl\"uppelberg}}, \bibinfo {author} {\bibfnamefont {R.}~\bibnamefont {Kovacik}}, \bibinfo {author} {\bibfnamefont {P.}~\bibnamefont {Kurz}}, \bibinfo {author} {\bibfnamefont {M.}~\bibnamefont {Lezaic}}, \bibinfo {author} {\bibfnamefont {G.~K.~H.}\ \bibnamefont {Madsen}}, \bibinfo {author} {\bibfnamefont {Y.}~\bibnamefont {Mokrousov}}, \bibinfo {author} {\bibfnamefont {A.}~\bibnamefont {Neukirchen}}, \bibinfo {author} {\bibfnamefont {M.}~\bibnamefont {Redies}}, \bibinfo {author} {\bibfnamefont {S.}~\bibnamefont {Rost}}, \bibinfo {author} {\bibfnamefont {M.}~\bibnamefont {Schlipf}}, \bibinfo {author} {\bibfnamefont {A.}~\bibnamefont {Schindlmayr}}, \bibinfo {author} {\bibfnamefont {M.}~\bibnamefont {Winkelmann}},\ and\ \bibinfo
  {author} {\bibfnamefont {S.}~\bibnamefont {Bl\"ugel}},\ }\href {https://doi.org/10.5281/zenodo.7576163} {\bibinfo {title} {{FLEUR}}},\ \bibinfo {howpublished} {Zenodo} (\bibinfo {year} {2023})\BibitemShut {NoStop}%
\bibitem [{\citenamefont {Mu}\ \emph {et~al.}(2019)\citenamefont {Mu}, \citenamefont {Hermann}, \citenamefont {Gorsse}, \citenamefont {Zhao}, \citenamefont {Manley}, \citenamefont {Fishman},\ and\ \citenamefont {Lindsay}}]{Mu2019}%
  \BibitemOpen
  \bibfield  {author} {\bibinfo {author} {\bibfnamefont {S.}~\bibnamefont {Mu}}, \bibinfo {author} {\bibfnamefont {R.~P.}\ \bibnamefont {Hermann}}, \bibinfo {author} {\bibfnamefont {S.}~\bibnamefont {Gorsse}}, \bibinfo {author} {\bibfnamefont {H.}~\bibnamefont {Zhao}}, \bibinfo {author} {\bibfnamefont {M.~E.}\ \bibnamefont {Manley}}, \bibinfo {author} {\bibfnamefont {R.~S.}\ \bibnamefont {Fishman}},\ and\ \bibinfo {author} {\bibfnamefont {L.}~\bibnamefont {Lindsay}},\ }\bibfield  {title} {\bibinfo {title} {Phonons, magnons, and lattice thermal transport in antiferromagnetic semiconductor {MnTe}},\ }\href {https://doi.org/10.1103/PhysRevMaterials.3.025403} {\bibfield  {journal} {\bibinfo  {journal} {Phys. Rev. Mater.}\ }\textbf {\bibinfo {volume} {3}},\ \bibinfo {pages} {025403} (\bibinfo {year} {2019})}\BibitemShut {NoStop}%
\bibitem [{\citenamefont {Mazin}(2023)}]{Mazin2023}%
  \BibitemOpen
  \bibfield  {author} {\bibinfo {author} {\bibfnamefont {I.~I.}\ \bibnamefont {Mazin}},\ }\bibfield  {title} {\bibinfo {title} {Altermagnetism in {MnTe}: Origin, predicted manifestations, and routes to detwinning},\ }\href {https://doi.org/10.1103/PhysRevB.107.L100418} {\bibfield  {journal} {\bibinfo  {journal} {Phys. Rev. B}\ }\textbf {\bibinfo {volume} {107}},\ \bibinfo {pages} {L100418} (\bibinfo {year} {2023})}\BibitemShut {NoStop}%
\end{thebibliography}
\end{document}